\documentclass{article}
\usepackage{algorithm}
\usepackage{algpseudocode}
\usepackage{amsmath,amsthm}
\usepackage{amsfonts}
\usepackage{array}
\usepackage{graphicx}
\usepackage{blkarray}
\usepackage{soul}
\usepackage{amssymb}
\usepackage{comment}
\usepackage[table]{xcolor}
\usepackage{tabularray}
\usepackage[margin=1in]{geometry}
\usepackage{natbib}

\newtheorem{example}{Example}
\newtheorem{proposition}{Proposition}
\usepackage{relsize}
\usepackage{authblk}
\newcommand{\mstar}{ \mathlarger{\star}}

\begin{document}
\title{Gaussian Process with dissolution spline kernel}

\author[1]{Fiona Murphy}
\author[2]{Marina Navas Bachiller}
\author[3]{Deirdre M. D'Arcy}
\author[4]{Alessio Benavoli}

\affil[1, 4]{School of Computer Science and Statistics, Trinity College Dublin, Ireland}
\affil[2, 3]{School of Pharmacy and Pharmaceutical Sciences, Trinity College Dublin, Ireland\\
SSPC, The Science Foundation Ireland Research Centre for Pharmaceuticals, Trinity College Dublin, Ireland.}

\maketitle

\begin{abstract}
In-vitro dissolution testing is a critical component in the quality control of manufactured drug products. The $\mathrm{f}_2$ statistic is the standard for assessing similarity between two dissolution profiles. However, the $\mathrm{f}_2$ statistic  has known limitations: it lacks an uncertainty estimate, is a discrete-time metric, and is a biased measure, calculating the differences between profiles at discrete time points. To address these limitations, we propose a Gaussian Process (GP) with a dissolution spline kernel for dissolution profile comparison. The dissolution spline kernel is a new spline kernel using logistic functions as its basis functions, enabling the GP to capture the expected monotonic increase in dissolution curves. This results in better predictions of dissolution curves. This new GP model reduces bias in the $\mathrm{f}_2$ calculation by allowing predictions to be interpolated in time between observed values, and provides uncertainty quantification. We assess the model's performance through simulations and real datasets, demonstrating its improvement over a previous GP-based model introduced for dissolution testing. We also show that the new model can be adapted to include dissolution-specific covariates. Applying the model to real ibuprofen dissolution data under various conditions, we demonstrate its ability to extrapolate curve shapes across different experimental settings.
\end{abstract}

\section{Introduction}
In the pharmaceutical industry, in-vitro dissolution or drug release testing is a critical analytical procedure used as a quality control measure for dosage forms. This testing primarily measures the rate and extent to which the Active Pharmaceutical Ingredient (API) is released from a drug product under standardised conditions. Understanding this release profile is essential, as it impacts the bioavailability of the drug, influencing its absorption into the bloodstream and, ultimately, its therapeutic efficacy. Dissolution  testing can be undertaken at several stages during the drug development process, such as a) for the characterisation of the API, b) during formulation development and optimisation and c) in routine quality control. When changes are made in the production process of pharmaceutical drug products, it must be demonstrated that quality persists from the pre-change product to the post-change product. In some cases, demonstration of similarity of dissolution profiles can result in the waiving of the requirement for clinical studies to demonstrate pre- and post-change (or test and reference product) equivalence. Therefore, demonstration of similarity of dissolution profiles is often a critical and valuable quality attribute. In order to carry out a dissolution study to demonstrate product similarity, the pre-change product is used as a reference, where product from the reference set is placed in a dissolution apparatus, and its percent dissolution is subsequently measured over time. This procedure is repeated for the post-change product, or test set. The dissolution curves of the reference and test sets are then compared to determine whether the changes in the production process have resulted in significant changes in the quality of the test set.

The standard method for comparing dissolution curves, as recommended by regulatory agencies such as \citet{fda_1997}, is the calculation of the similarity factor or $\mathrm{f}_2$ statistic, originally proposed by \citet{moore_flanner}. However, this similarity factor has several known limitations, which we will discuss in the next section. In particular, its validity is restricted to narrow conditions due to a lack of uncertainty quantification and an inherent bias determined by its discrete-time nature.

Various alternative approaches have been proposed to address each of its limitations. For instance, certain methods have been introduced to address the inherent bias through a bootstrap $f_2$ approach \citep{shah_1998, ma_assessment_2000}. Other proposed methods for dissolution comparison include multivariate-distance based methods \citep{tsong_statistical_1996, saranadasa_2001, saranadasa_multivariate_2005}, statistical testing and ANOVA-based methods \citep{mauger_1986, polli1997methods, wang2016statistical}, model-independent approaches \citep{podczeck1993comparison, polli1997methods}, and model-dependent approaches \citep{polli1997methods, yuksel2000comparison, adams_evaluation_2001, adams_non-linear_2002}. Recently, a Continuous Time Gaussian Process (CTGP) model, was introduced by \citet{pourmohamad_gaussian_2022} to estimate dissolution curves. Given that the dissolution process is continuous, employing a GP to model the dissolution curve allows for interpolation between discrete time points where data has been collected. Moreover, a GP provides a measure of uncertainty for the estimated dissolution profiles, facilitating a probabilistic assessment of the similarity between two curves. However, a drawback of this approach is that it does not account for the typically monotonic shape of a dissolution curve (the amount of drug released into the medium generally increases over time).

In this paper, we introduce a novel GP model tailored to capture the expected monotonic increase in the dissolution curve, which we term the \textit{Logistic Spline Gaussian Process} (LSGP) model. The LSGP model is derived by employing a q-fold Wiener kernel, $W_q(t, t')$ as the covariance function and substituting the linear function of time 
$t$ with a logistic function. This adaptation yields a \textit{dissolution spline kernel}, that is a spline kernel with logistic functions as the underlying basis functions.

Sampling from this posterior then enables us to derive a posterior distribution for the $\mathrm{f}_2$-similarity index, facilitating more reliable decisions regarding the similarity between the test and reference dissolution curves.
\begin{comment}
    In the original $\mathrm{f}_2$ equation, the potential bias of using the mean dissolution measurements across repeated samples is noted by \citet{fda_1997}, contributing to the recommendation that variation across samples at the same time point be less than $20\%$ for early data points, and $10\%$ for subsequent data points. Rather than extending this biased approach, which requires additional specifications for the datasets, the LSGP model uses predictions of the actual dissolution curves, rather than the mean of the GP, to reduce bias in the calculated $\mathrm{f}_2$ statistic. \hl{The LSGP model also utilizes informative priors based on expected parameter values given typical dissolution curve patterns.}
\end{comment}
By using simulations and real dissolution data,  we demonstrate that LSGP provides more reliable estimates of the dissolution curve, with reduced uncertainty compared to CTGP.

Furthermore, the LSGP model can be used to provide an alternative hypothesis testing procedure to assess the similarity between dissolution curves. For instance, we use LSGP to derive a Bayesian counterpart to the Multivariate Statistical Distance (MSD), recommended by \citet{fda_1997} for datasets that do not meet the validity criteria required for calculating the $\mathrm{f}_2$ similarity criterion.
Furthermore, we demonstrate that LSGP can be modified to incorporate covariates specific to dissolution experiments, such as the composition and volume of the medium used in in-vitro dissolution testing, the medium's viscosity, or the apparatus's medium agitation velocity. We apply the model to real data on the dissolution of ibuprofen under various experimental conditions, showing its ability to extrapolate the dissolution curve shape across different settings.

\section{Background on drug dissolution profile comparison}
The standard method for comparing dissolution curves, as recommended by regulatory agencies such as  \citet{fda_1997}, is the calculation of the similarity factor or $\mathrm{f}_2$ statistic, originally proposed by \citet{moore_flanner}. The $\mathrm{f}_2$ statistic is a measure of the similarity between the percent dissolution of two dissolution profiles over time.
To be more specific, let us consider a dissolution study with $n_R$ solid oral dosage forms from the reference product and $n_T$ dosage forms from the test product.
For a given
product, either reference or test, let $y^{(j)}_{\mstar}(t_i)$ denote the observed cumulative percentage dissolved for dosage unit $j$ at sampling time $t_i$ for product $\mstar$, where $i=1,\dots,p$, $j=1,\dots,n_\mstar$  and $\mstar \in \{R,T\}$.
For each product, the $n_\mstar$ dissolution time-series are averaged, resulting in $\overline{y}_{\mstar}(t) = \frac{1}{n_\mstar}\sum_{\substack{j=1}}^{n_\mstar} y^{(j)}_{\mstar}(t)$, which for $t=t_1,\dots,t_p$ is known as the average dissolution profile.

After calculating the average dissolution profile for the reference and test products, the $\mathrm{f}_2$ statistic is computed as follows: 

\begin{equation}
\label{eq:f2}
\mathrm{f}_2 = 50\log_{10}\Bigg(100 \Bigg(1 + \frac{1}{p} \sum\limits_{\substack{i=1}}^p w_i(\overline{y}_{R}(t_i) - \overline{y}_{T}(t_i))^2 \Bigg)^{-1/2}\Bigg) ,
\end{equation}

\noindent where $w_i$ are weights assigned at each time point to reflect a possibly heterogeneous dissolution profile. In the rest of the paper, we assume that the weights are set to 1, as is typically done.
If two dissolution profiles are identical, the $\mathrm{f}_2$ statistic will have a value of $50\log_{10}(100) = 100$ as there will be no difference between the average percent dissolution of the reference and test sets. However, it is acknowledged that this is highly unlikely to occur because of measurement noise, and there can be on average $10\%$ variation in dissolution measurements at the same time point across different samples \citep{shah_1998}. As a result, a threshold of 50 is set such that $\mathrm{f}_2 \geq 50$ defines similar dissolution profiles.

\subsection{Validity criteria}
In order to use the $\mathrm{f}_2$ statistic, regulatory agencies have established certain criteria for the dissolution data collected. If these criteria are not met, the data is not recommended for use in calculating the $\mathrm{f}_2$ statistic for the purpose of comparing dissolution profiles. The \citet{fda_1997} reports the following criteria for determining the validity of using a dataset in the calculation of the $\mathrm{f}_2$ statistic:

\begin{enumerate}
\item Percent dissolution measurements should be collected for a minimum of three time points.
\item Percent dissolution measurements of 12 individual units for each of the reference and test sets should be obtained to establish the respective dissolution profiles.
\item Time points at which measurements are taken should be identical across reference and test sets.
\item A maximum of only one measurement above 85$\%$ dissolution for each of the reference and test sets should be considered.
\item The percent coefficient of variation should not exceed 20\% at early time points, which reduces to 10\% for all subsequent time points.
\end{enumerate}

\noindent Some datasets that fail to meet these guidelines may still be valid for the purpose of comparing dissolution profiles. The exclusion of these otherwise valid datasets is one limitation of the $\mathrm{f}_2$ statistic. 

\subsection{Lack of uncertainty quantification} One of the main drawbacks of the $\mathrm{f}_2$ statistic is the lack of uncertainty quantification. In its original formulation, specified in \eqref{eq:f2}, the $\mathrm{f}_2$ statistic produces a point measure for declaring dissolution profiles as similar or dissimilar. However, when taking into account variability in the calculation, the range of possible $\mathrm{f}_2$ values may result in a differing decision than the point measure that \eqref{eq:f2} yields. Several methods have been designed to address this absence of uncertainty quantification, for instance by using bootstrapped confidence intervals \citep{shah_1998,  ma_assessment_2000}, where the performance of parametric versus non-parametric bootstrapping methods was assessed by \citet{m_islam_bootstrap_2018}. \citet{jamil_application_2024} recently noted that the bootstrap $\mathrm{f}_2$ method provides a more conservative assessment of dissolution profile similarity than the traditional $\mathrm{f}_2$ statistic.
Other approaches to uncertainty quantification are based on  Bayesian probabilistic models \citep{novick_2015,ng2018gamma,pourmohamad_gaussian_2022}. We will discuss the model presented in \citep{pourmohamad_gaussian_2022} in Section \ref{sec:CTGP}. 

\subsection{Inherent bias}
Another limitation of the $\mathrm{f}_2$ metric is that it is a biased estimator of $\mathrm{f}_{2, integral}$, the value of $\mathrm{f}_2$ given the true underlying function of the dissolution curve. This was first observed by \citet{shah_1998}, and then elaborated upon by \citet{pourmohamad_gaussian_2022} with an example of $\mathrm{f}_2$ as a biased estimator.

\begin{figure}
    \centering
    \includegraphics[width=0.45\textwidth]{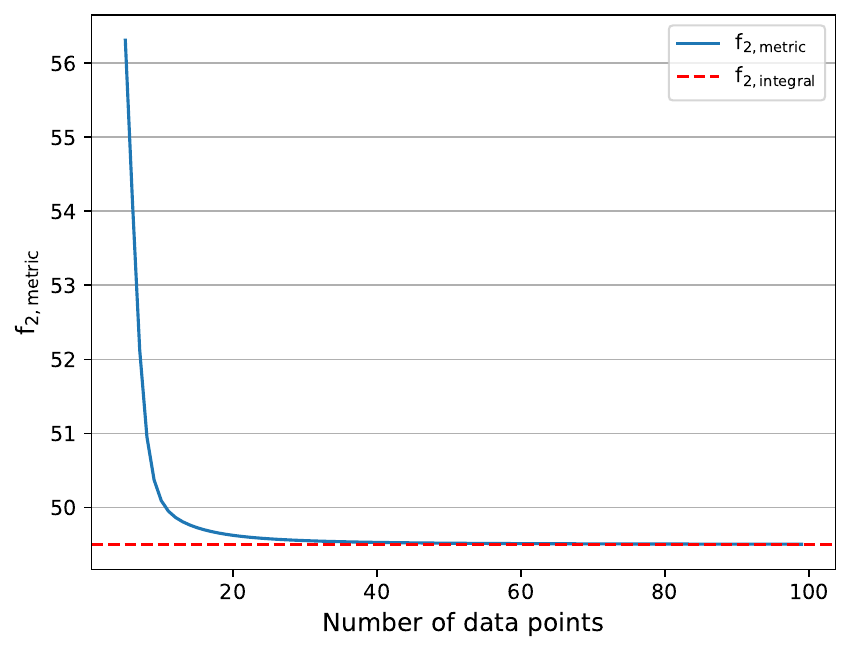}
     \includegraphics[width=0.45\textwidth]{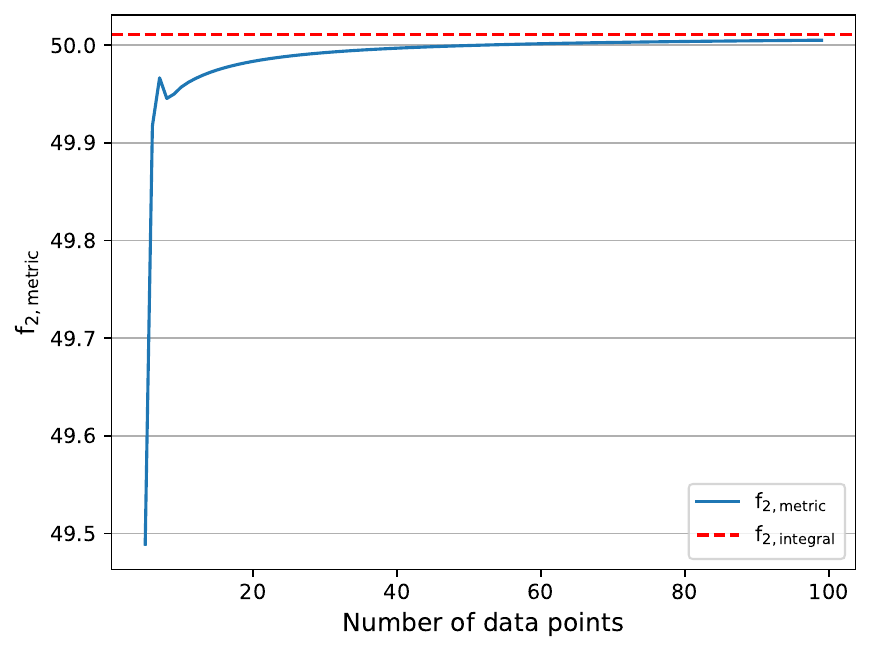}
    \caption{(Left) The blue line shows the value of $\mathrm{f}_{2,metric}$ as a function of the number of data points $p$ overestimating the value of $\mathrm{f}_{2, integral} = 49.50$ (red dashed line). (Right) The $\mathrm{f}_{2,metric}$ is shown in blue for a case where it underestimates $\mathrm{f}_{2, integral} = 50.01$ (red dashed line).}
    \label{fig:bias}
\end{figure}

Using the true underlying functions of the dissolution curves for the reference and test products, $f_R(t)$ and $f_T(t)$ for $t = 1, 
\dots, p$, the $\mathrm{f}_2$ metric can be written as 
\begin{equation}
    \mathrm{f}_{2,metric} = 50\log_{10} \left(100 \left(1 + \Delta\right)^{-1/2} \right),
\end{equation}
where $
\Delta= \tfrac{1}{p} \sum_{\substack{i=1}}^p (f_{R}(t_i) - f_{T}(t_i))^2$. Since the dissolution process is continuous in time, assuming equally spaced sampling times $t_1,\dots,t_p$, then  $\tfrac{(t_p-t_1)p}{p-1}\Delta$ corresponds to a left-rule Riemann sum approximation, that is:
\begin{equation}
\tfrac{(t_p-t_1)p}{p-1}\Delta= \frac{t_p-t_1}{p-1}\sum\limits_{\substack{i=1}}^p (f_{R}(t_i) - f_{T}(t_i))^2 \approx\int\limits_{t_1}^{t_p}(f_R(t) - f_T(t))^2dt,
\end{equation}
whose worst-case approximation error is $\tfrac{M(t_p-t_1)^2}{2(p-1)}$,  where $M$ is the maximum value of $|\tfrac{d}{dt}(f_R(t) - f_T(t))|$
over the interval $[t_1,t_p]$.
Therefore, $\mathrm{f}_{2,metric}$ is a biased approximation of
\begin{equation}
    \mathrm{f}_{2,integral} = 50\log_{10} \left(100 \left(1 + \frac{1}{t_p-t_1}\int\limits_{t_1}^{t_p}(f_R(t) - f_T(t))^2dt\right)^{-1/2} \right) .
\end{equation}
To provide a concrete understanding of the value of this bias, we computed the $\mathrm{f}_2$ metric for an increasing number $p$ of equally spaced sampling points, using a logistic function as a dissolution profile:

\begin{equation}
    f_\mstar(t)=\frac{\alpha_1^\mstar}{1 + \alpha_2^\mstar e^{-\beta^\mstar t}} ~~\text{ for }~~t=t_1,\dots t_p, ~~\mstar \in \{R,T\}, 
\end{equation}

\noindent with $t_1 = 10$ to $t_p = 60$. Figure \ref{fig:bias} shows the $\mathrm{f}_{2,metric}$ calculated using \eqref{eq:f2}, for a  number of sampling points $p=5,\dots,100$. In particular, for the plot on the left we used the parameters $\alpha^R_1 = 70.91,\alpha^R_2 = 100, \beta^R = 0.403$ and $\alpha^T_1 = 70.69, \alpha^T_2 = 99.98, \beta^T = 0.292$, resulting in $\mathrm{f}_{2, integral}=49.5$ (red dashed line). The blue line shows the value of the $\mathrm{f}_{2,metric}$ as a function of $p$ showing  an overestimate of the $\mathrm{f}_{2, integral}$. %
For $5 \leq p \leq 10$, the value of the $\mathrm{f}_{2,metric}$ is above $50$ and thus would result in the dissolution profiles being declared similar, although the true value of the $\mathrm{f}_{2, integral}$ should result in a decision of dissimilarity. 

For the plot on the right we used the parameters $\alpha^R_1 = 60.55, \alpha^R_2 = 90, \beta^R = 0.228$ and $\alpha^T_1 = 75, \alpha^T_2 = 100, \beta^T = 0.19$  leading to $\mathrm{f}_{2,integral}=50.01$ (red dashed line). For $5 \leq p \leq 52$, the value of the $\mathrm{f}_{2,metric}$ is below $50$. This would result in a decision of dissimilarity, when the true decision should be the opposite.

This motivated \citet{pourmohamad_gaussian_2022} to introduce the use of Gaussian Processes (GPs) to estimate the dissolution curves directly. The use of GPs allows for the dissolution process to be interpolated between discrete data points and provides an estimate of $\mathrm{f}_{2, integral}$ which is more accurate than the value computed using the $\mathrm{f}_{2, metric}$. We will discuss this approach in more detail in Section \ref{sec:CTGP}.

\begin{comment}
uses a hierarchical Gaussian Process (GP) model to generate a probability distribution of the dissolution curves given observed data. Samples of the curves are drawn from the psoterior probability distribution via MCMC sampling in order to calculate the mean and variance of the $\mathrm{f}_2$ statistic in order to assess uncertainty. Additionally, the use of the GP allows for the dissolution process to be interpolated between discrete data points, which more closely approximates the value of $\mathrm{f}_{2, True}$ than the value of $\mathrm{f}_2$ which uses a limited amount of observed data in its calculation.
\end{comment}

\subsection{Multivariate Statistical Distance (MSD)}
\label{sec:MSD}

An alternative approach recommended by the \citet{fda_1997} for a dataset which violates the validity criteria needed for the calculation of the $\mathrm{f}_2$ similarity is to evaluate the Multivariate Statistical Distance (MSD) between the dissolution data of the reference and test products \citep{tsong_statistical_1996}. The MSD is a measurement of the Mahalanobis distance between the reference and test data and is computed as: 

\begin{equation}
d_{R,T} = \sqrt{(\overline{\bf y}_{R} - \overline{\bf y}_{T})^\top S^{-1}(\overline{\bf y}_{R} - \overline{\bf y}_{T})} ,
\end{equation}
 where $\overline{\bf y}_{\mstar} = \frac{1}{n}\sum_{j=1}^n {\bf y}^{(j)}_{\mstar}$ with ${\bf y}^{(j)}_{\mstar}=[ y_{\mstar}^{(j)}(t_{1}),y_{\mstar}^{(j)}(t_{2}),\dots,y_{\mstar}^{(j)}(t_{p})]^\top$  being the vector including  the cumulative percent dissolution data for unit $j$ at times $t_1,\dots,t_p$. The matrix $S$ is the pooled covariance matrix,  defined as $S = 0.5(S_R + S_T)$, where $S_\mstar= \frac{1}{n-1}\sum_{j=1}^n({\bf y}_{\mstar}^{(j)}-\overline{\bf y}_{\mstar})({\bf y}_{\mstar}^{(j)}-\overline{\bf y}_{\mstar})^\top$ is the sample covariance matrix. 
 
 As described in \citep{tsong_statistical_1996}, the procedure for comparing dissolution sets with the MSD is based on $10\%$ variation in dissolution measurements being considered acceptable. First, a maximum similarity limit, or the maximum distance that can exist between the reference and test products to be declared similar, is calculated via

\begin{equation}
\label{msd_limit}
d_{limit} = \sqrt{{\bf v}^\top S^{-1} {\bf v}} ,
\end{equation}

\noindent where ${\bf v} = [10, 10, \dots, 10]^\top$ is a $p \times 1$ (number of time points) vector of the maximum variation accepted between dissolution measurements. Then, a $90\%$ confidence region is evaluated for $d_{R,T}$, the MSD between the reference and test sets,
\begin{equation}
    \label{msd_original_cr}
        CR = K\left(\left((\boldsymbol{\mu}_R - \boldsymbol{\mu}_T) - (\overline{\bf y}_R-\overline{\bf y}_T)\right)^\top S^{-1}\left((\boldsymbol{\mu}_R - \boldsymbol{\mu}_T) - (\overline{\bf y}_R - \overline{\bf y}_T)\right)\right) \leq F_{p, 2n - p - 1, 1 - \alpha} ,
\end{equation}
\begin{equation}
    K = \frac{n^2}{2n} \Bigg (\frac{2n - p - 1}{(2n - 2)p} \Bigg) ,
\end{equation}

\noindent where ${\mu}_\mstar = [\mu_\mstar(t_1), \mu_\mstar(t_2), \dots, \mu_\mstar(t_p)]$ are the population means of the reference and test products such that $\mu_R - \mu_T$ represents the possible differences between the reference and test products at each time point. The number of individual percent dissolution curves from the reference or test products is represented by $n$, and $K$ is a scaling factor. 

The confidence interval for the true difference between the reference and test products is determined by identifying the two points contained within the CR that are the minimum and maximum distances from the origin (point at which there is no difference between the reference and test products). Evaluating the Mahalanobis distances from the origin at these two points results in the confidence interval 
\begin{equation}
    (d_{R, T}^{l}, d_{R,T}^u) ,
\end{equation}
which can be compared to $d_{limit}$ to determine similarity. If the upper bound of the confidence interval is less than or equal to the maximum similarity limit, $d_{R,T}^u \leq d_{limit}$, the dissolution profiles are determined to be similar. 

Several works have proposed $T^2$ tests for equivalence applied to the Mahalanobis distance calculation described in \citet{tsong_statistical_1996} in order to use statistical hypothesis testing to make a decision on the similarity of the dissolution profiles \citep{hoffelder_comparison_2018, saranadasa_multivariate_2005, saranadasa_2001}. The MSD method has acknowledged limitations, such as the possibility of having variation between the dissolution profiles at one time point greater than the $10\%$ variability limit, yet lower than the global similarity limit \citep{ema_msd_limit_report}. In this case, the MSD would not consider the large difference at one time point to be significant in determining dissimilarity. Differences in release at some time points and not others could indicate a differently-shaped dissolution profile. The shape of the dissolution profile is directly relevant to the release kinetics of the dosage form, and a change in shape could indicate a different mechanism of release of the API, in particular for prolonged- or extended-release dosage forms. Understanding the mechanism of API release from a dosage form is a core prerequisite for determining appropriately discriminating dissolution test conditions, pointing to the underlying importance of the shape of the dissolution profile.  

It is noted that Mahalanobis distance-based methods for dissolution profile comparison are not recommended by the \citet{ema_msd_limit_report}. Additionally, \citet{mangas-sanjuan_assessment_2016} advocate the usage of the bootstrap $\mathrm{f}_2$ procedure over Mahalanobis distance-based methods of comparison for datasets with large variability, finding that the Mahalanobis distance methods exhibited lower specificity than the bootstrap $\mathrm{f}_2$. 

\section{Modelling dissolution curves}
Since the dissolution process is continuous, using a GP to estimate a dissolution curve allows for the interpolation of values between discrete time points where data has been collected. This approach results in a more accurate dissimilarity assessment as it eliminates the bias inherent in the original $\mathrm{f}_2$ metric, which relies on discrete measurements for a continuous process. Moreover, GPs provide a measure of uncertainty of their estimates, enabling more informed decision-making.
The use of GPs for   modelling a dissolution curve was proposed by \cite{pourmohamad_gaussian_2022}. We will review their  Continuous Time Gaussian Process (CTGP) model in the next section. 

The CTGP model does not incorporate a known characteristic of dissolution testing: the generally monotonic
nature of cumulative dissolution curves. To better capture the behavior of dissolution curves, in Section \ref{sec:ourmodel} we propose a new model that incorporates this knowledge of the monotonicity of the dissolution process by using what we called a \textit{dissolution spline kernel}.

\subsection{Continuous Time Gaussian Process (CTGP)}
\label{sec:CTGP}

 The CTGP model is  defined as follows: 
\begin{align}
\label{eq:CTGP1}
    y^{(j)}(t) &\sim GP (f(t), k(t, t';\boldsymbol{\theta})),\ \ j = 1, \dots, n, \\
\label{eq:CTGP2}
    f(t) &\sim GP(0, h(t, t';\boldsymbol{\theta})),
\end{align}

\noindent where $y^{(j)}(t)$, which is the observed cumulative percentage dissolved
for dosage unit $j$ at time $t$,  is assumed to be GP distributed with   mean function $f(t)$ and kernel $ k(t, t';\boldsymbol{\theta})$ for every   $j = 1, \dots, n$. The  function $f(t)$ is also assumed to be GP distributed with zero mean function and kernel $h(t, t';\boldsymbol{\theta})$. The vector $\boldsymbol{\theta}=[\sigma,\tau,\phi,\psi]^\top$ includes the hyperparameters of the kernels:

\begin{equation}
 k(t, t';\boldsymbol{\theta})=\sigma ^{2}\left(1+{\tfrac {{\sqrt {3}}|t-t'|}{\phi }}\right)e^{ \left(-{\frac {{\sqrt {3}}|t-t'|}{\phi }}\right)}, ~~~  h(t, t';\boldsymbol{\theta})=\tau^{2}e^{-\frac{(t-t')^2}{2\psi^2}},
\end{equation}
which are a  Mat{\'e}rn $3/2$ and, respectively, a Square Exponential kernel. In \citep{pourmohamad_gaussian_2022}, for fixed  $\phi,\psi$, the posterior for the hyperparameters $\sigma^2,\tau^2$ is computed  via Gibbs sampling assuming independent inverse gamma priors. The hyperparameters $\phi,\psi$ are kept either fixed or inferred via Metropolis-Hastings.

\begin{comment}
\indent The predictive posterior mean function computed at time points $\mathbf{t}^*=[t^*_1,\dots,t^*_r]^\top$ is: 

\begin{equation}
    \boldsymbol{\mu_{t^*}} | \boldsymbol{\mu} \sim N(\boldsymbol{h}_{t^*}^T\boldsymbol{H}^{-1}\boldsymbol{\mu}, \boldsymbol{H_{t^*}} - \boldsymbol{h}_{t^*}^T\boldsymbol{H}^{-1}\boldsymbol{h}_{t^*}) ,
\end{equation}

\noindent where $\boldsymbol{H}_{t^*}$ is a covariance matrix of the unobserved time points and $\boldsymbol{h}_{t^*}$ is the cross covariance between unobserved and observed time points. Additionally, inverse gamma priors are placed on the parameters of the covariance functions for both levels of the hierarchical GP via 

\begin{eqnarray}
    \sigma^2 | \boldsymbol{\mu}, \boldsymbol{Y} &\sim& IG \Bigg( \alpha + np/2, \beta + \frac{1}{2}\sum^n_{i=1}(\boldsymbol{Y}_i - \boldsymbol{\mu})'\boldsymbol{K}(\phi)^{-1}(\boldsymbol{Y}_i - \boldsymbol{\mu})\Bigg)\\
    \tau^2 | \boldsymbol{\mu} &\sim& IG \Bigg (\alpha + p/2, \beta + \frac{1}{2}\boldsymbol{\mu}'\boldsymbol{H}(\psi)^{-1}\boldsymbol{\mu}\Bigg) ,
\end{eqnarray}
\end{comment}
Two independent CTGP models are fitted for the reference and test product data. Posterior samples for $f_R(t),f_T(t)$ from these two models are used  to approximate   the $\mathrm{f}_{2,integral}$.  Specifically, $m$ posterior samples are generated at the  time points $t_1,\dots,t_p,t_1^*,\dots,t_r^*$ (where $t_1^*,\dots,t_r^*$ are the time points at which the GP interpolates the dissolution process) and the vectors $\mathbf{f}^{(i)}_\mstar=[f^{(i)}_\mstar(t_1),\dots,f^{(i)}_\mstar(t_p),f^{(i)}_\mstar(t_1^*),\dots,f^{(i)}_\mstar(t_r^*)]^\top$ for $\mstar \in \{R,T\}$ are used to approximate $\mathrm{f}_{2,integral}$  as follows:
\begin{equation}
    \mathrm{f}^{(i)}_2 = 50\log_{10}\left(100\left(1 + \frac{1}{p + r}||\mathbf{f}^{(i)}_{R} - \mathbf{f}^{(i)}_{T}||^2\right)^{-1/2}\right),
\end{equation}
for each sample $i=1,\dots,m$. The posterior samples $\{\mathrm{f}^{(i)}_2\}_{i=1}^m$ are used for hypothesis testing regarding the dissimilarity of the dissolution curves through the $\mathrm{f}_{2,metric}$. \cite{pourmohamad_gaussian_2022} also evaluate the test proposed by \cite{novick_2015}, where the test statistic
\begin{equation}
    \delta^{(i)} = \max(|\mathbf{f}^{(i)}_{R} - \mathbf{f}^{(i)}_{T}|)
\end{equation}
is calculated for each sample $i$ over $m$ posterior draws of $f_R(t)$ and $f_T(t)$. The maximum variation permitted for a reference and test product to be considered similar is $\delta < 15$, and so posterior draws of $f_R(t)$ and $f_T(t)$ allow a probabilistic determination of whether this additional condition for similarity is met.

\begin{example}
\label{ex:1}
To illustrate the model, we present in Figure \ref{fig:ctgp} the predictions generated by the CTGP model given two simulated datasets. The hyperparameters of the CTGP model used for the simulations were those reported in \citep{pourmohamad_gaussian_2022}. The variance parameters $\sigma^2$ and $\tau^2$ were estimated via Gibbs sampling using an inverse gamma prior with parameters $\alpha = 10$ and $\beta = 3$. The shape parameter of the kernel $k(t, t';\boldsymbol{\theta})$ is $1.5$, while the range parameter, $\phi$, is $5$. The range parameter, $\psi$, of the Square Exponential kernel for the mean function, $h(t, t';\boldsymbol{\theta})$, is $25$.

The first simulated dataset is generated by a logistic function defined as 
\begin{align}
    \label{logistic-fx}
    y_{\mstar}(t_i) = \frac{\alpha_1^\mstar}{1 + \alpha_2^\mstar e^{-\beta^\mstar t_i}} + \epsilon_i, ~~i \in \{1 \dots p\} ,
    \epsilon_i \sim N(0, \sigma^2) ,
\end{align}
where $y_{\mstar}(t_i)$ is the dissolution curve for the reference or test product, $\mstar \in \{R, T\}$, $\alpha_1^\mstar$, $\alpha_2^\mstar$, and $\beta^\mstar$ are parameters to be varied depending on the scenario, and $\epsilon_i$ is noise following a multivariate normal distribution with zero mean and variance $\sigma^2$. In this example, the parameters have values $\alpha_1^R=\alpha_1^T=100$, $\alpha_2^R=75$, $\alpha_2^T=80$, $\beta^R=0.19$, $\beta^T=0.215$, and $\sigma^2=5$. Data was generated at time points $t \in \{10, 20, 30, 40, 50, 60\}$ for twelve individual curves, upon which the model was trained and predictions were subsequently generated. The second simulated dataset is generated using the Higuchi model \citep{higuchi} 
\begin{align}
    \label{higuchi-fx}
    y_{\mstar}(t_i) = \sqrt{\omega_\mstar t_i} + \epsilon_i, ~~i \in \{1 \dots p\} , \epsilon_i \sim N(0, \sigma^2) , 
\end{align}
\noindent where the $\omega_\mstar$ is a parameter varied across each scenario. In this scenario, the parameter values were $\omega_R =110$, $ \omega_T=79$, and $\sigma^2=5$. 

The data relevant to the logistic and Higuchi models are representative of two very different release mechanisms for pharmaceutical dosage forms. The logistic model data represent the shape of a typical profile from a disintegrating immediate-release dosage form e.g. an immediate-release oral tablet. The Higuchi data, with release proportional to the square root of time is typical for diffusion of a drug through a matrix, which can occur in some controlled release tablets or, as in the reference used in this work, diffusion of a drug through an ointment base. Thus, these two examples present very different profile shapes which are nonetheless pertinent to different but relevant pharmaceutical release mechanisms.
\begin{figure}
\includegraphics[width=0.5\textwidth]{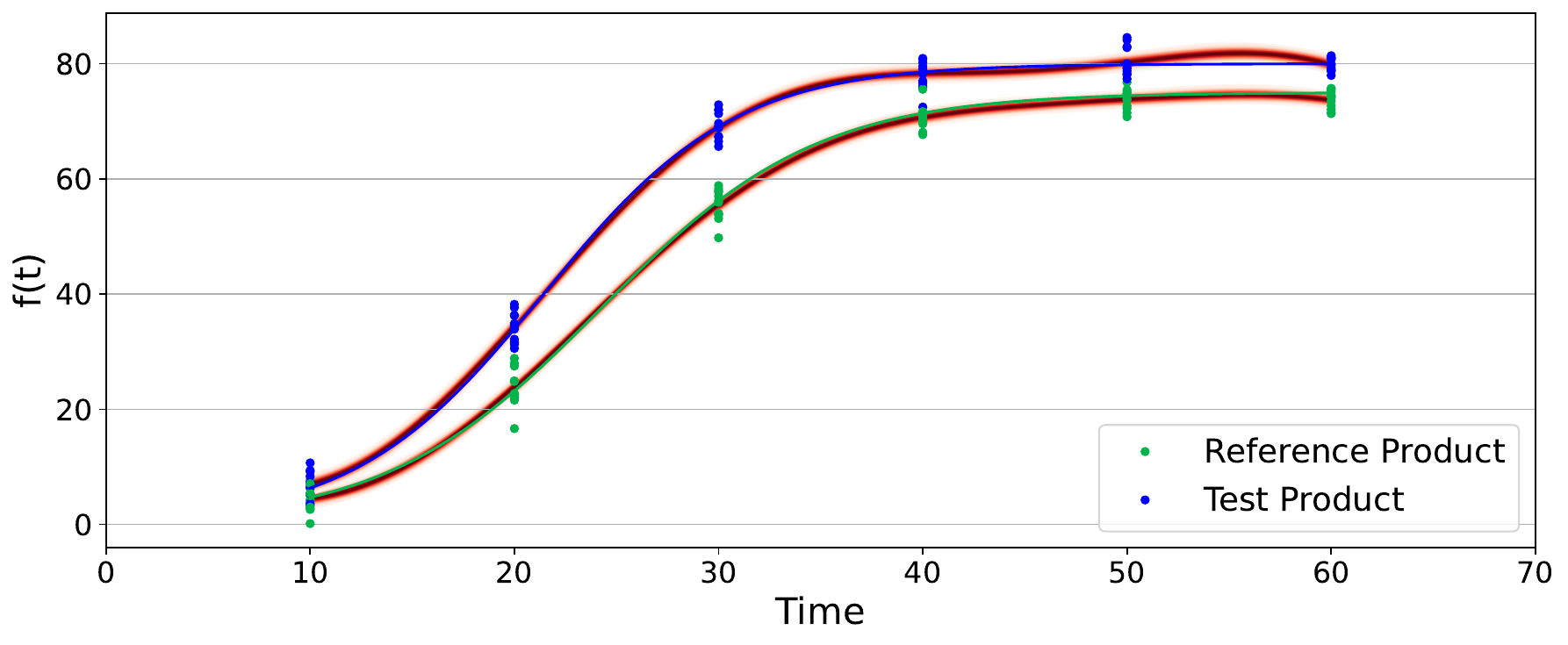}
\includegraphics[width=0.5\textwidth]{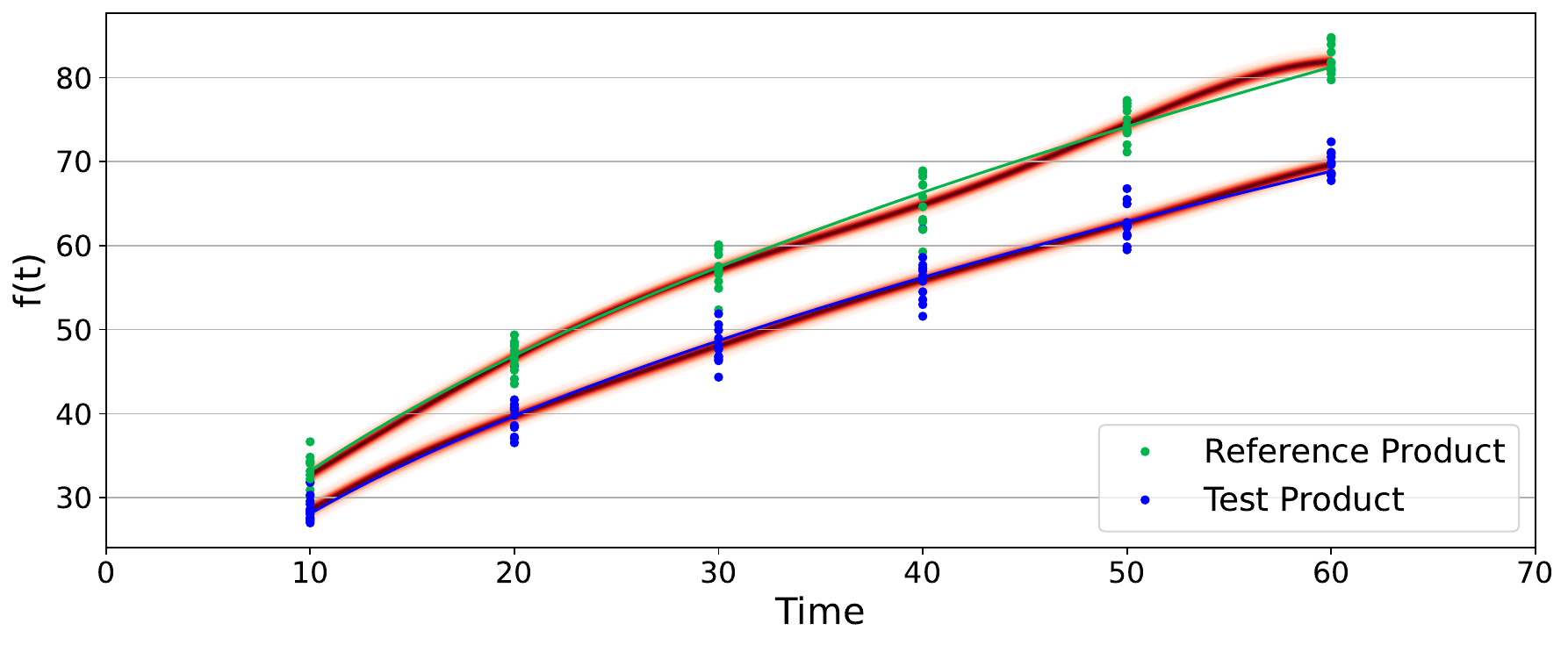}
\caption{Predictions $f_R(t)$, $f_T(t)$ and $95\%$ credible regions, shown in red, generated by the CTGP model for the reference and test product of the logistic data (left) and Higuchi data (right), where the true underlying functions are shown in blue or green. The blue and green points represent the noisy cumulative dissolution percentages obtained from the two data-generating models for $12$ dosage units (repetitions) at time $t=10,20,\dots,60$.}
\label{fig:ctgp}
\end{figure}

We additionally illustrate the CTGP model's prediction performance in Figure \ref{fig:ctgp-exc-30} by presenting the predictions for the logistic and Higuchi data, where the model was trained excluding the observations at time points $t=\{30, 40\}$. The true underlying functions are shown in blue and green. The dissolution curve is less accurately estimated by the CTGP model with the removal of the observations at two time points  and fails to capture its monotonicity.
This is because the posterior mean of a Gaussian process with a stationary kernel reverts to the prior mean (which is zero in CTGP) as the distance between data points increases, due to the diminishing influence of individual data points on the predictions. This property is modulated by the lengthscale parameter of the kernel, which determines how quickly the influence of a data point decays with distance.
In the next section, we will introduce a new model that overcomes this issue.
\begin{figure}
\includegraphics[width=0.5\textwidth]{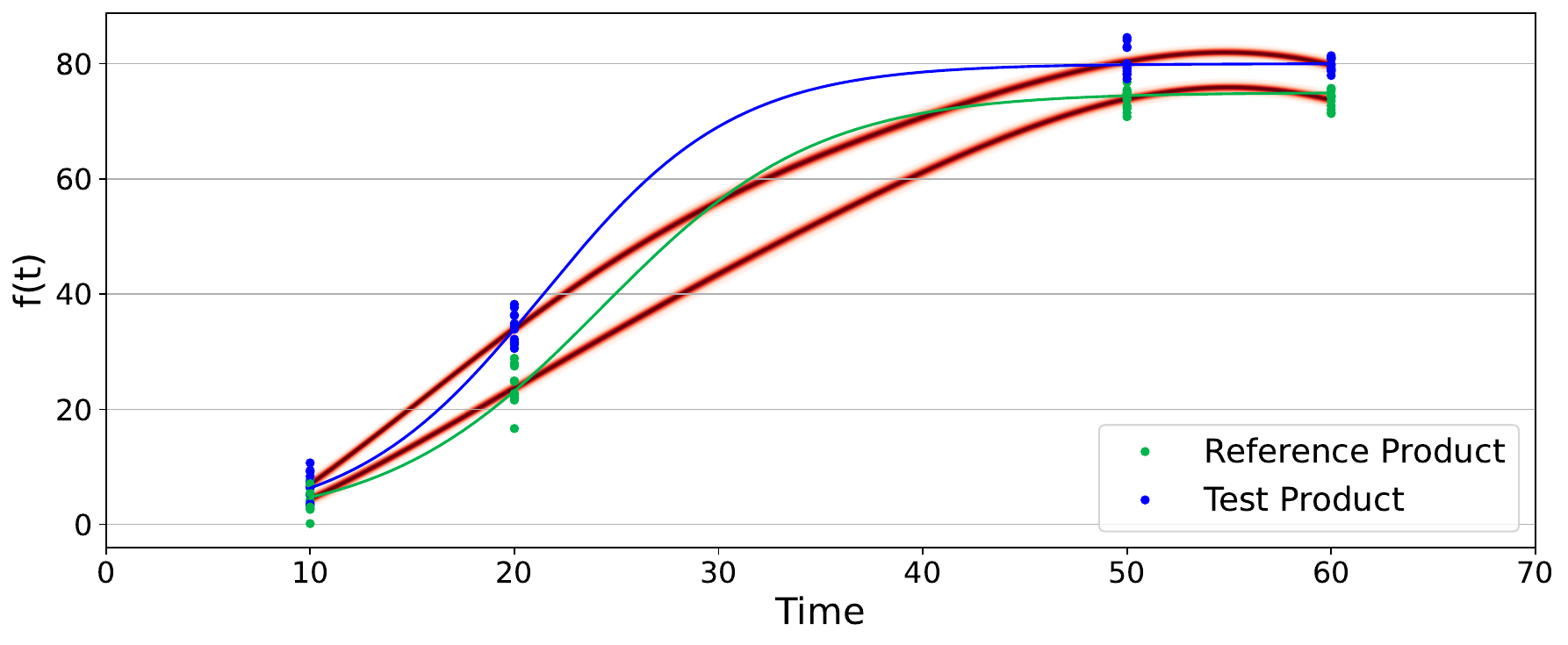}
\includegraphics[width=0.5\textwidth]{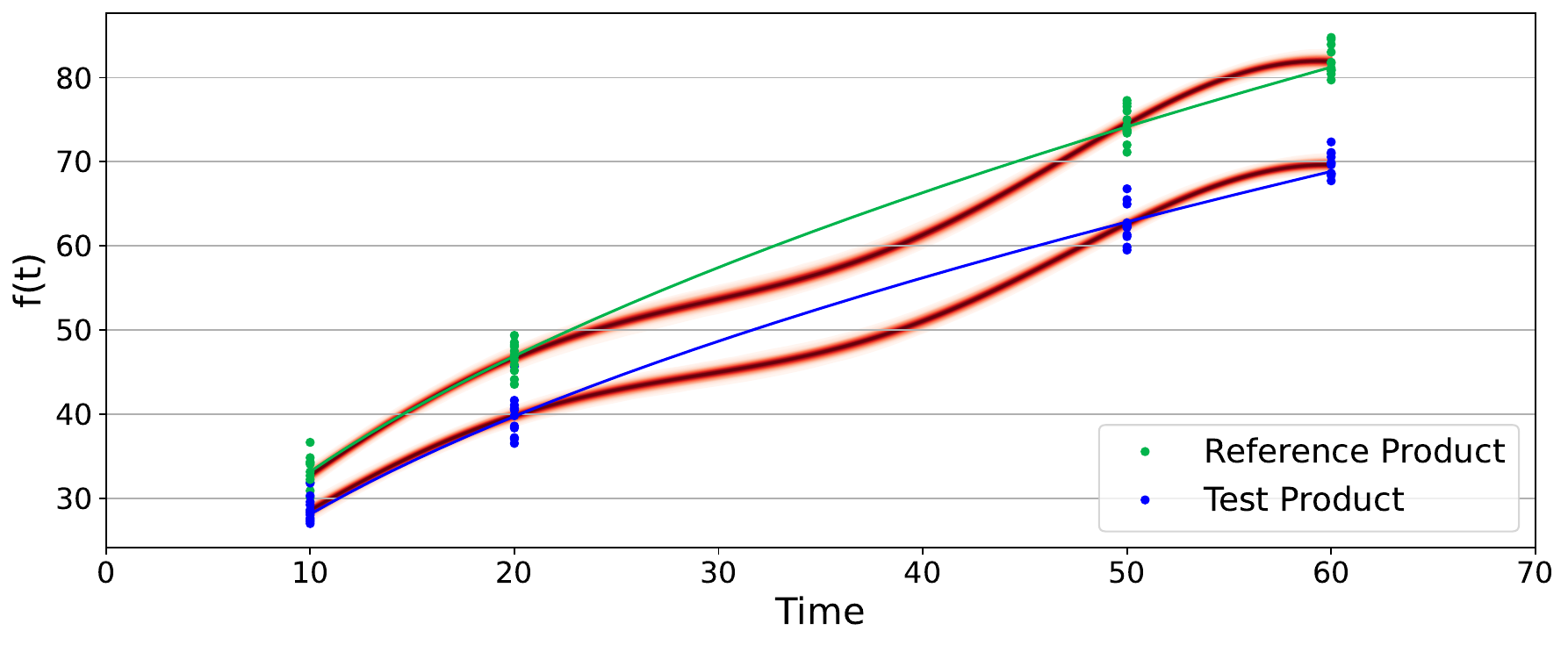}
\caption{Predictions $f_R(t)$, $f_T(t)$ and $95\%$ credible regions, shown in red, for the CTGP model on logistic data (left), and Higuchi data (right). The models were trained excluding observations at $t = \{30, 40\}$, and the true underlying functions are shown in blue or green.}
\label{fig:ctgp-exc-30}
\end{figure}
\end{example}

\subsection{Logistic Spline Gaussian Process (LSGP)}
\label{sec:ourmodel}
\begin{comment}
An alternative to this hierarchical approach of embedding a non-parametric model within another non-parametric model is to use prior knowledge of the domain to embed a parametric function within a non-parametric model. 
 Using a GP allows for the dissolution process to be interpolated at unobserved time points, and provides flexibility within the model to fit different underlying functions given the data. 
\end{comment}
 To capture the generally monotonic behavior of dissolution curves, we propose a new model, the Logistic Spline Gaussian Process (LSGP), that includes a  parametric  function in the mean and kernel of the GP. In particular, we make the assumption that dissolution curves are typically monotonically increasing, and  follow a logistic function. The proposed model is: 
\begin{eqnarray}
\label{eq:your}
y^{(j)}(t) &=& f(t)+w^{(j)}(t), \text{ for } j=1,\dots,n,~t=t_1,\dots,t_p,\\ 
\label{eq:four}
f(t) &\sim& GP(\mu(t;\boldsymbol{\theta}), k(t, t'; \boldsymbol{\theta})),\\
\label{eq:ourmean} 
\mu(t; \boldsymbol{\theta}) &=& \frac{\alpha_1}{1 + \alpha_2e^{-\beta t}} , \\
\label{eq:ourcov}
k(t, t'; \boldsymbol{\theta}) &=& \tau^2 W_q\left(\frac{\alpha_1}{1 + \alpha_2e^{-\beta t}}, \frac{\alpha_1}{1 + \alpha_2e^{-\beta t'}}\right) ,
\end{eqnarray}
where the observed cumulative dissolution $y^{(j)}(t)$ differ from  the  cumulative dissolution $f(t)$ by additive Gaussian noise $w^{(j)}(t)$, with zero mean and covariance:
$$
E[w^{(i)}(t)w^{(j)}(t')]=\sigma^2(t;\boldsymbol{\theta})\delta(t-t')\delta(i-j),
$$
where $\delta(a-a')$ is the Kronecker delta\footnote{When $a=a'$, $\delta(a-a')=1$ and zero otherwise.} and $\sigma^2(t;\boldsymbol{\theta})$ is the  time-dependent variance. In other words, the noise is assumed to be independent across different dosage units and time instants. We model $\log \sigma^2(t;\boldsymbol{\theta})= a+bt$,  which typically describes the heteroskedastic behavior we observe in the noise in dissolution experiments. Note that although there is a difference between the noise model assumed in CTGP \eqref{eq:CTGP1} and our noise model, the primary distinction between the two models lies in the prior kernel. Indeed, in our model, the  function $f(t)$ is GP distributed with mean $\mu(t; \boldsymbol{\theta})$ and kernel $k(t, t'; \boldsymbol{\theta})$, which is a scaled $q$-fold  Wiener kernel. We denote the vector of hyperparameters  with $\boldsymbol{\theta}=[\alpha_1,\alpha_2,\beta,\tau,a,b]^\top$, where $\alpha_1,\alpha_2,\beta \in \mathbb{R}^{+}$ and $a,b \in \mathbb{R}$.

We recall that the $q$-fold integrated Wiener process is defined as \citep{shepp1966radon}:
\begin{align}
W_q(t)&= \int_{0}^1 \frac{(t-u)_+^{q-1}}{(q-1)!}Z(u)du,
\end{align}
for $t\in (0,1)$ and $q=1,2,\dots$, where $(\cdot)_+$ is the positive part operator and $Z(u)$ denotes a Gaussian white noise process. 
The $q$-fold  Wiener kernel is the covariance of the relative integrated Wiener process, which is equal to:
\begin{align}
W_q(t,t')&= \int_{0}^1 \frac{(t-z)_+^{q-1}}{(q-1)!} \frac{(t'-z)_+^{q-1}}{(q-1)!} dz=\int_{0}^{\min\{t,t'\}} \frac{(t-z)^{q-1}}{(q-1)!} \frac{(t'-z)^{q-1}}{(q-1)!} dz.
\end{align}
For $q=1$ and $q=2$, we obtain the linear and, respectively, cubic spline kernels:
\begin{align}
\label{eq:wiener1}
 W_1(t,t')&=v,   \\
 W_2(t,t')&=\frac{v^3}{3}-\frac{v^2}{2}(t+t')+tt'v=\frac{v^3}{3}+\frac{v^2}{2}|t-t'|,
\end{align}
with $v=\min\{t,t'\}$ \citep{wahba1990spline}.

We propose replacing the linear function $t$ in  $W_q(t,t')$ with the logistic function (as shown in \eqref{eq:ourcov}). This leads to a class of kernels we called \textit{dissolution spline kernels}. These kernels inherit all the approximation capabilities of spline curves, but the underlying basis functions  are logistic functions. 
In particular, we will consider $q=2$.\footnote{Note that the covariance function $W_q(t,t')$ corresponds to a Gaussian process that is Mean-Square (MS) continuous and $2q-2$ MS differentiable.} The presented model \eqref{eq:four}--\eqref{eq:ourcov}  draws inspiration from the work of \cite{pillonetto_new_2010} on dynamic system identification. While their approach uses an exponentially decaying function within a Wiener kernel for impulse response estimation in linear time-invariant systems, our model incorporates a logistic function to capture the cumulative dissolution profile observed in dissolution testing.

\begin{figure}
    \centering
    \includegraphics[width=10cm]{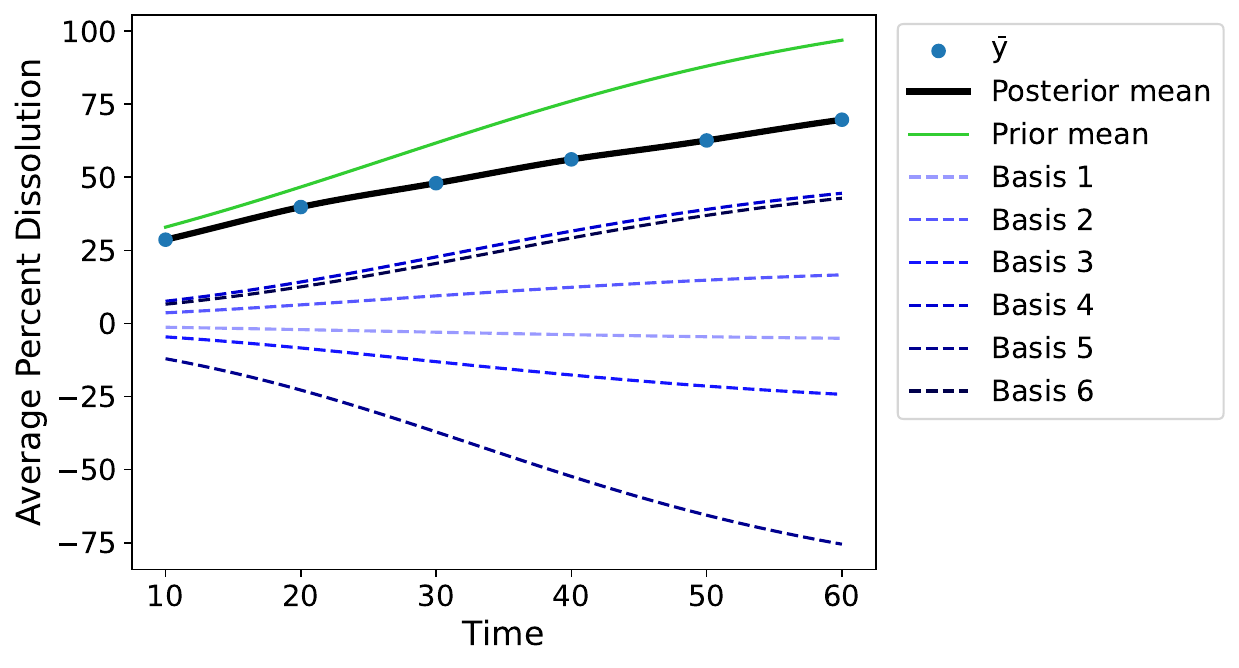}
    \caption{The posterior mean (in black) $m_p(t;\boldsymbol{\theta})$  is decomposed into its components: the prior mean (in green) and the six basis functions (one for each observation). The observations $\bar{\bf y}$ are shown as a scatter plot (blue points).}
    \label{fig:piecewise}
\end{figure}

Denoting the mean of observed cumulative percentage dissolved for the $n$ dosage units with $\bar{\bf y} = \frac{1}{n}\sum_{j=1}^n \mathbf{y}^{(j)}$ where $\mathbf{y}^{(j)}=[y^{(j)}(t_1),\dots,y^{(j)}(t_p)]^\top$, we have that the posterior of $f$, given the hyperparameters $\boldsymbol{\theta}$, is a GP. 
\begin{proposition}
\label{prop:1}
The  posterior distribution of $f$ is a GP:
\begin{equation}
p(f(t)|\{y^{(j)}(t_1),\dots,y^{(j)}(t_p)\}_{j=1}^{n},\boldsymbol{\theta}),=GP(m_p(t;\boldsymbol{\theta}), K_p(t,t';\boldsymbol{\theta})),
\end{equation}
with mean and kernel function equal to:
\begin{align}
\label{eq:postmean}
m_p(t;\boldsymbol{\theta})&=\mu(t)+K(t, {\bf t};\boldsymbol{\theta})V_{\boldsymbol{\theta}}^{-1}({\bf \bar{y}}-\mu({\bf t}))\\
\label{eq:postcov}
K_p(t,t';\boldsymbol{\theta})&=K(t,t';\boldsymbol{\theta})- K(t, {\bf t};\boldsymbol{\theta})V_{\boldsymbol{\theta}}^{-1}K({\bf t}, t';\boldsymbol{\theta}),
\end{align}
where $\mathbf{t}=[t_1,\dots,t_p]$, $V_{\boldsymbol{\theta}}=K({\bf t}, {\bf t};\boldsymbol{\theta})+\frac{1}{n}D(\mathbf{t},\mathbf{t};{\boldsymbol{\theta}})$, 
with
\begin{equation}
    \label{eq:D}
D(\mathbf{t},\mathbf{t};\boldsymbol{\theta})=\begin{bmatrix}
    \sigma^2(t_1;\boldsymbol{\theta}) & & \\
    & \ddots & \\
    & & \sigma^2(t_p;\boldsymbol{\theta})
  \end{bmatrix},
\end{equation}
and $K(\mathbf{t},\mathbf{t};\boldsymbol{\theta})$ denotes the $p\times p$ matrix of the kernel evaluated at all pairs $\{(t_i,t_j)\}_{i,j=1}^p$ and  similarly for the other matrices $K(t,t';\boldsymbol{\theta}),K(t,\mathbf{t};\boldsymbol{\theta}),K(\mathbf{t},t';\boldsymbol{\theta})$.

The  posterior distribution of $y$ is also a GP:
\begin{equation}
\label{eq:postprop1}
p(y(t)|\{y^{(j)}(t_1),\dots,y^{(j)}(t_p)\}_{j=1}^{n},\boldsymbol{\theta})=GP(m_p(t;\boldsymbol{\theta}), K_p(t,t';\boldsymbol{\theta})+\sigma^2(t;\boldsymbol{\theta})\delta(t-t')).
\end{equation}
The marginal likelihood of the observations given the hyperparameters is:
\begin{align}
\nonumber
p(\{y^{(j)}(t_1),\dots,y^{(j)}(t_p)\}_{j=1}^{n}|\boldsymbol{\theta})&=\int p(\{y^{(j)}(t_1),\dots,y^{(j)}(t_p)\}_{j=1}^{n}|f(\mathbf{t}),\boldsymbol{\theta})p(f(\mathbf{t})|\boldsymbol{\theta})df(\mathbf{t}),\\
\label{eq:marginalikelihood}
&=\tfrac{1}{(2\pi)^\frac{p(n-1)}{2}n^\frac{p}{2}|D(\mathbf{t},\mathbf{t};\boldsymbol{\theta})|^\frac{n-1}{2}}N({\bar{\bf y}};\mu(\mathbf{t};\boldsymbol{\theta}),V_{\boldsymbol{\theta}})e^{\frac{-\sum_{j=1}^n (\mathbf{ y}^{(j)})^T(D(\mathbf{t},\mathbf{t};\boldsymbol{\theta}))^{-1}\mathbf{y}^{(j)}+n\bar{\bf y}^T (D(\mathbf{t},\mathbf{t};\boldsymbol{\theta}))^{-1}\bar{\bf y}}{2}}.
\end{align}
\end{proposition}
The proof of Proposition \ref{prop:1} is in Appendix. To understand the proposed model, it is worth noticing  that the  the posterior mean in \eqref{eq:postmean} can be rewritten as:
\begin{align}
\nonumber
m_p(t;\boldsymbol{\theta})=\frac{\alpha_1}{1 + \alpha_2e^{-\beta t}}
+\sum_{i=1}^p \gamma_i \Bigg(&\tfrac{1}{3}\left(\min\left\{\tfrac{\alpha_1}{1 + \alpha_2e^{-\beta t}},\tfrac{\alpha_1}{1 + \alpha_2e^{-\beta t_i}}\right\}\right)^3\\
\label{eq:postmean2}
&-\tfrac{1}{2}\left(\min\left\{\tfrac{\alpha_1}{1 + \alpha_2e^{-\beta t}},\tfrac{\alpha_1}{1 + \alpha_2e^{-\beta t_i}}\right\}\right)^2\left|\tfrac{\alpha_1}{1 + \alpha_2e^{-\beta t}}-\tfrac{\alpha_1}{1 + \alpha_2e^{-\beta t_i}}\right|\Bigg),
\end{align}
where  $\boldsymbol{\gamma}$ is the vector of coefficients $V_{\boldsymbol{\theta}}^{-1}({\bf \bar{y}}-\mu({\bf t}))$. This 
 highlights that the posterior is a  piecewise cubic logistic function. We illustrate this formula in Figure \ref{fig:piecewise}, which  shows the predicted posterior mean computed from observations at six different time instants, with the six average values $\bar{\bf y}$ 
  reported in the figure as a scatter plot. The figure also includes the corresponding prior mean (in green) and the six basis functions, scaled by 0.03 for visualization purposes. The sum of the prior mean and basis function gives the posterior mean (in black).

\begin{example}
\label{ex:2}
To illustrate the GP with dissolution spline kernel, we present in Figure \ref{fig:spline-preds} the predictions generated by this model for the reference and test product of the simulated logistic data (\ref{logistic-fx}) used in Example \ref{ex:1} from $t = 10$ to $t = 60$, along with the $95\%$ credible region. We show both the noiseless predictions, $f_\mstar(t)$, as well as the predictions for the observed dissolution profile $y_\mstar(t)$. The hyperparameters for this model were estimated by calculating the maximum a posteriori estimate, which will be discussed in the next section. The calculated hyperparameters for the test group were $\alpha_1 \approx 79.52$, $\alpha_2 \approx 88.72$, $\beta \approx 0.21$, $\tau^2 \approx 0.0001$, $a \approx 4.99$, and $b = 0$. The hyperparameters for the reference group were $\alpha_1 \approx 73.79$, $\alpha_2 \approx 93.09$, $\beta \approx 0.19$, $\tau^2 \approx 0.0001$, $a \approx 5.15$, and $b = 0$. The true underlying logistic function is shown in blue or green, demonstrating that the model closely predicts the true underlying function. 

Additionally, we present the predictions generated by the LSGP model for the simulated data following the Higuchi model (\ref{higuchi-fx}) used in Example \ref{ex:1}. The hyperparameters estimated given the Higuchi data for the test group were $\alpha_1 \approx 80.76$, $\alpha_2 \approx 2.77$, $\beta \approx 0.046$, $\tau^2 \approx 0.0001$, $a \approx 4.83$, and $b = 0$, and for the reference group were $\alpha_1 \approx 112.88$, $\alpha_2 \approx 4.15$, $\beta \approx 0.054$, $\tau^2 \approx 0.0025$, $a \approx 4.96$, and $b = 0$. Although the true underlying function of the data differs from the parametric function used in the GP model, the LSGP model is still able to closely predict the true underlying function of the data. 

We further illustrate the prediction performance of the LSGP model in Figure \ref{fig:spline-exc-30}, which shows the predictions for the logistic and Higuchi data, where the model was trained excluding the observations at time points $t=\{30, 40\}$. The predictions generated by the LSGP model still closely fit the true underlying function, shown in green in the plots. 
In particular, unlike the CTGP model predictions shown in Figure \ref{fig:ctgp-exc-30}, we do not observe mean reversion to zero. This is due to the non-stationarity of the kernel used in the LGSP model, which additionally allows us to incorporate the knowledge that dissolution curves are generally monotonic.

\begin{figure}
\includegraphics[width=0.5\textwidth]{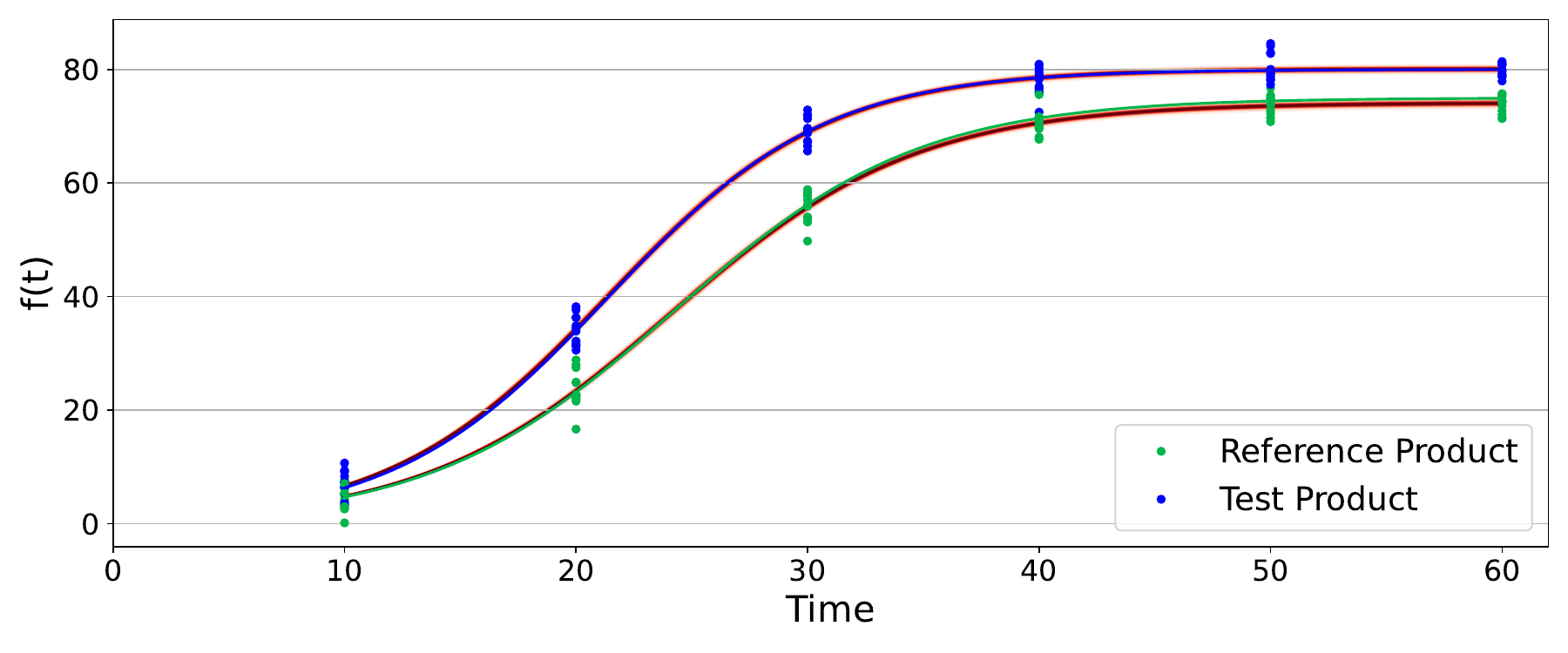}
\includegraphics[width=0.5\textwidth]{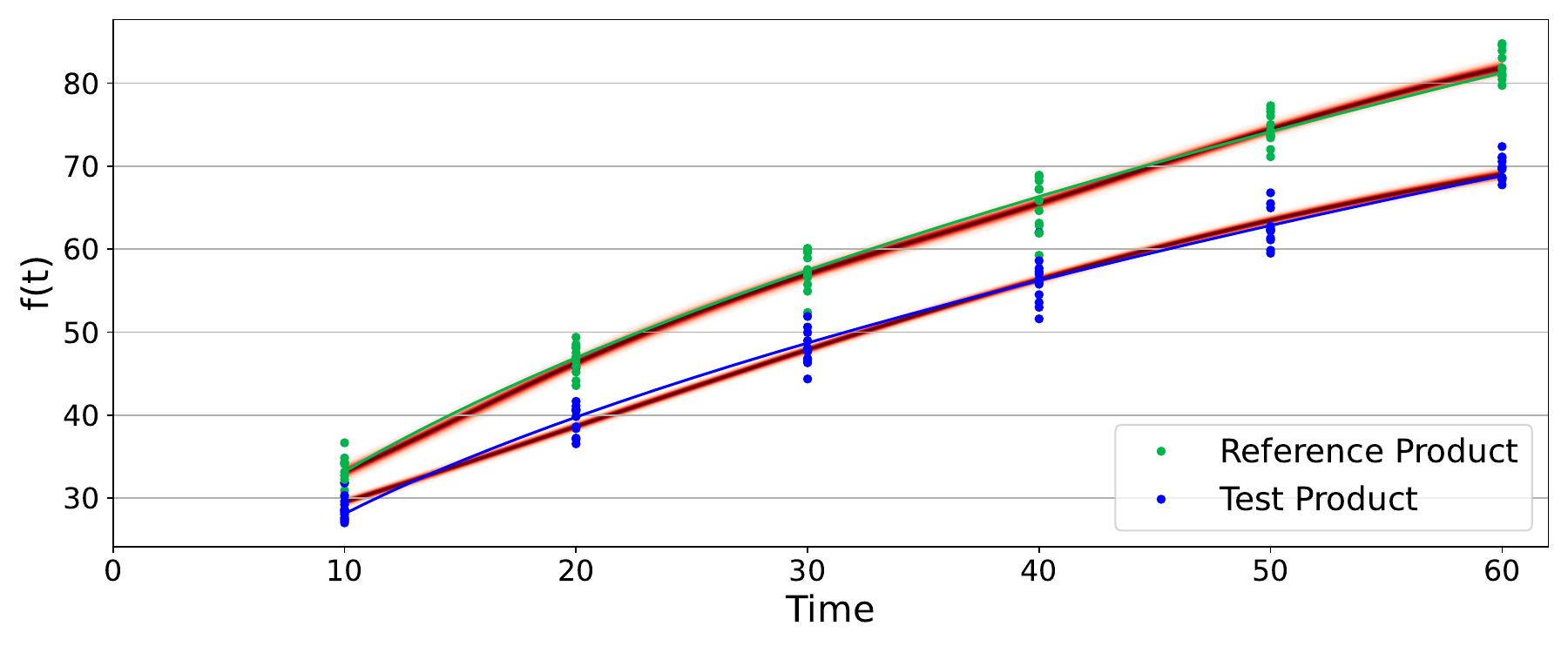}
\includegraphics[width=0.5\textwidth]{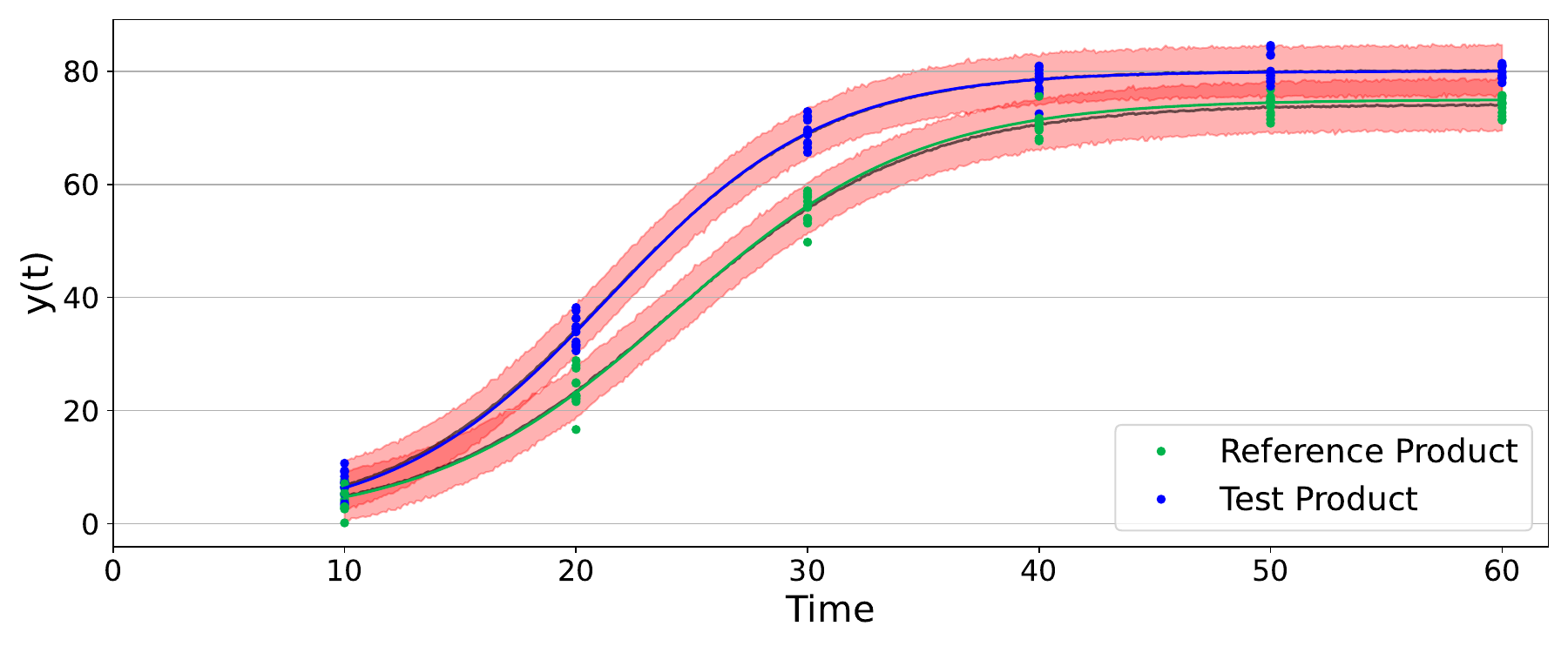}
\includegraphics[width=0.5\textwidth]{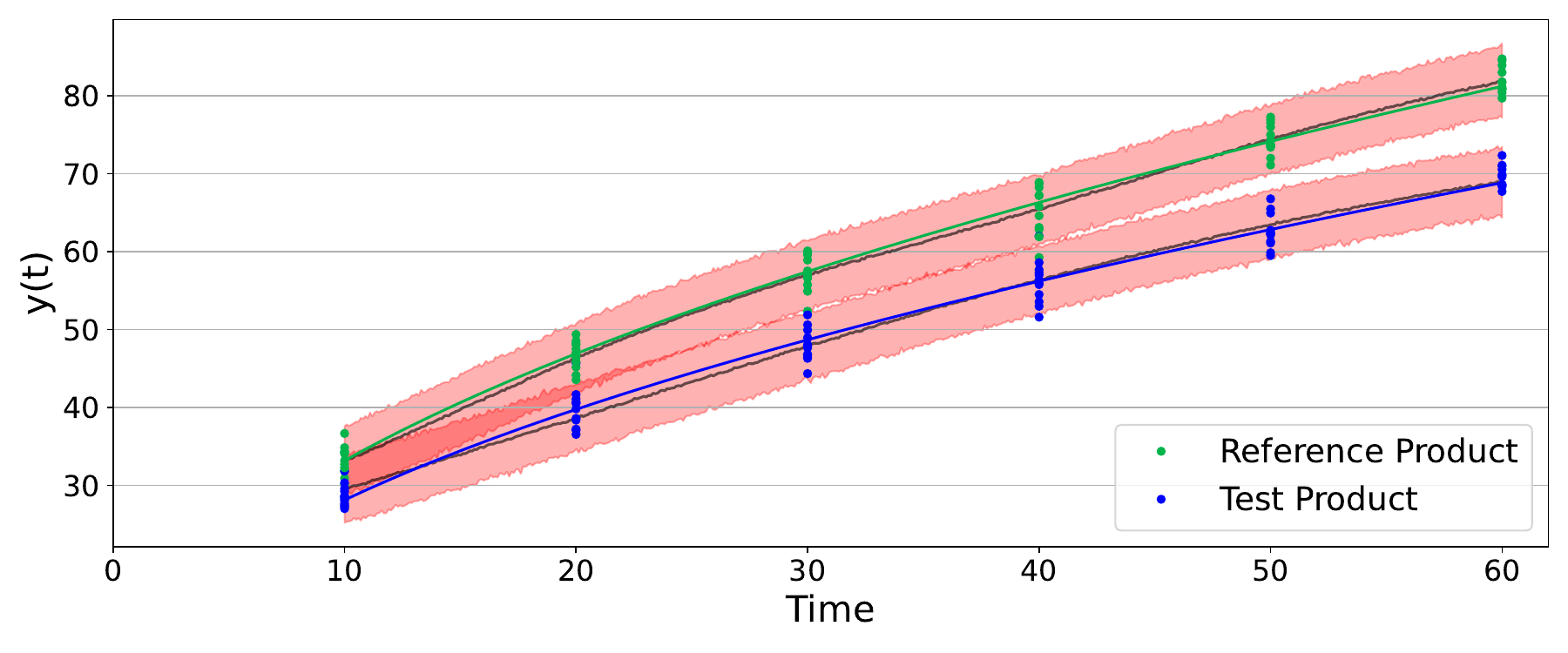}
\caption{(Top) Predictions $f_R(t)$, $f_T(t)$ and $95\%$ credible regions, shown in red, generated by the LSGP model for the reference and test product of the logistic model data (left) and Higuchi model data (right), where the true underlying functions are shown in blue and green. (Bottom) Predictions for the observed dissolution profile $y_R(t)$ and $y_T(t)$, with $95\%$ credible regions, shown in red, and the mean shown in black.}
\label{fig:spline-preds}
\end{figure}

\begin{figure}
\includegraphics[width=0.5\textwidth]{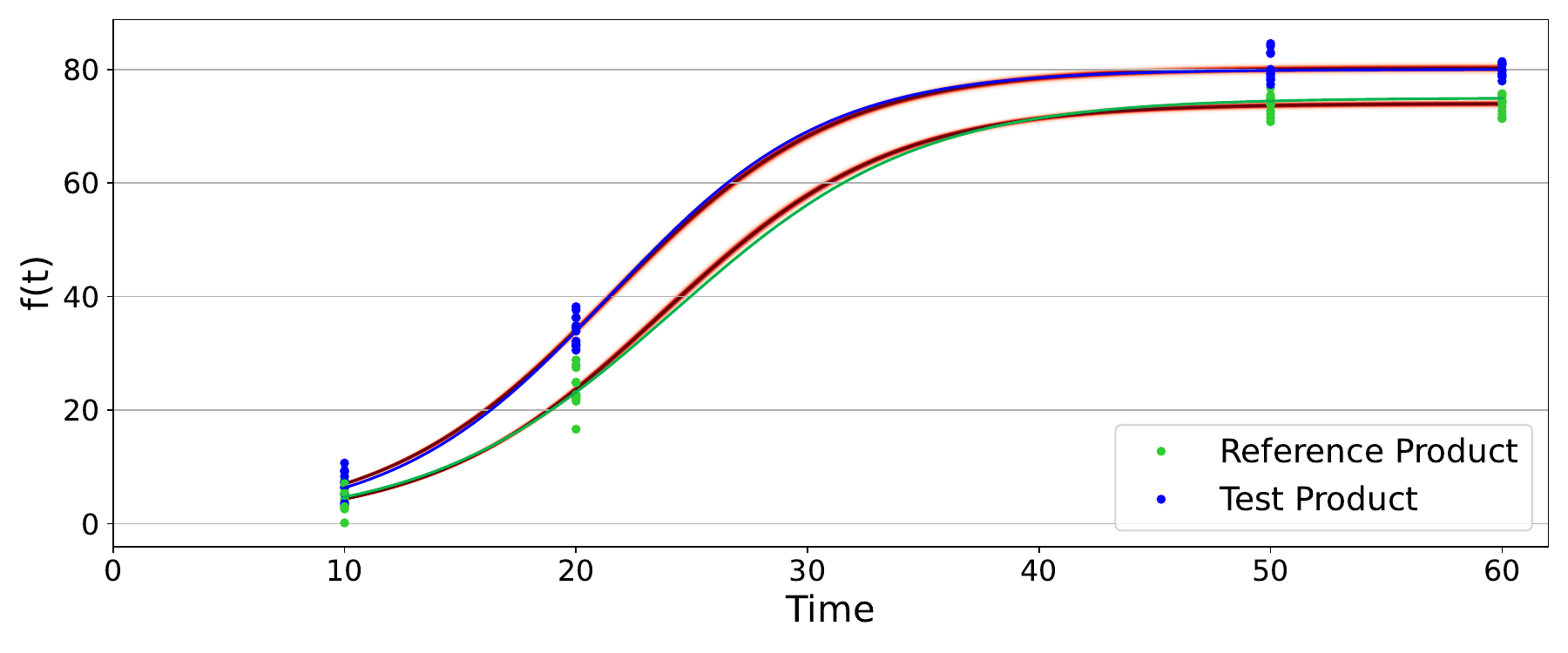}
\includegraphics[width=0.5\textwidth]{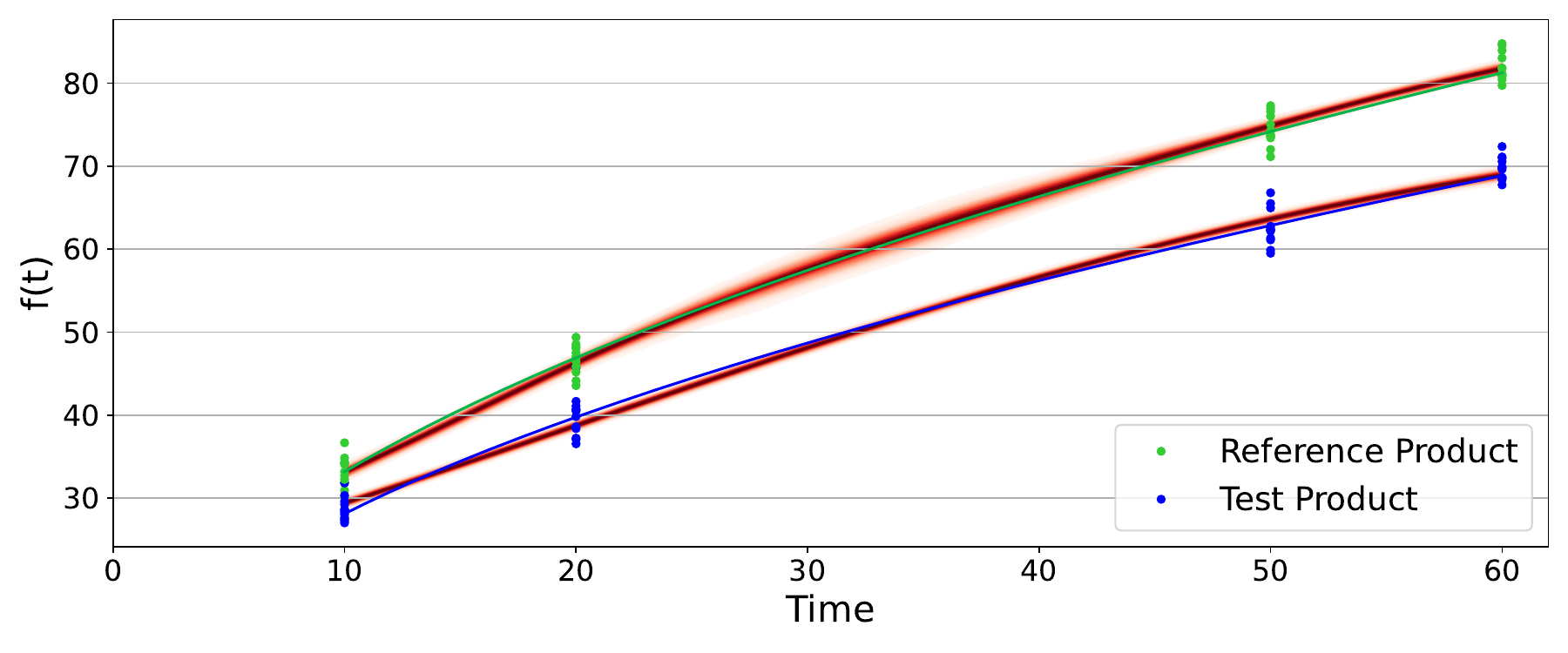}
\includegraphics[width=0.5\textwidth]{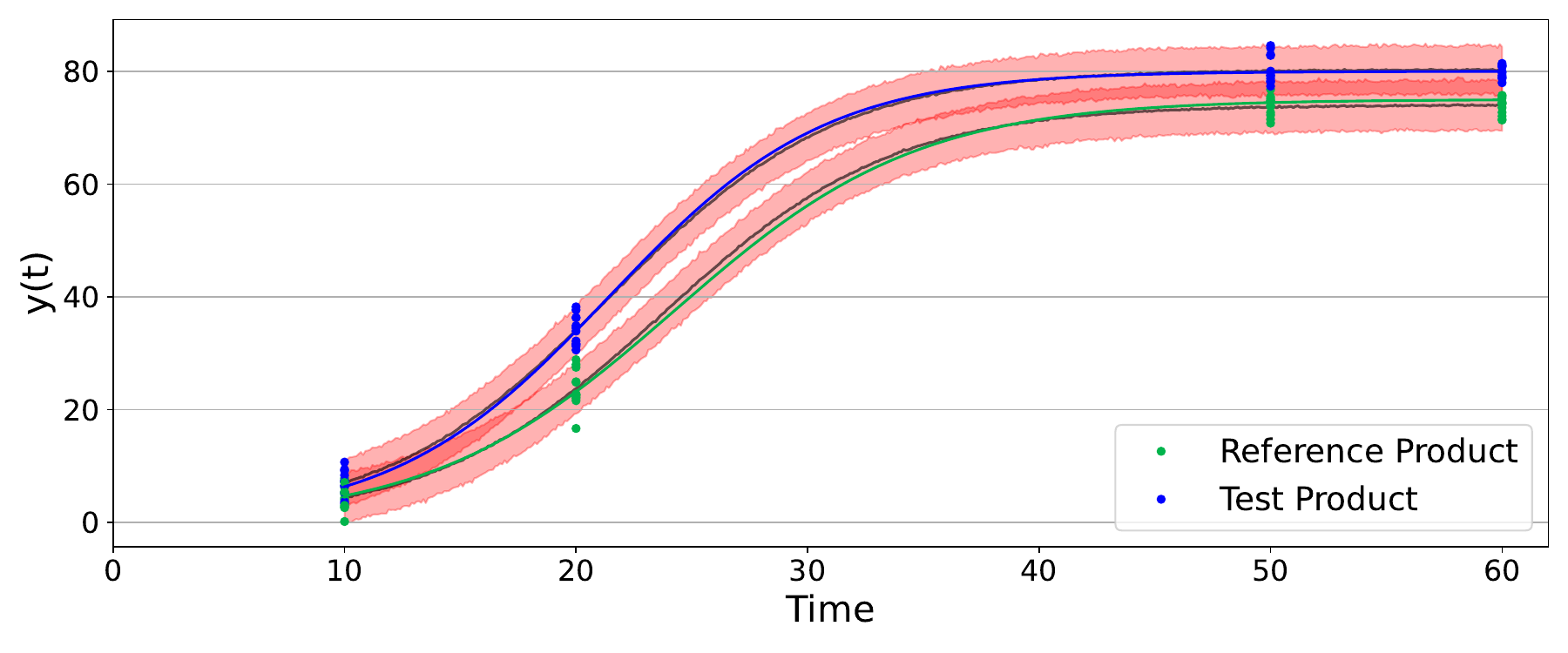}
\includegraphics[width=0.5\textwidth]{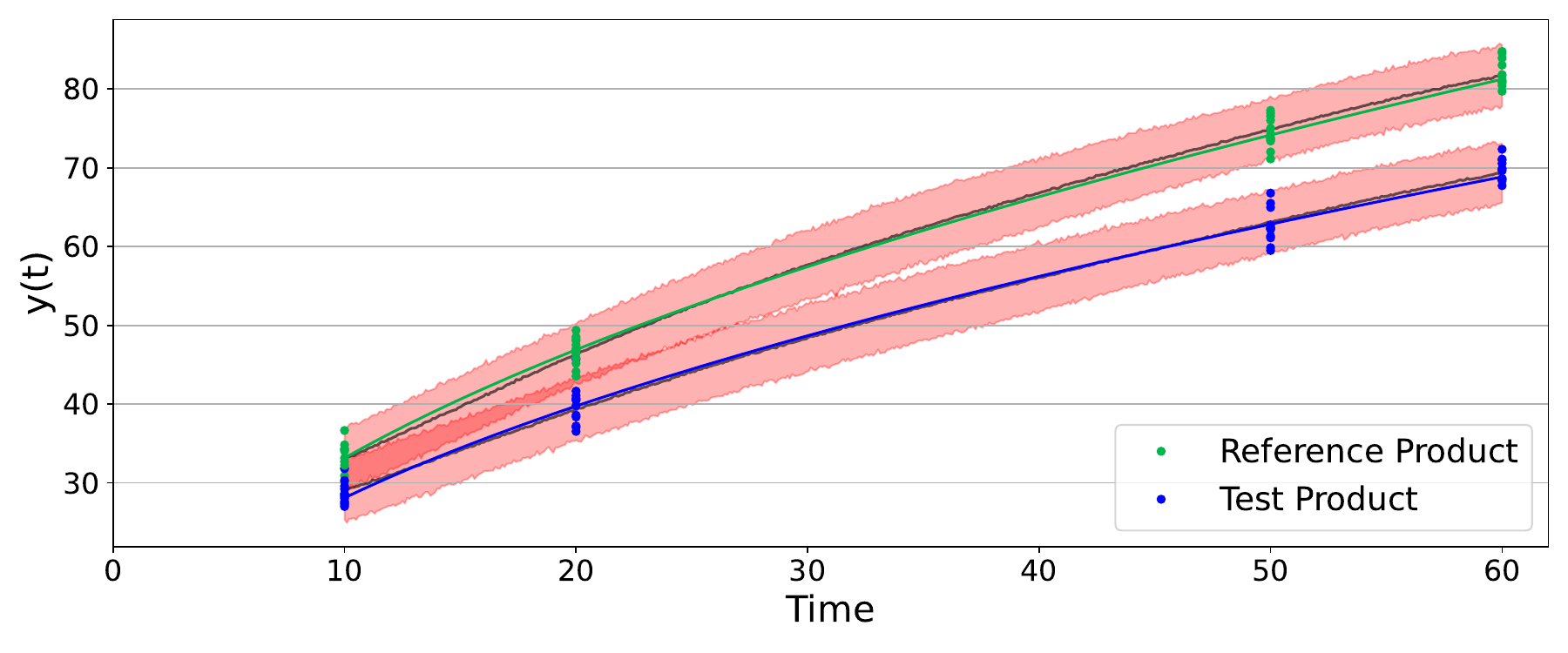}
\caption{(Top) Predictions $f_R(t)$, $f_T(t)$ and $95\%$ credible region, shown in red, for the LSGP model on logistic data (left), and Higuchi data (right). Model was trained excluding observations at $t =\{30, 40\}$, and the true underlying functions are shown in blue and green. (Bottom) Predictions including additive noise, $y_R(t)$ and $y_T(t)$, with $95\%$ credible regions, shown in red, and the mean shown in black.}
\label{fig:spline-exc-30}
\end{figure}

\end{example}

\subsubsection{Remarks} LSGP introduces a strong prior bias toward a monotonic increasing dissolution curve by incorporating a logistic function (with constrained coefficients) in both the mean and kernel. However, this does not prevent the model from fitting a non-monotonic dissolution profile (since, for instance, some of the coefficients $\gamma_i$ in \eqref{eq:postmean2} can be negative). We chose not to constrain the GP to be  monotonic (which could be done using the methodologies discussed by \cite{riihimaki2010gaussian,wang2016estimating,agrell2019,da2020gaussian,golchi2015monotone,lopez2018finite,lopez2022high,maatoukhal-04084865,lin2014bayesian,andersen2018non,benavolilinearly2024}) because drug concentration may temporarily decrease due to factors such as supersaturation and precipitation, or chemical degradation (e.g., caused by changes in temperature or pH).
Imposing monotonicity in dissolution testing was proposed by \cite{ng2018gamma}, using a Gamma process. This approach guarantees monotonicity because the Gamma process inherently has non-negative, independent increments. However, as previously discussed, non-monotonic dissolution profiles can still occur in dissolution testing; therefore, this model is not general enough to capture the full range of dissolution behaviors.

LSGP is defined using a logistic function in both the mean and kernel. However, other monotonic functions, such as the Higuchi model in \eqref{higuchi-fx}, which describes the diffusion of a drug through a matrix, could also be used. LSGP can be adapted to incorporate the Higuchi model or easily extended to to include the span of both the logistic and Higuchi function. Nevertheless, as demonstrated in Example \ref{ex:2} and in Section \ref{sec:numexp}, LSGP already performs very satisfactorily across a variety of scenarios.

\subsubsection{Hyperparameters estimation}
 \label{sec:hypest}
Given the marginal likelihood \eqref{eq:marginalikelihood} and a prior  $p(\boldsymbol{\theta})$,  the hyperparameters  $\boldsymbol{\theta}$ can be estimated using two methods: calculating the Maximum A Posteriori (MAP), defined as
\begin{align}
\hat{\boldsymbol{\theta}}_{MAP}= \arg\max_{\boldsymbol{\theta}} p(\{y^{(j)}(t_1),\dots,y^{(j)}(t_p)\}_{j=1}^{n}|\boldsymbol{\theta})p(\boldsymbol{\theta}),
\end{align}
 or by approximating the posterior $p(\boldsymbol{\theta}|\{y^{(j)}(t_1),\dots,y^{(j)}(t_p)\}_{j=1}^{n})$ via Markov Chain Monte Carlo (MCMC) methods.  
We follow the first approach in this paper, as it allows us to exploit the conjugacy of GPs  with the likelihood to analytically derive the posterior (given the hyperparameters). Since the likelihood (and the joint) is not globally concave, we use  multiple restarts (sampling the starting value of $\boldsymbol{\theta}$ from the prior) to numerically optimise the joint.

We assume the following prior\footnote{Note that, the probability density function of the log-normal distribution of the real variable $x>0$ with location parameter $m$ and scale $s>0$ is $p(x|m,s) = \frac{e^{-(\log(x) - m)^2/2s^2}}{x s \sqrt{2\pi}}$. The PDF of the Half-Cauchy distribution of the real variable $x \geq m$ is $p(x|m,s) = \frac{2}{\pi s(1 + (x - m)^2/s^2)}$.} on the hyperparameters $p(\boldsymbol{\theta})=\prod_{i=1}^6p(\theta_i)$ with $\alpha_i \sim \text{Log-Normal}(m_i, s_i)$ for $i=1,2$, $\beta \sim \text{Log-Normal}(m_3, s_3)$, $a \sim \text{Log-Normal}(m_4, s_4)$, $b \sim \text{Log-Normal}(m_5, s_5)$  and $\tau^2 \sim $ HalfCauchy(5) with 
[$m_1,m_2,m_3,m_4,m_5$] $=$ [$\log(76.56)$, $\log(100)$, $\log(0.196)$,
$\bar{a}$,$\bar{b}$] and [$s_1,s_2,s_3,s_4,s_5$] $=$ [$3.0$, $3.0$, $3.0$, $1.25$, $1.25$].

The priors for the parameters of the mean function, $\alpha_1, \alpha_2,$ and $\beta$, were selected by numerically assessing the average of these values given a range of possible dissolution curve patterns. These average values were set for the mean $m_i$  of each hyperparameter, while the standard deviation $s_i$ was empirically determined  to ensure that the prior predictive distribution of the implied dissolution curve aligns with realistic dissolution profiles (maximum percent dissolution of about $100\%$).

The priors for $a$ and $b$ are determined using an empirical Bayes approach. Let  \\ $\phi_\mstar(t_i) =\log\left(\frac{1}{n_\mstar}\sum_{j=1}^{n_\mstar}(y^{(j)}_{\mstar}(t_i)-\overline{y}_{\mstar}(t_i))^2\right)$ represent the  sample variance at each time point $t_i$. Then, the means for the prior of $a$ and $b$ are defined using the following sample-based estimator:
$m_4= {\bar a}_{\mstar}$, $m_5= {\bar b}_{\mstar}$ with

$$
\begin{aligned}&\qquad {\bar a}_{\mstar}={\frac {\sum _{i=1}^{p}\phi_\mstar(t_i) \sum _{i=1}^{p}t_{i}^{2}-\sum _{i=1}^{p}t_{i}\sum _{i=1}^{p}t_{i}\phi_\mstar(t_i) }{p\sum _{i=1}^{p}t_{i}^{2}-(\sum _{i=1}^{p}t_{i})^{2}}},\vspace{0.2cm}\\&\qquad {\bar b}_{\mstar}={\frac {n\sum _{i=1}^{p}t_{i}\phi_\mstar(t_i)-\sum _{i=1}^{p}t_{i}\sum _{i=1}^{p}\phi_\mstar(t_i)}{p\sum _{i=1}^{p}t_{i}^{2}-(\sum _{i=1}^{p}t_{i})^{2}}},&\qquad \end{aligned}
$$
for $\mstar \in \{R,T\}$.  The scale parameters $s_4$ and $s_5$ are set to $1.25$, based on prior predictive checks.

 \subsubsection{Dissolution curve comparisons}
 \label{sec:ourht}
To compare dissolution curves, we fit two  independent GP models  for the reference and test product data. Posterior samples for $f_R(t),f_T(t)$ from these two models can be used  to compute a distribution for $\mathrm{f}_{2,integral}$ or to perform other statistical tests for decision making. 

The posterior distribution of $\mathrm{f}_{2,integral}$ can be computed from the posterior samples as follows:
\begin{equation}
\label{eq:f2our}
    E[\mathrm{f}_{2,integral}|\text{data}] \approx \frac{1}{m} \sum\limits_{i=1}^m \mathrm{f}^{(i)}_2 = \frac{1}{m} \sum\limits_{i=1}^m 50\log_{10}\left(100\left(1 + \frac{1}{ r}||\mathbf{f}^{(i)}_{R} - \mathbf{f}^{(i)}_{T}||^2\right)^{-1/2}\right),
\end{equation}
 for each sample $i=1,\dots,m$,  where $\mathbf{f}^{(i)}_\mstar=[f^{(i)}_\mstar(t_1^*),\dots,f^{(i)}_\mstar(t_r^*)]^\top$ for $\mstar \in \{R,T\}$.\footnote{The LSGP model draws posterior samples at $r$ time-points equally spaced in the interval $[t_1,t_p]$ for profile comparison, whereas the CTGP model add to these $r$ points additional $p$ points located at $t_1,t_2,\dots,t_p$ to approximate the $\mathrm{f}_{2,integral}$. For large $r$, there is no practical difference between these two approaches.}
The posterior samples  $\{\mathrm{f}^{(i)}_2\}_{i=1}^m$ can also be used for  computing the \textit{probability of similarity} of the two dissolution curves:
$$
P(\mathrm{f}_{2,integral}\geq 50) \approx \frac{1}{m}\sum_{i=1}^m I_{[50,100]}(\mathrm{f}^{(i)}_2).
$$

 We can leverage the Gaussianity of the posterior for fixed hyperparameters to approximate this test. 
Assume the hyperparameters are estimated through MAP inference,  that is we have that $\boldsymbol{\theta}_\mstar=\hat{\boldsymbol{\theta}}_\mstar$ for $\mstar \in \{R,T\}$. 
Then, from \eqref{eq:postcov}--\eqref{eq:postmean} for fixed hyperparameters $\boldsymbol{\theta}_R,\boldsymbol{\theta}_T$, a-posteriori we have that
\begin{equation}
\label{eq:jointGP}
    \begin{bmatrix}
    f_R({\bf t})\\
    f_T({\bf t})
    \end{bmatrix}\sim N\left(    \begin{bmatrix}
    m^{(R)}_p({\bf t};\boldsymbol{\theta}_R)\\
    m^{(T)}_p({\bf t};\boldsymbol{\theta}_T)
    \end{bmatrix},\begin{bmatrix}
    K^{(R)}_p({\bf t},{\bf t};\boldsymbol{\theta}_R) & 0\\
    0& K^{(T)}_p({\bf t},{\bf t};\boldsymbol{\theta}_T)
    \end{bmatrix}\right),
\end{equation}
where $m^{(\mstar)}_p({\bf t};\boldsymbol{\theta}_\mstar),K^{(\mstar)}_p({\bf t},{\bf t};\boldsymbol{\theta}_\mstar)$ are the posterior mean and covariance of the GP computed at the time ${\bf t}=[t_1,\dots,t_p]$ for reference and test. The diagonal covariance results from fitting two independent GPs on the reference and test data.
In this case, the test \eqref{eq:f2our} can be  performed by computing samples from the above independent normal distributions. 

Alternatively, we can perform a test similar to the MSD test to decide whether two functions $f_R(t)$ and $f_T(t)$ are similar.
From \eqref{eq:jointGP}, we have that
\begin{equation}
   f_R({\bf t}) -  f_T({\bf t}) \sim N\left( m^{(R)}_p({\bf t};\boldsymbol{\theta}_R)- m^{(T)}_p({\bf t};\boldsymbol{\theta}_T),K^{(R)}_p({\bf t},{\bf t};\boldsymbol{\theta}_R)+K^{(T)}_p({\bf t},{\bf t};\boldsymbol{\theta}_T)\right).
\end{equation}

\noindent As such, the confidence region for  ${\bf z}=f_R({\bf t}) -  f_T({\bf t})$ becomes 

\begin{equation}
\label{eq:ourMSD}
    CR = ({\bf z} - (m^{(R)}_p({\bf t};\boldsymbol{\theta}_R)- m^{(T)}_p({\bf t})))^T S^{-1}({\bf z} - (m^{(R)}_p({\bf t};\boldsymbol{\theta}_R)- m^{(T)}_p({\bf t})))\leq \chi^2_\nu(1-\delta) ,
\end{equation}
with $S = K^{(R)}_p({\bf t},{\bf t};\boldsymbol{\theta}_R)+K^{(T)}_p({\bf t},{\bf t};\boldsymbol{\theta}_T)$. By evaluating this test, we can say that the functions $f_R(t)$ and $f_T(t)$ are similar with posterior probability $1 - \delta$. Following \citep{tsong_statistical_1996}, we can solve  
\begin{align}
&\max {\bf z}^\top S^{-1} {\bf z}\\
\nonumber
&s.t.\\
&({\bf z} - (m^{(R)}_p({\bf t};\boldsymbol{\theta}_R)- m^{(T)}_p({\bf t})))^T S^{-1}({\bf z} - (m^{(R)}_p({\bf t};\boldsymbol{\theta}_R)- m^{(T)}_p({\bf t})))= \chi^2_\nu(1-\delta).
\end{align}
This optimisation  problem can be solved analytically using Lagrange multipliers.

Denoting the above maximum as $d^u_{R,T}$, we compare this distance to the maximum similarity limit $
    d_{limit}$.
 In order to evaluate this test with a less restrictive maximum similarity limit than \eqref{msd_limit}, in the numerical experiments we compute the maximum similarity limit at each individual time point $t = t_1,\dots,t_p$ and use the minimum limit across all $p$ time points as the global limit, that is where $d_{limit} = \min_{i=1,\dots,p} {\bf v}_i^\top S^{-1} {\bf v}_i$ where ${\bf v}_i$ is a $p \times 1$ vector which is equal to $10$ in the i-th element and zero otherwise. We declare the two profiles to be similar if  $d^u_{R,T} \leq d_{limit}$.

Note that both tests based on \eqref{eq:f2our} and \eqref{eq:ourMSD} should ideally be conducted by estimating the hyperparameters via MCMC within a fully Bayesian framework. However, we have leveraged the Gaussianity of the posterior \eqref{eq:jointGP} for fixed hyperparameters (equal to the MAP estimate) to efficiently compute approximations of these tests.

\subsubsection{Including Covariates}
\label{sec:covariates_noise}
The LSGP model also supports incorporating additional information to aid its predictions by including covariates $\{x_i\}_{i=1}^{n_c}$ that are specific to the dissolution experiments. Covariates can be incorporated into the parametric function that is used for the mean and kernel of the model as

\begin{align}
    \frac{e^{\beta_0 + \sum_{i=1}^{n_c} \beta_i x_i}}{1 + e^{\gamma_0 +  \sum_{i=1}^{n_c} \gamma_i x_i - bt}} , ~~    b = e^{\delta_0 + \sum_{i=1}^{n_c} \delta_i x_i } ,
\end{align}
where $\beta_i,\gamma_i,\delta_i \in \mathbb{R}$ are parameters of the model and estimated via MAP. For instance, possible covariates can be the composition and volume of medium used in the in-vitro dissolution testing, the viscosity of medium,
or the velocity of medium agitation in the apparatus used in the dissolution experiment. In Section \ref{sec:covariatesexp}  we apply this model to real dissolution data involving the dissolution of ibuprofen under various medium-type and viscosity conditions, intended as a high-level representation of the gastrointestinal environment. Such changes to the dissolution media can be considered to perform biopredictive dissolution tests.

\section{Numerical experiments}
\label{sec:numexp}

\subsection{Simulations}

The LSGP model was tested on simulated data from two different dissolution scenarios and compared to the CTGP model proposed in \citep{pourmohamad_gaussian_2022}. In the first simulation, we assumed the underlying dissolution curve is a logistic function, as   in Equation \eqref{logistic-fx}. Data was generated for the reference and test products at time points $t \in \{10, 20, 30, 40, 50, 60\}$ for twelve individual curves. The parameters of the ground-truth dissolution curves  ($\alpha_1^\mstar$, $\alpha_2^\mstar$, and $\beta^\mstar$ for $\mstar \in \{R, T\}$) are varied according to the experiments in \citep{pourmohamad_gaussian_2022} to create different ground-truth values for $\mathrm{f}_{2,integral}$, while the variance of the noise is set to either $1$ or $5$ to assess the impact of different amounts of noise in the data on the model performance. For each different ground-truth value of $\mathrm{f}_{2,integral}$, we ran $100$ Monte Carlo (MC) simulations differing in the realisation of noise. We made the assumption that noise across time points is uncorrelated. In the second scenario, we simulated the data using the Higuchi model, defined in \eqref{higuchi-fx}, as the ground truth. While in the first scenario the function form of the dissolution curve (a logistic function) is the same as the parametric function embedded into the mean and covariance function of the LSGP model, in the second scenario the Higuchi function allows us to test the ability of the LSGP model to predict patterns different from the selected embedded parametric function. 

When testing the CTGP model, the parameters were estimated as described in \citep{pourmohamad_gaussian_2022}. The variance parameters $\sigma^2$ and $\tau^2$ were estimated via Gibbs sampling using an inverse gamma prior with parameters $\alpha = 10$ and $\beta = 3$. The shape parameter and range parameter, $\phi$, of the kernel $k(t, t';\boldsymbol{\theta})$, are $1.5$ and $5$, respectively, while the range parameter, $\psi$, of the Square Exponential kernel for the mean function, $h(t, t';\boldsymbol{\theta})$, is $25$.

For LGTP, we estimated the hyperparameters via MAP, as described in Section \ref{sec:hypest}. We computed $\mathrm{f}_{2,\mathrm{integral}}$ following the procedure in Section \ref{sec:ourht}, using $r=500$ time points between $t=10$ and $t=60$ and $m=1000$ samples.

\subsubsection{$\mathrm{f}_2$ Evaluation}

\begin{figure}
    \centering
    \includegraphics[width=0.45\textwidth]{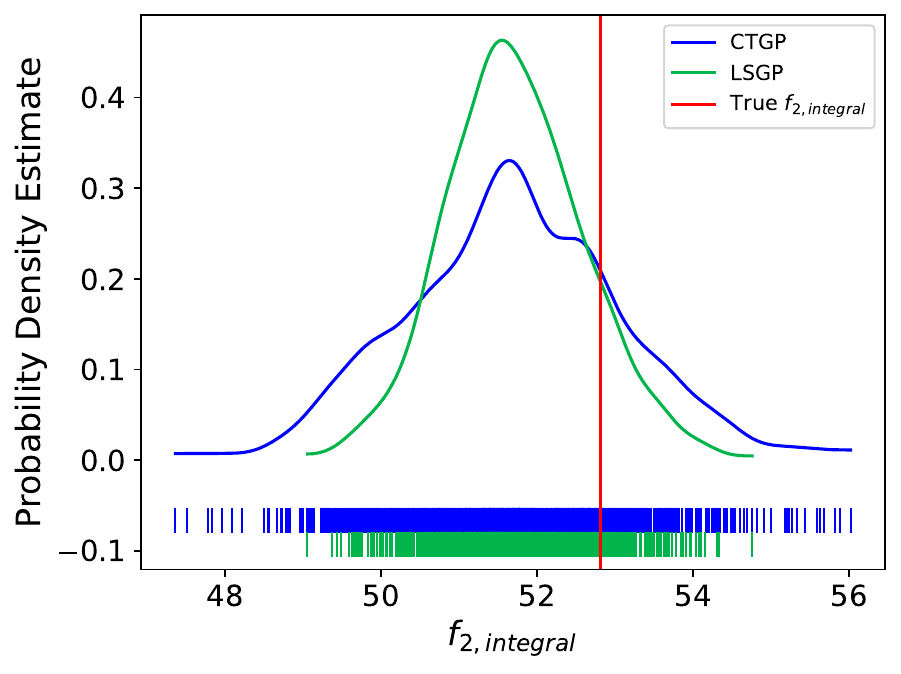}
    \includegraphics[width=0.45\textwidth]{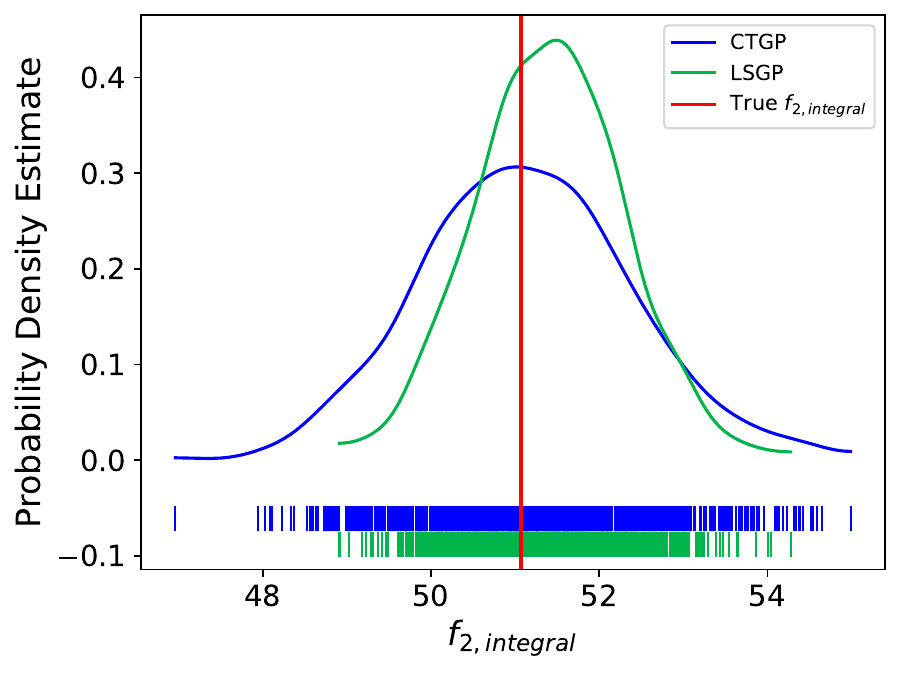}
    \caption{(Left) Distribution of $\mathrm{f}_{2, integral}$ calculated across 1000 samples of posterior predictive distribution from 1 MC run for both the LSGP model and CTGP model. The observed data was generated by the logistic model, where the true $\mathrm{f}_{2, integral} = 52.81$ and is shown in red, and the variance of the noise is 5. (Right) The observed data was generated by the Higuchi model, where the true $\mathrm{f}_{2, integral} = 51.07$ and is shown in red, and the variance of the noise is 5.}
    \label{fig:f2_distb_sim}
\end{figure}

 We assessed the accuracy of the $\mathrm{f}_{2,integral}$ estimation for both  LSGP  and CTGP.

\begin{table}[!h]
  \caption{\label{f2_log}Average $\mathrm{f}_{2,integral}$ and variance for data generated from the logistic model across 100 MC simulations. The variance is reported in parentheses.}
\centering
  \begin{tabular}{|c|c|c|c|c|c|}
    \hline
    \bf{Ground-truth} & \bf{$\beta^{R}, \alpha_2^{R}$} & \bf{$\beta^{T}, \alpha_2^{T}$} & \bf{Variance} & \bf{CTGP} & \bf{LSGP}\\[5pt]
    \hline
    $\mathrm{f}_{2, integral} = 36.74$ & (0.14, 75) & (0.19, 80) & 1 & 36.72 (0.04) & 36.74 (0.04) \\
    \hline
    $\mathrm{f}_{2, integral} = 48.19$ & (0.19, 75) & (0.226, 80) & 1 & 48.27 (0.15) & 48.25 (0.15) \\
    \hline
    $\mathrm{f}_{2, integral} = 48.19$ & (0.19, 75) & (0.226, 80) & 5 & 48.16 (0.63) & 48.16 (0.62) \\
    \hline
    $\mathrm{f}_{2, integral} = 52.81$ & (0.19, 75) & (0.215, 80) & 1 & 52.83 (0.21) & 52.80 (0.20) \\
    \hline
    $\mathrm{f}_{2, integral} = 52.81$ & (0.19, 75) & (0.215, 80) & 5 & 52.64 (0.85) & 52.64 (0.77) \\
    \hline
    $\mathrm{f}_{2, integral} = 76.10$ & (0.19, 75) & (0.202, 75) & 1 & 75.86 (1.53) &  75.82 (1.63)\\
    \hline
    $\mathrm{f}_{2, integral} = 76.10$ & (0.19, 75) & (0.202, 75) & 5 & 75.32 (8.32) & 75.87 (9.72)\\
    \hline
    $\mathrm{f}_{2, integral} = 100$ & (0.19, 75) & (0.19, 75) & 1 & 96.79 (1.07) & 98.35 (0.46) \\
    \hline
    \end{tabular}
\end{table}

\begin{table}[!h]
\caption{\label{f2_hig}Average $\mathrm{f}_{2,integral}$ and variance for data generated from Higuchi model across 100 MC simulations. The variance is reported in parentheses.}
\centering
  \begin{tabular}{|c|c|c|c|c|c|}
    \hline
    \bf{Ground-truth} & \bf{$\omega_R$} & \bf{$\omega_T$} & \bf{Variance} & \bf{CTGP} & \bf{LSGP}\\[5pt]
    \hline
    $\mathrm{f}_{2, integral} = 45$ & 110 & 70 & 1 & 45.03 (0.11) & 45.03 (0.10) \\
    \hline
    $\mathrm{f}_{2, integral} = 49.60$ & 110 & 77 & 1 & 49.58 (0.14) & 49.57 (0.13)\\
    \hline
    $\mathrm{f}_{2, integral} = 49.60$ & 110 & 77 & 5 & 49.50 (0.60) & 49.55 (0.56) \\
    \hline
    $\mathrm{f}_{2, integral} = 51.07$ & 110 & 79 & 1 & 50.98 (0.14) & 50.97 (0.14)\\
    \hline
    $\mathrm{f}_{2, integral} = 51.07$ & 110 & 79 & 5 & 50.95 (0.70) & 51.04 (0.69) \\
    \hline
    $\mathrm{f}_{2, integral} = 67.35$ & 110 & 95 & 1 & 67.31 (0.69) & 67.19 (0.63) \\
    \hline
    $\mathrm{f}_{2, integral} = 100$ & 110 & 110 & 1 & 97.05 (0.47) & 97.21 (0.35)\\
    \hline
  \end{tabular}
\end{table}

Table \ref{f2_log} shows the average and variance (across the 100 MC simulations) of the posterior mean  of $E[f_{2,integral}|\text{data}]$    for each scenario generated from the logistic function for both the LSGP model and the CTGP model. Table \ref{f2_hig} reports the same quantities for each scenario generated from the Higuchi model. Both LSGP and CTGP compute an estimate of the $\mathrm{f}_{2,integral}$ that closely matches the ground-truth value. Additionally, both models have similarly low variances across the MC simulations. 
This shows that the both the models provide consistent results across different MC runs.

Although the models return similar posterior means for $\mathrm{f}_{2,integral}$, they differ in the posterior distribution.
 Figure \ref{fig:f2_distb_sim} shows the distribution of $\mathrm{f}_{2,integral}$ calculated by the LSGP model and CTGP model given the logistic and Higuchi data in one MC run, where the generating parameters were $\alpha_1^R = 100$, $\alpha_2^R = 75$, $\beta^R = 0.19$, $\alpha_2^T = 80$, $\beta^T = 0.215$, and $\sigma^2=5$ for the logistic data and $\omega_R=110$, $\omega_T=79$, and $\sigma^2=5$ for the Higuchi data. These density plots were generated from 1000 posterior sample estimates of $\mathrm{f}_{2,integral}$. In Figure \ref{fig:f2_distb_sim}, we observe that the LSGP model has lower variability in the calculation of the $\mathrm{f}_{2,integral}$ than the CTGP model, while both models are able to capture the true $\mathrm{f}_{2, integral}$ value given the generating parameters of the simulated dataset (the true value is well within the credible intervals of the distribution). For the given MC run, the CTGP model computes a probability of similarity of $87.6\%$ for the logistic data, while the LSGP model computes a probability of similarity of $97.7\%$. For the Higuchi data, the CTGP model computes a probability of similarity of $82.1\%$, while the LSGP model determines similarity with a probability of $94.4\%$
 
\subsubsection{Predictive Performance}
As highlighted in the previous sections, the primary difference between the CTGP and the LSGP lies in the dissolution spline kernel. This kernel enables us to account for the typical monotonicity of dissolution curves, resulting in more accurate predictions with reduced uncertainty.
To corroborate this statement, we assessed the Continuous Rank Probability Score (CRPS) \citep{jolliffe_forecast_2003}, defined as 
\begin{equation}
    CRPS(F, y(t)) = \int^\infty_{-\infty}(F(f(t)) - H(f(t) \geq y(t)))^2dt ~,
\end{equation}
where $F(f(t))$ is the CDF over values of $f(t)$ and $H(f(t) \geq y(t))$ is the Heaviside step function with value $1$ when $f(t)$ exceeds the observation $y(t)$. The CRPS thus compares a distribution of predicted values to a singular observation, and given discrete samples from the distribution of predicted values, it can be computed as \citep{zamo_estimation_2018}:
\begin{equation}
    CRPS(m, y(t)) = \frac{1}{m}\sum^m_{i=1}|f^{(i)}(t) - y(t)| - \frac{1}{2m^2}\sum^m_{i=1}\sum^m_{j=1}|f^{(i)}(t) - f^{(j)}(t)| ~,
\end{equation}
where $m$ is the number of posterior samples drawn of $f(t)$. We conducted a Leave-One-Out cross-validation (LOO) experiment, where we excluded all dissolution data at a single time point from the model's input data for each time point $t \in {10, 20, 30, 40, 50, 60}$. We then computed the posterior prediction for $f(t)$ from both models at the excluded time point $t$. The CRPS was computed for the LSGP model and the CTGP model for each of the $p=6$ time points and then averaged across these time points for each of the $100$ MC simulations. Table \ref{crps_log} reports the average CRPS and variance computed across $100$ MC simulations for the logistic model, while Table \ref{crps_hig} reports the average and variance  for the Higuchi model. The CRPS for the LSGP model is significantly  lower than that of the CTGP model, demonstrating that there is less uncertainty in our predictions. These results agree with what is shown in Figures \ref{fig:ctgp-exc-30} and \ref{fig:spline-exc-30}, where the LSGP model predicts the underlying function more accurately than the CTGP model.%

\begin{table}[!h]
  \caption{\label{crps_log} LOO average and variance of CRPS for the logistic function scenario.}
\centering
  \begin{tabular}{|c|c|c|c|}
    \hline
    \bf{$(\beta^{T}, \alpha_2^{T})$} & \bf{Variance} & \bf{CTGP} & \bf{LSGP}\\[5pt]
    \hline
    (0.19, 80) & 1 & 4.58 (0.35) & 0.79 (0.31) \\
    \hline
    (0.226, 80) & 1 & 7.35 (0.44) & 0.74 (0.01) \\
    \hline
    (0.226, 80) & 5 & 6.77 (1.46) & 1.70 (0.04)\\
    \hline
    (0.215, 80) & 1 & 7.37 (0.50) & 0.73 (0.01) \\
    \hline
    (0.215, 80) & 5 & 7.01 (1.42) & 1.70 (0.04) \\
    \hline
    (0.202, 75) & 1 & 5.61 (0.35) & 0.73 (0.01) \\
    \hline
    (0.202, 75) & 5 & 5.53 (1.57) & 1.70 (0.06)\\
    \hline
    (0.19, 75) & 1 & 4.27 (0.29) & 0.77 (0.02) \\
    \hline
  \end{tabular}
\end{table}

\begin{table}[!h]
\caption{\label{crps_hig} LOO average and variance of CRPS for the Higuchi model scenario.}
\centering
  \begin{tabular}{|c|c|c|c|}
    \hline
    \bf{$\omega_T$} & \bf{Variance} & \bf{CTGP} & \bf{LSGP}\\[5pt]
    \hline
    70 & 1 & 5.62 (0.24) & 1.36 (0.05)\\
    \hline
    77 & 1 & 4.94 (0.25) & 1.38 (0.05)\\
    \hline
    77 & 5 & 4.96 (0.57) & 2.26 (0.16)\\
    \hline
    79 & 1 & 4.86 (0.23) & 1.41 (0.04)\\
    \hline
    79 & 5 & 4.96 (0.61) & 2.21 (0.14)\\
    \hline
    95 & 1 & 3.68 (0.32) & 1.50 (0.08) \\
    \hline
    110 & 1 & 3.15 (0.37) & 1.56 (0.49)\\
    \hline
  \end{tabular}
\end{table}

\subsubsection{MSD Evaluation}
One advantage of a model like LSGP, which directly estimates the dissolution curve, is its ability to perform various hypothesis tests. Typically, each hypothesis test assesses different notions of similarity. For example, we can use the posterior samples from LSGP to conduct an MSD-based test, as discussed in Section \ref{sec:hypest}.

The MSD-based test for comparison of dissolution profiles was also evaluated for the LSGP model across 100 MC simulations. For each simulation, a decision of similarity was determined whenever $d^u_{R,T} \leq d_{limit}$, and percentage of similarity decisions was consequently computed by counting the number of times the dissolution profiles were deemed similar over the course of the 100 MC simulations. Tables \ref{msd_log} and \ref{msd_hig} report the  percentage of  similarity decisions determined by computing the MSD over 100 MC simulations for the logistic and Higuchi data, respectively. The maximum difference in percentage dissolution between the true underlying dissolution curves of the reference and test groups across the $p$ data points is also reported to demonstrate the basis for this test's decision of similarity for each scenario. 

The results of this similarity test show an inconsistency with the results of the $\mathrm{f}_2$ test for similarity. In some scenarios, the decision of similarity differs from the $\mathrm{f}_2$ test for comparison, as these tests are based on different criteria for similarity. For example, in Table \ref{msd_hig}, for the case where $\mathrm{f}_{2, integral} = 67.35$, the probability of similarity determined by the MSD test is 0. The $\mathrm{f}_{2, integral}$ would determine the dissolution profiles to be similar, while the MSD test concludes dissimilarity, reflecting how the different definitions of similarity applied by each test can produce differing conclusions. One advantage of using the LSGP model for dissolution testing, however, is that the shape of the dissolution profiles is directly estimated by the model and can be visualised, providing an additional assessment of the dissolution profile similarity to the result of a hypothesis test.

\begin{table}[!h]
\centering
  \caption{\label{msd_log} The probability of similarity for the logistic data calculated by the LSGP model by computing the MSD across 100 MC simulations and counting the number of times a decision of similarity was made.}
  \begin{tabular}{|c|c|c|c|c|c|c|}
    \hline
    \bf{Scenario} & \bf{$\beta^{R}, \alpha_2^{R}$} & \bf{$\beta^{T}, \alpha_2^{T}$} & \bf{Variance} & Max Difference & \bf{\% decisions}\\[5pt]
    \hline
    $\mathrm{f}_{2, integral} = 36.74$ & (0.14, 75) & (0.19, 80) & 1 & $29.93$ & 0 \\
    \hline
    $\mathrm{f}_{2, integral} = 48.19$ & (0.19, 75) & (0.226, 80) & 1 & $15.64$ & 0 \\
    \hline
    $\mathrm{f}_{2, integral} = 48.19$ & (0.19, 75) & (0.226, 80) & 5 & $15.64$ & 0\\
    \hline
    $\mathrm{f}_{2, integral} = 52.81$ & (0.19, 75) & (0.215, 80) & 1 & $12.88$ &  0\\
    \hline
    $\mathrm{f}_{2, integral} = 52.81$ & (0.19, 75) & (0.215, 80) & 5 & $12.88$ &  0\\
    \hline
    $\mathrm{f}_{2, integral} = 76.10$ & (0.19, 75) & (0.202, 75) & 1 & $4.61$ & 100\\
    \hline
    $\mathrm{f}_{2, integral} = 76.10$ & (0.19, 75) & (0.202, 75) & 5 & $4.61$ & 97\\
    \hline
    $\mathrm{f}_{2, integral} = 100$ & (0.19, 75) & (0.19, 75) & 1 & $0$ & 99 \\
    \hline
  \end{tabular}
\end{table}

\begin{table}[!h]
\caption{\label{msd_hig}The probability of similarity for the Higuchi data calculated by the LSGP model by computing the MSD across 100 MC simulations and counting the number of times a decision of similarity was made.}
\centering
  \begin{tabular}{|c|c|c|c|c|c|c|}
    \hline
    \bf{Scenario} & \bf{$\omega_R$} & \bf{$\omega_T$} & \bf{Variance} & Max Difference & \bf{\% decisions}\\[5pt]
    \hline
    $\mathrm{f}_{2, integral} = 45$ & 110 & 70 & 1 & $16.43$ & 0 \\
    \hline
    $\mathrm{f}_{2, integral} = 49.60$ & 110 & 77 & 1 & $13.27$ & 0 \\
    \hline
    $\mathrm{f}_{2, integral} = 49.60$ & 110 & 77 & 5 & $13.27$ & 38 \\
    \hline
    $\mathrm{f}_{2, integral} = 51.07$ & 110 & 79 & 1 &$12.39$ & 0 \\
    \hline
    $\mathrm{f}_{2, integral} = 51.07$ & 110 & 79 & 5 & $12.39$ & 51 \\
    \hline
    $\mathrm{f}_{2, integral} = 67.35$ & 110 & 95 & 1 & $5.74$ & 0 \\
    \hline
    $\mathrm{f}_{2, integral} = 100$ & 110 & 110 & 1 & $0$ & 100\\
    \hline
  \end{tabular}
\end{table}

\begin{comment}
as we see that the equality test fails to make the correct decision of similarity for the scenarios where $f_{2,integral} = 52.81$ for the logistic data, and where $f_{2,integral} = 51.07$, $\sigma^2  = 1$ and $f_{2,integral} = 67.35$ for the Higuchi data.
\end{comment}

\subsection{Case Studies}

The model was also tested on  two real dissolution datasets from \citep{ocana_using_2009}. The first dataset contains a reference and test product that differ by the inclusion of a tensioactive substance in the dissolution experiment, while the second dataset contains reference and test product with only a small change in the medium such that similarity is expected between the two groups. Data points were collected at nominal time points $t = 1, 2, \dots, 8$ for $n=12$ repetitions. These datasets are reported in Appendix Tables \ref{real_diss_1} and \ref{real_diss_2}. 
For the CTGP model, the parameters $\tau^2$ and $\sigma^2$ were estimated via Gibbs sampling from posterior inverse gamma distributions, using the default values for the inverse gamma priors as described in \citep{pourmohamad_gaussian_2022}. Additionally, the shape and range parameters, $\phi$ and $\psi$, of the covariance functions of the CTGP model were estimated via Metropolis-Hastings, as proposed in \citep{pourmohamad_gaussian_2022}.

To assess the accuracy of the  predictions of the two models on the real data, we again used the CRPS metric computed via LOO. 

\begin{table}[!h]
  \caption{\label{crps_real_data}LOO average and variance  CRPS for the LSGP model and CTGP model across the reference and test group for the two real dissolution datasets.}
\centering
  \begin{tabular}{|c|c|c|}
    \hline
    \bf{Dataset} & \bf{CTGP} & \bf{LSGP}\\[5pt]
    \hline
    Dataset 1 & 3.87 (0.01) & 2.74 (0.08) \\
    \hline
    Dataset 2 & 4.69 (4.09) & 2.65 (0.00) \\
    \hline
  \end{tabular}
\end{table}

Table \ref{crps_real_data} reports the average CRPS and variance across the reference and test group for both dissolution data sets as calculated by the LSGP model and CTGP model. We see that the CRPS for the LSGP model is less than that of the CTGP model, suggesting less uncertainty in the predictions of the LSGP model. Additionally, the $\mathrm{f}_{2, integral}$ was calculated between the reference and test group within each dataset for the LSGP model and the CTGP model; these results are shown in Table \ref{f2_real_data}.

\begin{table}[!h]
  \caption{\label{f2_real_data}Posterior mean of  $E[\mathrm{f}_2|\text{data}]$ computed by the LSGP model and CTGP model for the two datasets.}
\centering
  \begin{tabular}{|c|c|c|}
    \hline
    \bf{Dataset} & \bf{CTGP} & \bf{LSGP}\\[5pt]
    \hline
    Dataset 1 & 52.20 & 52.37 \\
    \hline
    Dataset 2 & 79.45 & 78.50 \\
    \hline
  \end{tabular}
\end{table}

\begin{figure}
    \centering
    \includegraphics[width=0.45\textwidth]{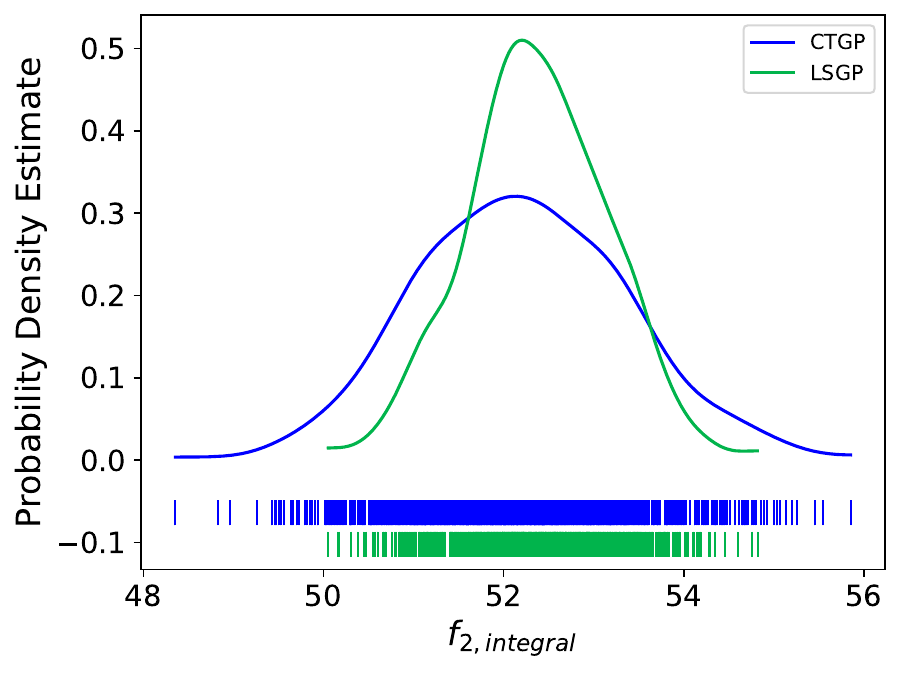}
    \includegraphics[width=0.45\textwidth]{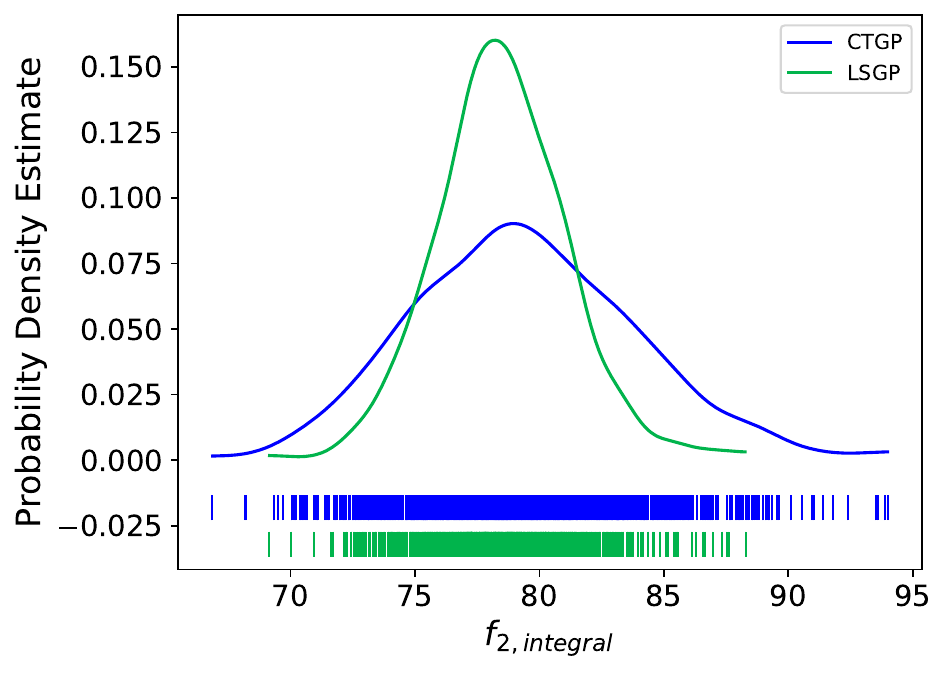}
    \caption{(Left) The posterior distribution for $\mathrm{f}_{2,integral}$  for the LSGP model and the CTGP model using data from Dataset 1. (Right) using data from Dataset 2.}
    \label{real_data_ctgp_f2}
\end{figure}

We also show  the posterior distribution of $\mathrm{f}_{2,integral}$ 
 computed for both datasets for the LSGP model and the CTGP model in Figure \ref{real_data_ctgp_f2} to show the posterior uncertainty for the two models. Note that, there is less uncertainty in the LSGP estimate of $\mathrm{f}_{2,integral}$  than in the CTGP model.

\begin{figure}
        \centering
         \includegraphics[width=0.3\textwidth]{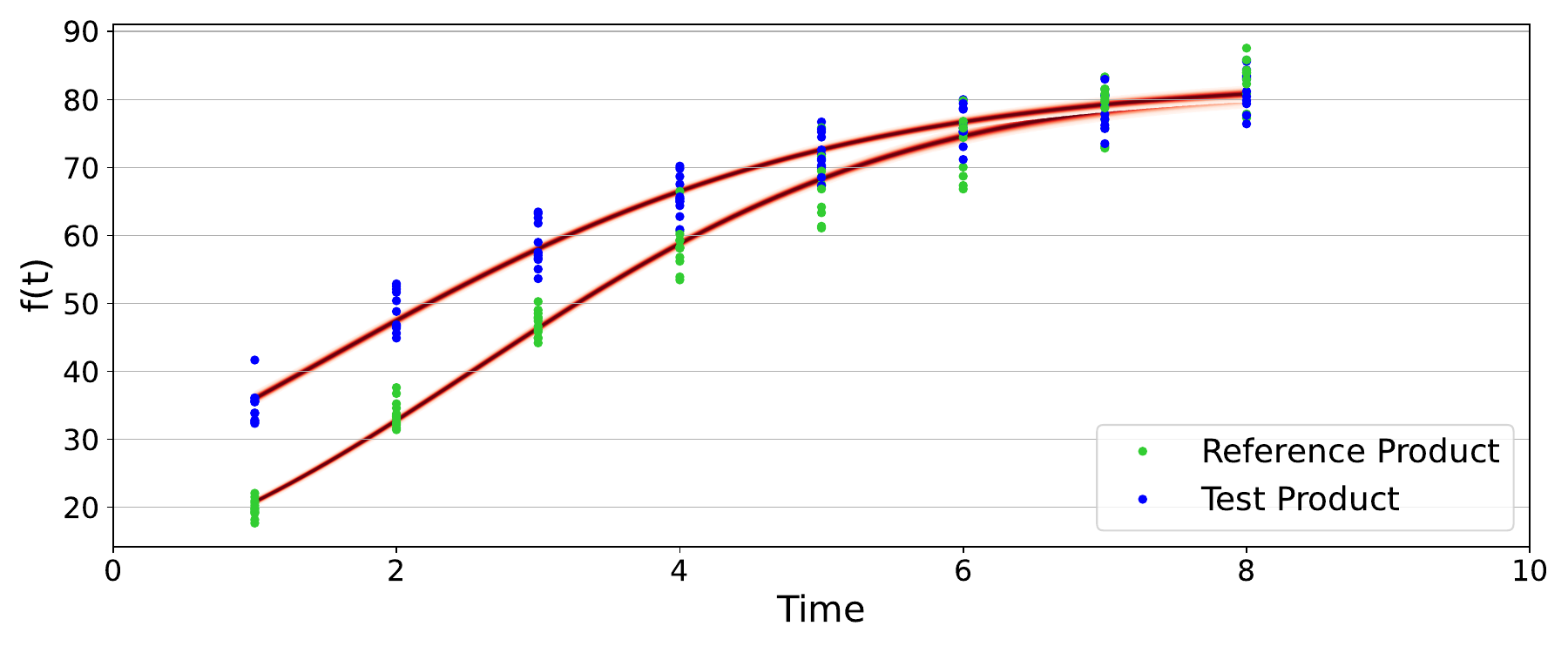}
         \includegraphics[width=0.3\textwidth]{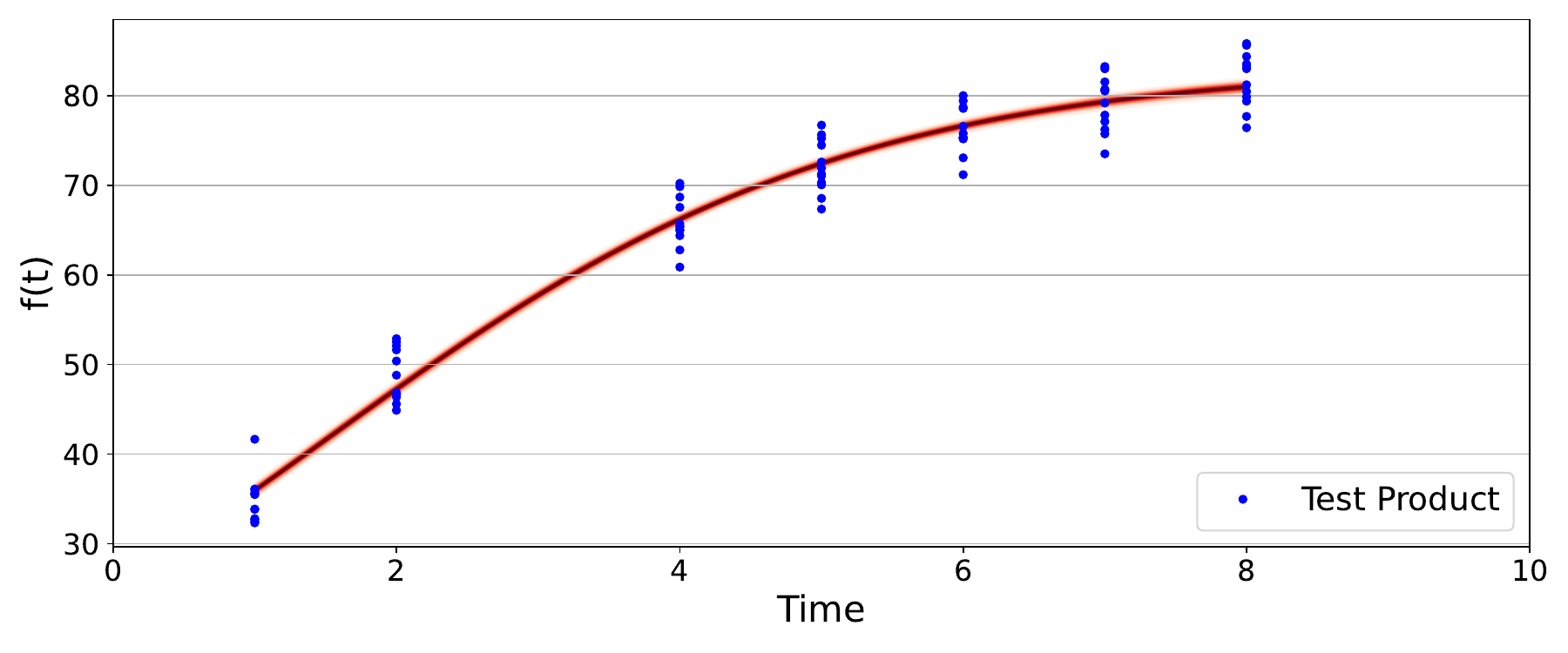}
         \includegraphics[width=0.3\textwidth]{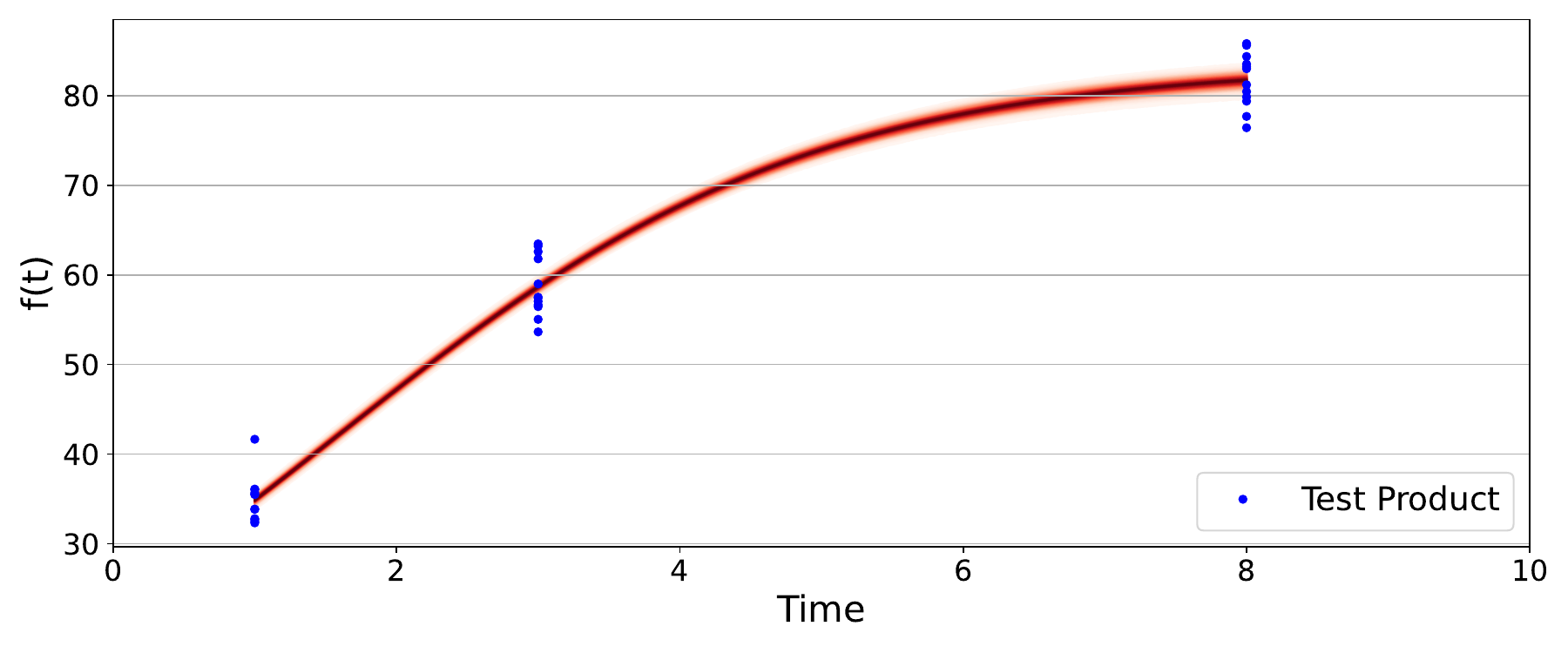}
        \includegraphics[width=0.3\textwidth]{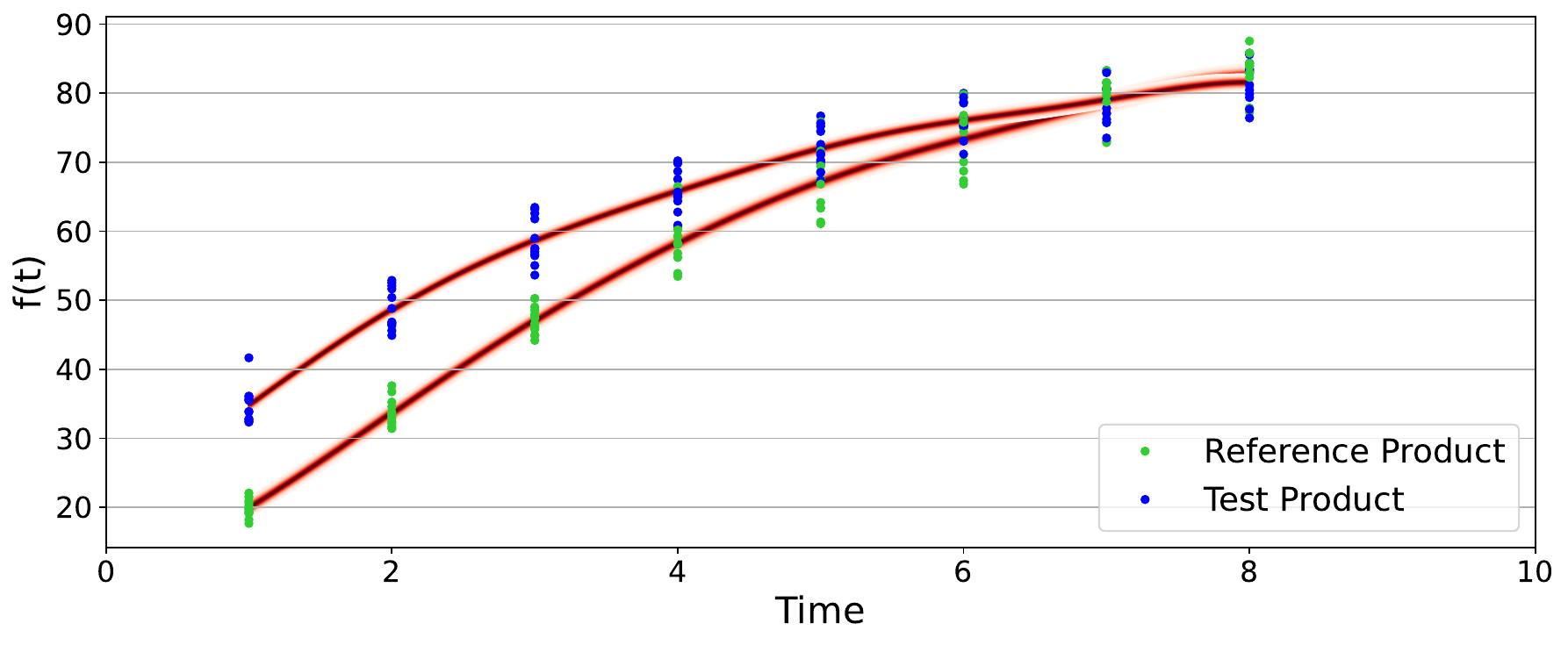}
        \includegraphics[width=0.3\textwidth]{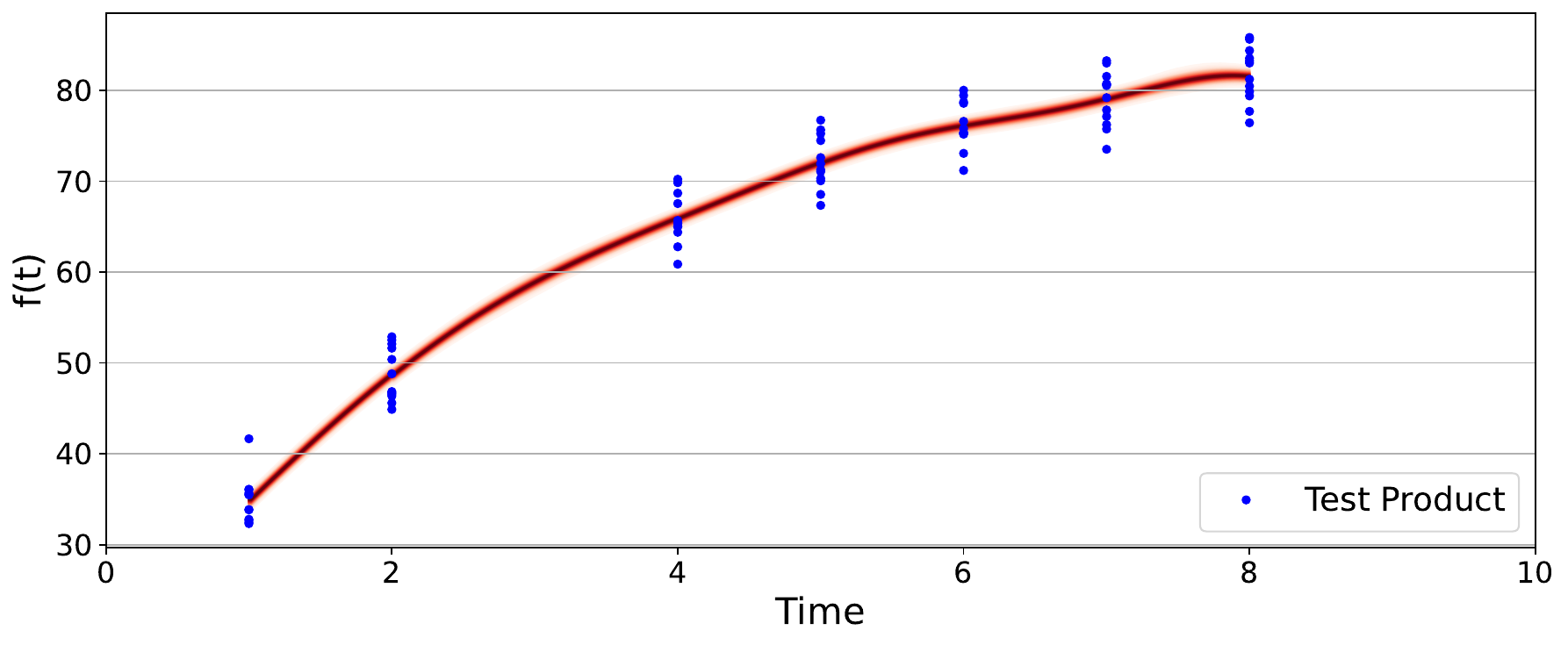}
        \includegraphics[width=0.3\textwidth]{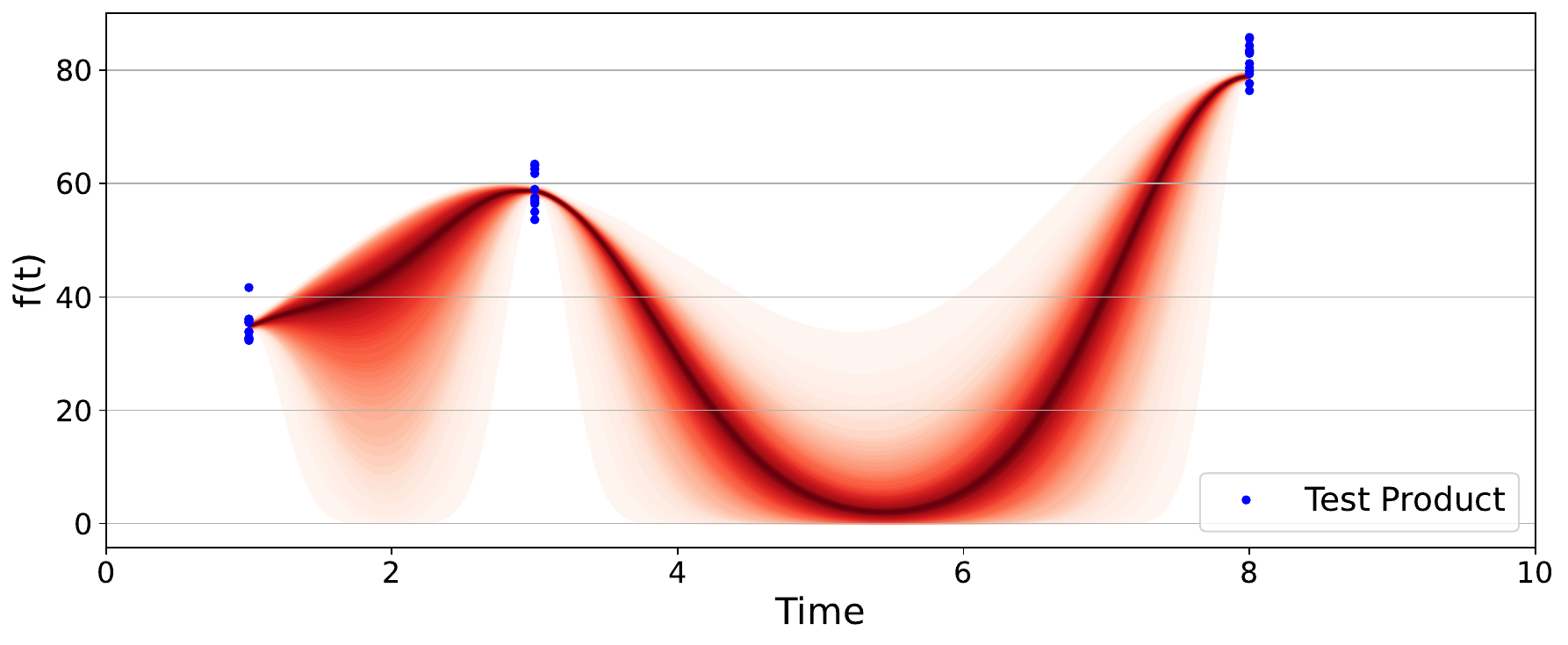}
        \caption{(Top: Left to Right) The predictions generated by the LSGP model using real dissolution dataset 1 and all time points to train the model; model trained excluding $t=3$; model trained with only $t = 1, 3, 8$.  (Bottom) CTGP model on dissolution dataset 1.}
        \label{fig:realdatapost}
    \end{figure}

To confirm that this is the case, in Figure \ref{fig:realdatapost} we plot the predictions of the LSGP model and the CTGP model for Dataset 1.
At the top, from left to right, we shows the posterior predictions for LSGP for both the reference and test group computed in 3 different case:
(left)  the model was fitted on all of the observations; (center) the model was fitted excluding the observation at $t=3$; (right) the model was fitted using only the data at  $t = 1, 3, 8$.    At the bottom, from left to right, we report the same plots for CTGP.

In the plots on the right, we observe that the CTGP model tends to revert to the GP mean of zero between observations, leading to a dissolution curve that diverges from the shape of the true underlying function. It is important to note that the lengthscales for the kernels in CTGP were estimated using the Metropolis-Hastings MCMC method, as described in \citep{pourmohamad_gaussian_2022}. This highlights how the estimation of the lengthscales can significantly affect the performance of the model.

In contrast, the LSGP model preserves the data's monotonically increasing trend, even when multiple data points are excluded, yielding an approximated dissolution curve that more closely resembles the true underlying function.

\subsection{Dissolution of ibuprofen under various conditions}
\label{sec:covariatesexp}
In this section, we demonstrate how incorporating covariates into the LSGP model enables extrapolation of the dissolution curve profile across different experimental settings.  We consider real dissolution data involving the dissolution of ibuprofen under various medium-type and viscosity conditions, intended as a high-level representation of the gastrointestinal environment.

Dissolution data from ibuprofen particles were collected using the United States Pharmacopeia (USP) paddle dissolution apparatus \citep{usp_2020}, as described in \citep{navas-bachiller_vitro_2023}. Media used are Phosphate Buffer (PB), with a pH of 6.8, similar to the small intestine, at which ibuprofen is highly soluble, and dilute Hydrochloric acid (HCl), at a pH of approximately 1.2, similar to the fasted-state gastric environment, where ibuprofen is poorly soluble. The viscosity of the media was adjusted using Hydroxypropyl methyl cellulose (HPMC), to three different levels. Level 1, 0.7 mPa.s, is the viscosity of water at 37C (body temperature), and thus the viscosity of most simple aqueous dissolution media used in dissolution testing. Level 2 viscosity media were adjusted to 1.4 mPa.s using HPMC, which is the viscosity of milk (an ingredient sometimes used in generating media to represent the gastric fed state). Level 3 viscosity media, at 5.5 mPa.s, represents a mid-range point of measured viscosity of gastric aspirates at a shear rate of 50/s \citep{pedersen_characterization_2013}. These variations in pH and solubility are aimed to be a high-level representation of the environment in the gastrointestinal tract, and such adjustments to dissolution media can be considered in attempting to develop biopredictive dissolution tests. We examined the inclusion of the covariates by modelling real dissolution data for several different experiments, each with their own set of covariates, as shown in Table \ref{tab:covariates}. Each experiment used a unique set of covariates including medium, rotations per minute (rpm), viscosity, and viscosity enhancing agent (VEA), with measurements for $n=3$ dosage units across $p=9$ or $p=10$ time points.

\begin{table}[!h]
\caption{Description of each set of covariates used for the dissolution experiments of the real data.}
\centering
  \begin{tabular}{|c|c|c|c|c|c|c|}
    \hline
    \multicolumn{7}{|c|}{\bf{Dissolution Data Covariates}} \\[5pt]
    \hline
    Exp Id & Substance & Apparatus & Medium & Velocity/RPM & Viscosity & Viscosity Enhancing Agent \\[5pt]
    \hline
    1 & Ibuprofen & Paddle & PB & 50 rpm & 0.7 mPa.s  & None  \\
    \hline
    2 & Ibuprofen & Paddle & PB & 50 rpm & 1.4 mPa.s  & HPMC  \\
    \hline
    3 & Ibuprofen & Paddle & PB & 50 rpm & 5.5 mPa.s & HPMC  \\
    \hline
    4 & Ibuprofen & Paddle & PB & 100 rpm & 0.7 mPa.s  & None  \\
    \hline
    5 & Ibuprofen & Paddle & PB & 100 rpm & 1.4 mPa.s & HPMC  \\
    \hline
    6 & Ibuprofen & Paddle & PB & 100 rpm & 5.5 mPa.s  & HPMC  \\
    \hline
    7 & Ibuprofen & Paddle & HCl & 50 rpm & 0.7 mPa.s  &None  \\
    \hline
    8 & Ibuprofen & Paddle & HCl & 50 rpm & 1.4 mPa.s  & HPMC  \\
    \hline
    9 & Ibuprofen & Paddle & HCl & 50 rpm & 5.5 mPa.s & HPMC  \\
    \hline
    10 & Ibuprofen & Paddle & HCl & 100 rpm & 0.7 mPa.s & None  \\
    \hline
    11 & Ibuprofen & Paddle & HCl & 100 rpm & 1.4 mPa.s  & HPMC  \\
    \hline
    12 & Ibuprofen & Paddle & HCl & 100 rpm & 5.5 mPa.s  & HPMC  \\
    \hline
    \end{tabular}
    \label{tab:covariates}
\end{table}

The LSGP model's hyperparameters are fitted on all experiments simultaneously, enabling LSGP to learn how  the dissolution patterns depend on the covariates. For each experiment, we estimate distinct noise parameters, meaning that only the parameters of the logistic function are shared across experiments. We observe in several experiments the data following a heteroskedastic noise. In some experiments, higher noise variance is observed at earlier time points, decreasing as time increases. In other experiments, the noise variance increases as time increases. These different behaviors may be the result of the low number of repetitions (only 3) and variability in time introduced by changes of the covariates, such as medium. However, LSGP is able to adapt to these noise patterns, estimating the underlying function of these experiments appropriately.

\begin{figure}
        \centering
        \includegraphics[width=0.45\textwidth]{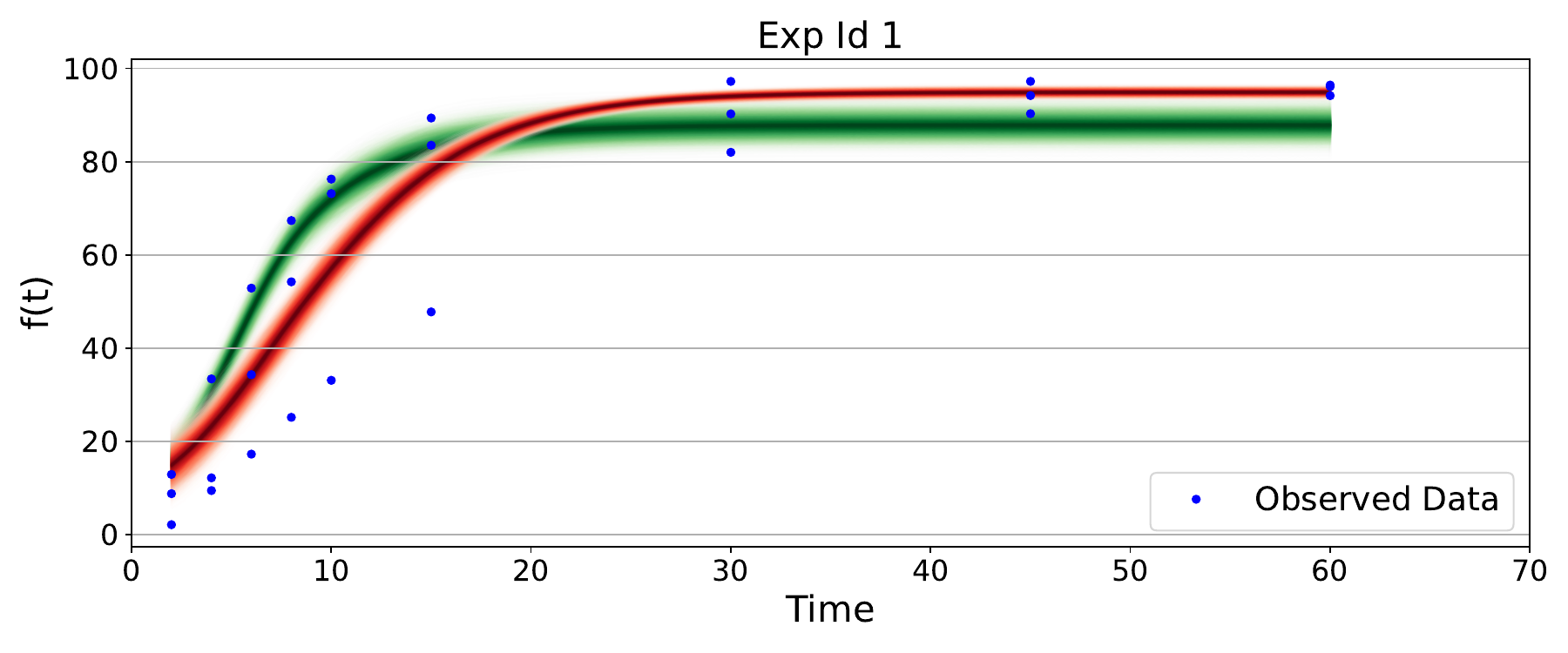}
        \includegraphics[width=0.45\textwidth]{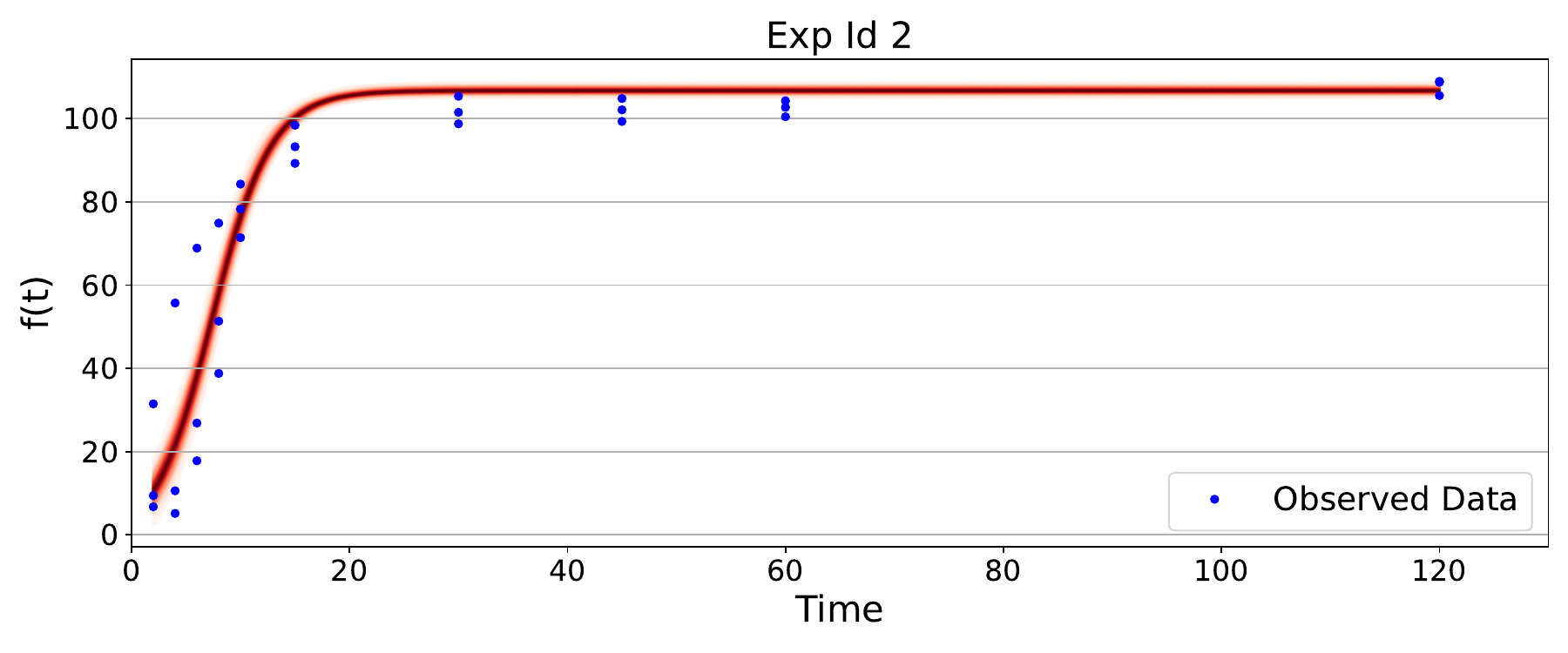}
        \includegraphics[width=0.45\textwidth]{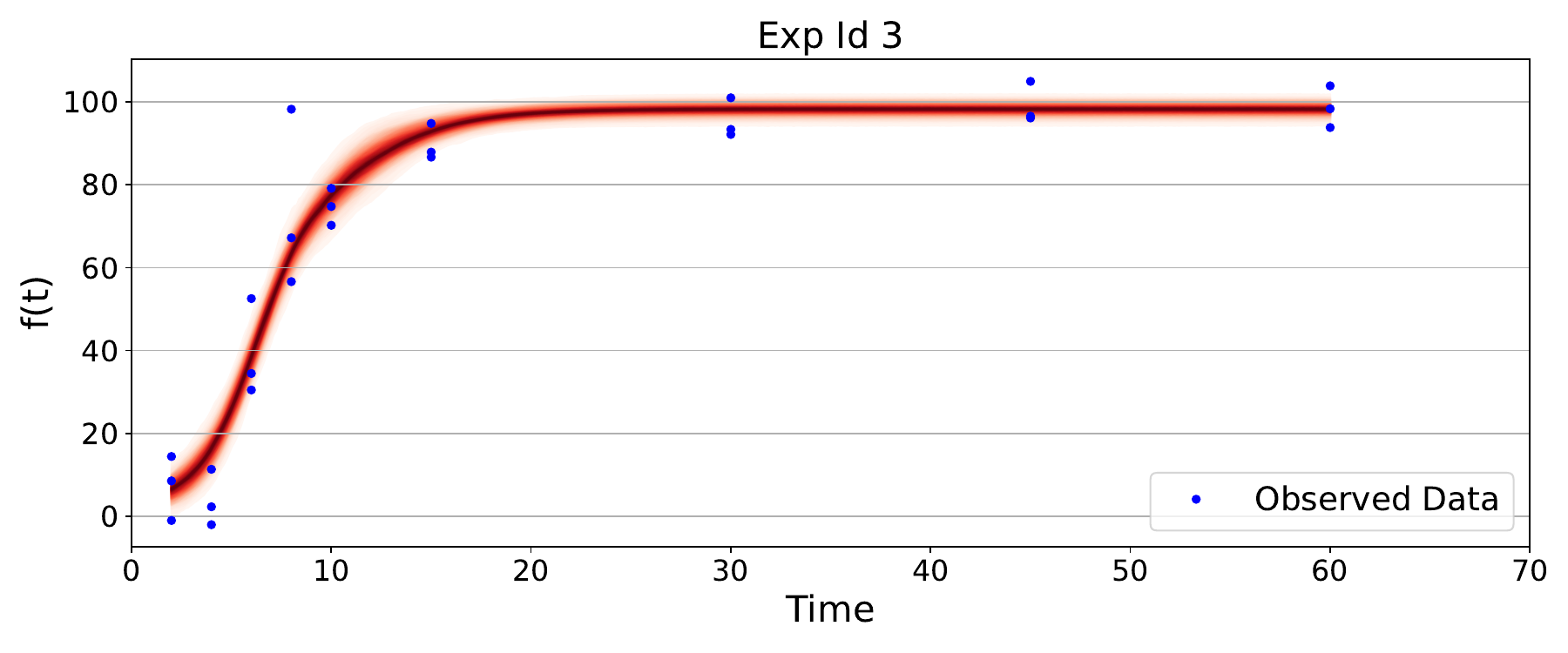}
        \includegraphics[width=0.45\textwidth]{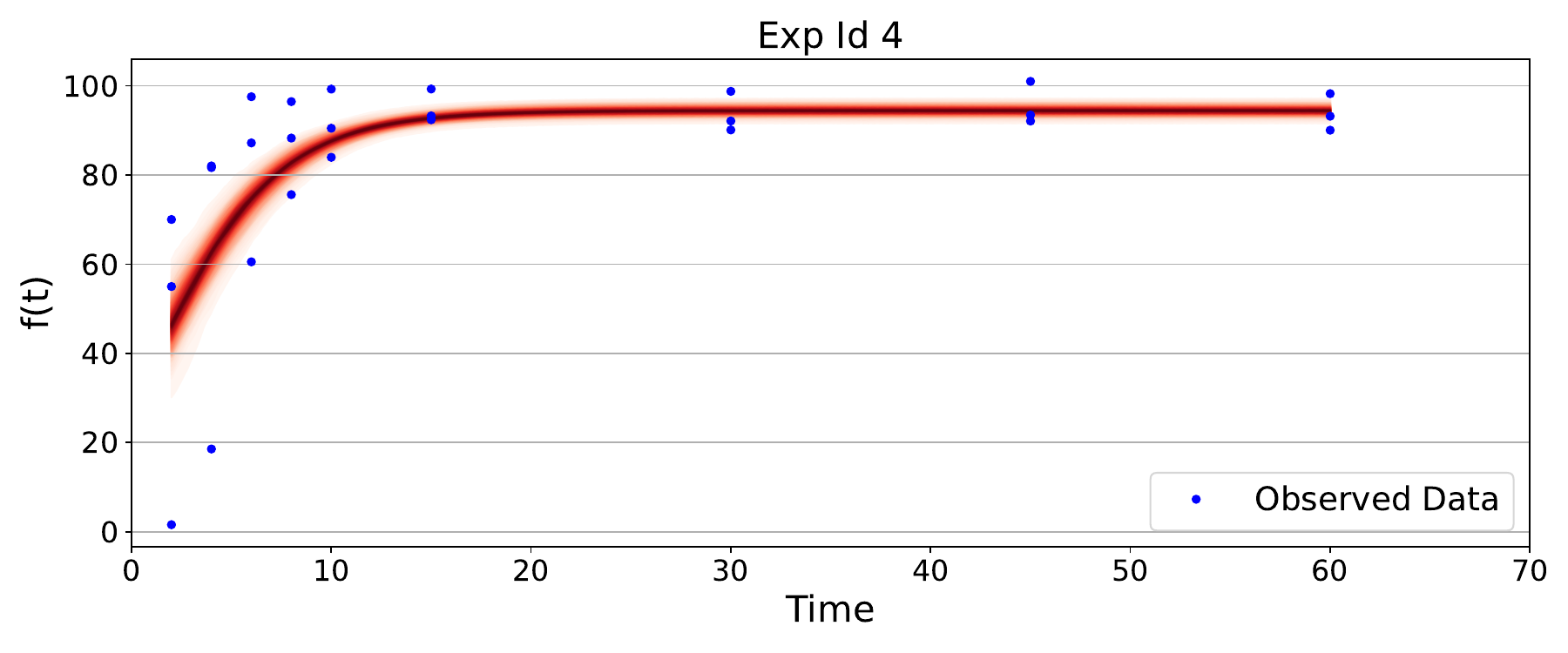}
        \includegraphics[width=0.45\textwidth]{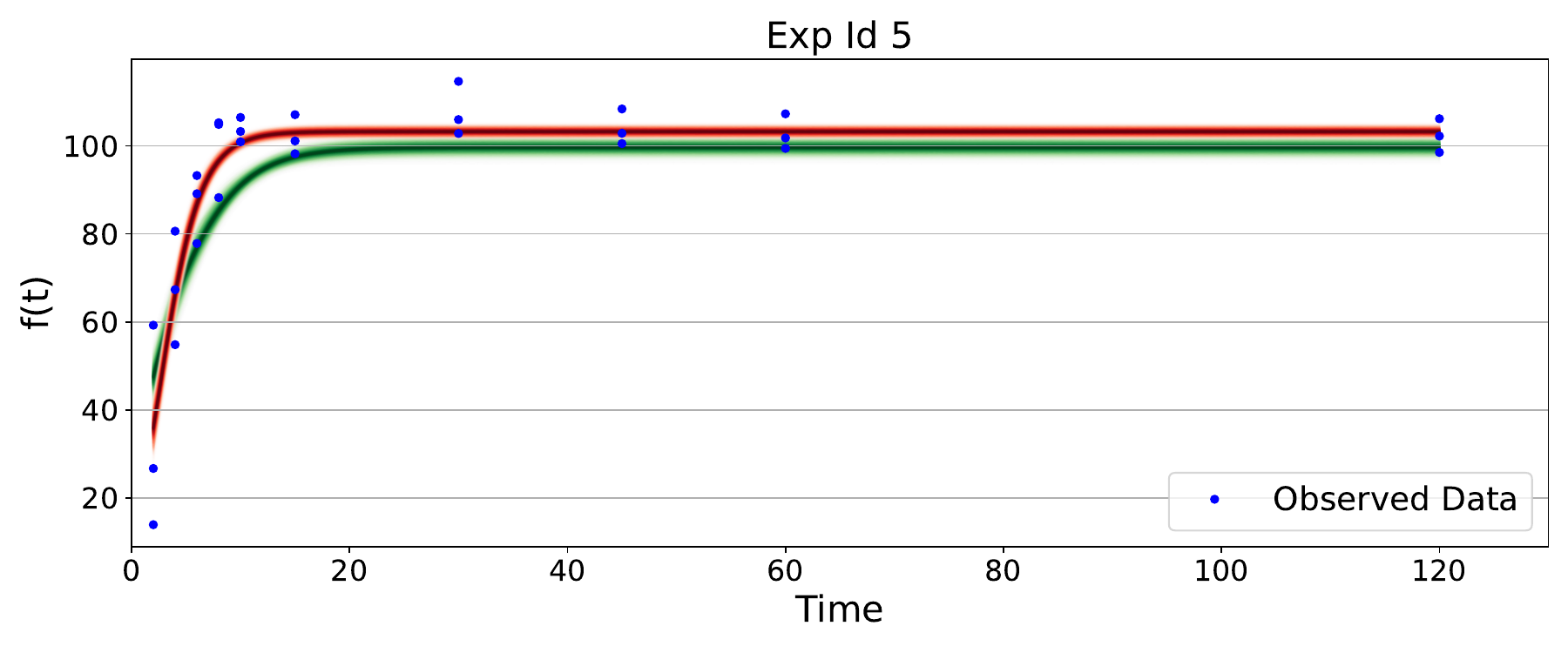}
        \includegraphics[width=0.45\textwidth]{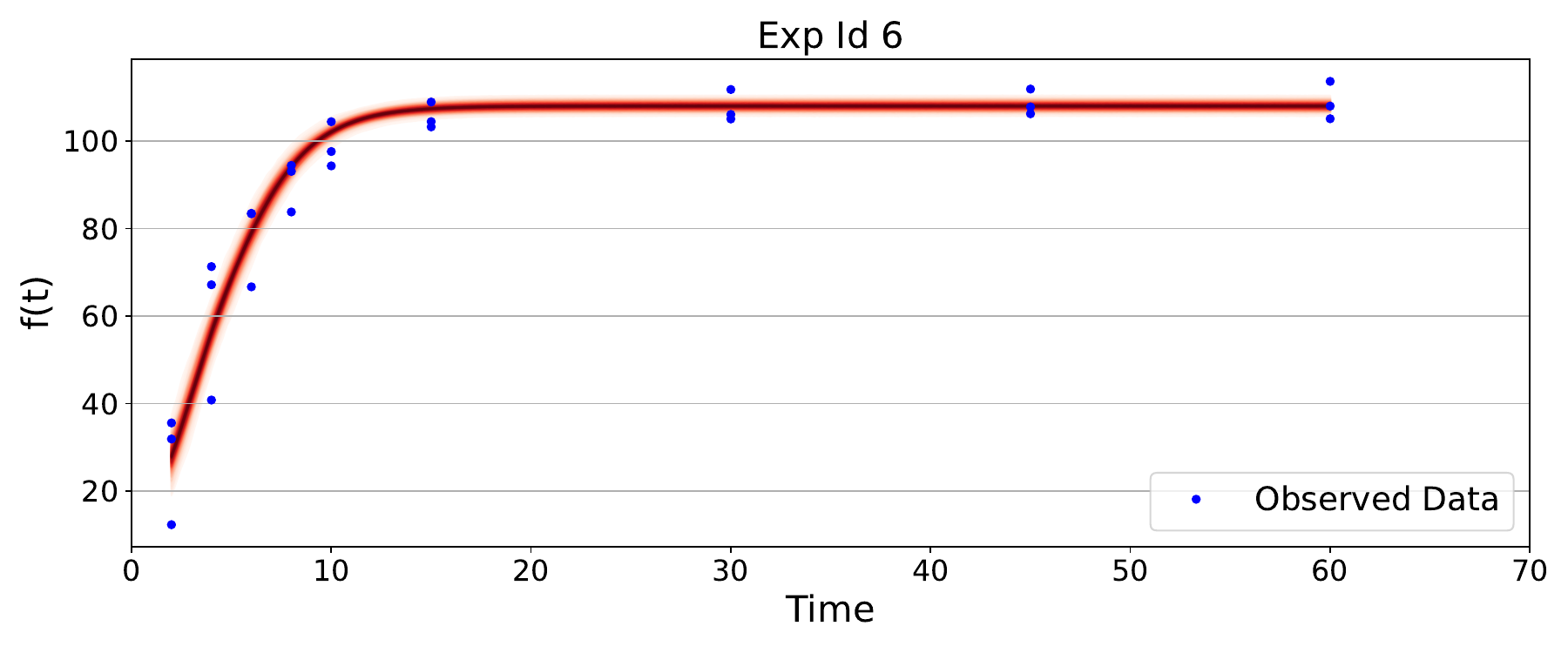}
        \includegraphics[width=0.45\textwidth]{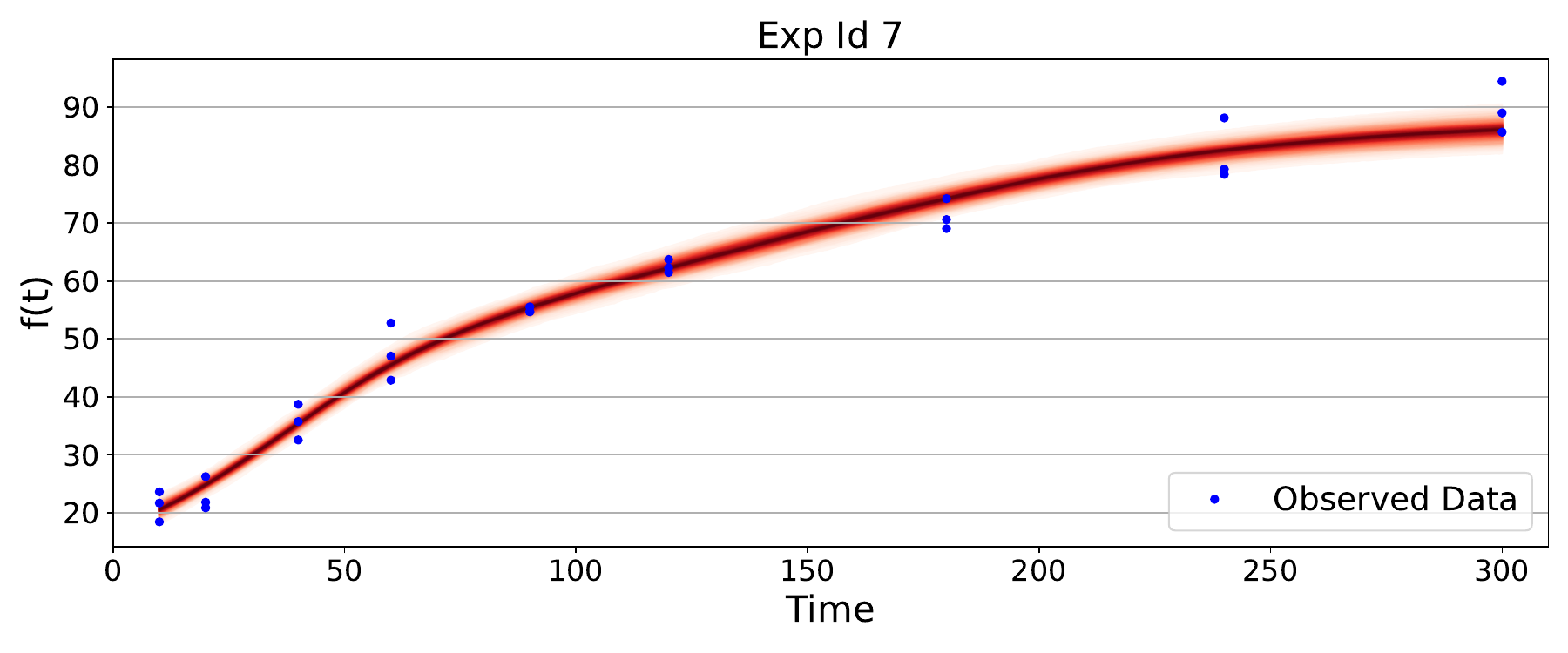}
        \includegraphics[width=0.45\textwidth]{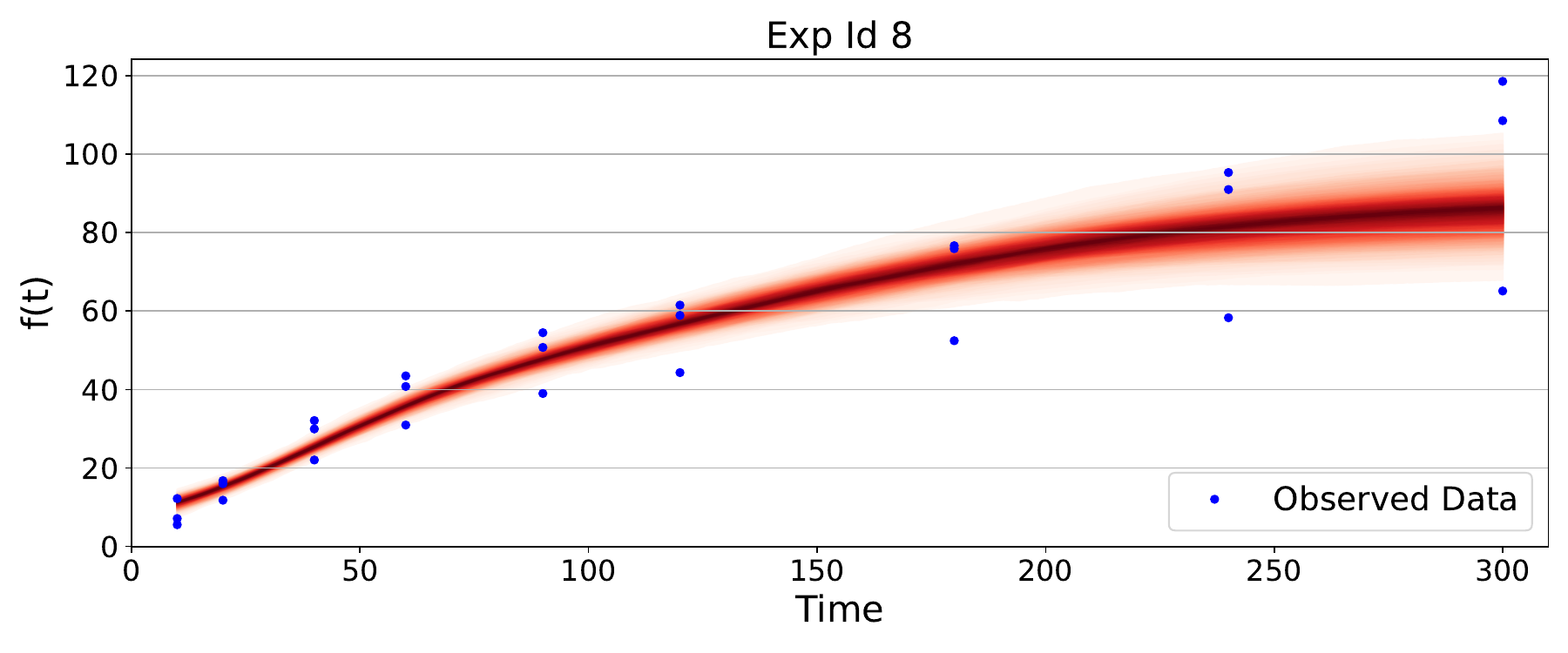}
        \includegraphics[width=0.45\textwidth]{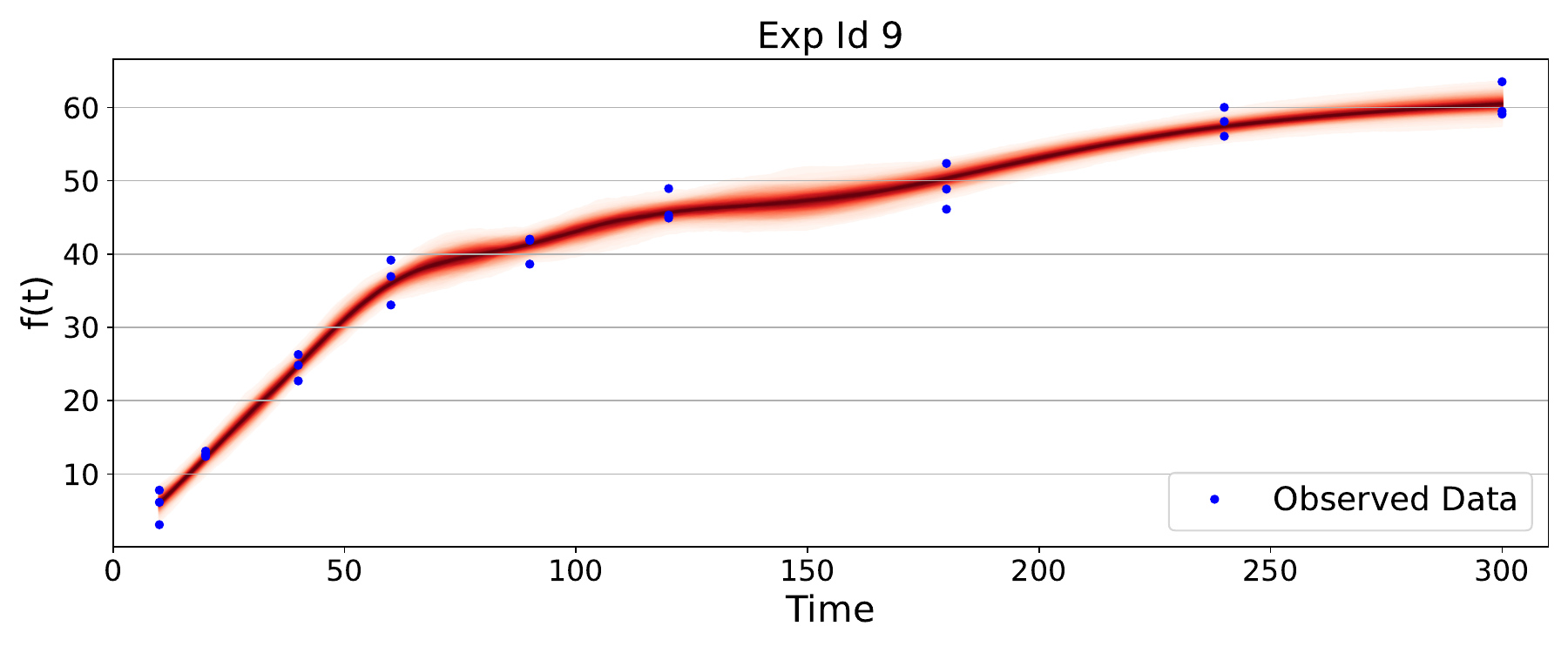}
        \includegraphics[width=0.45\textwidth]{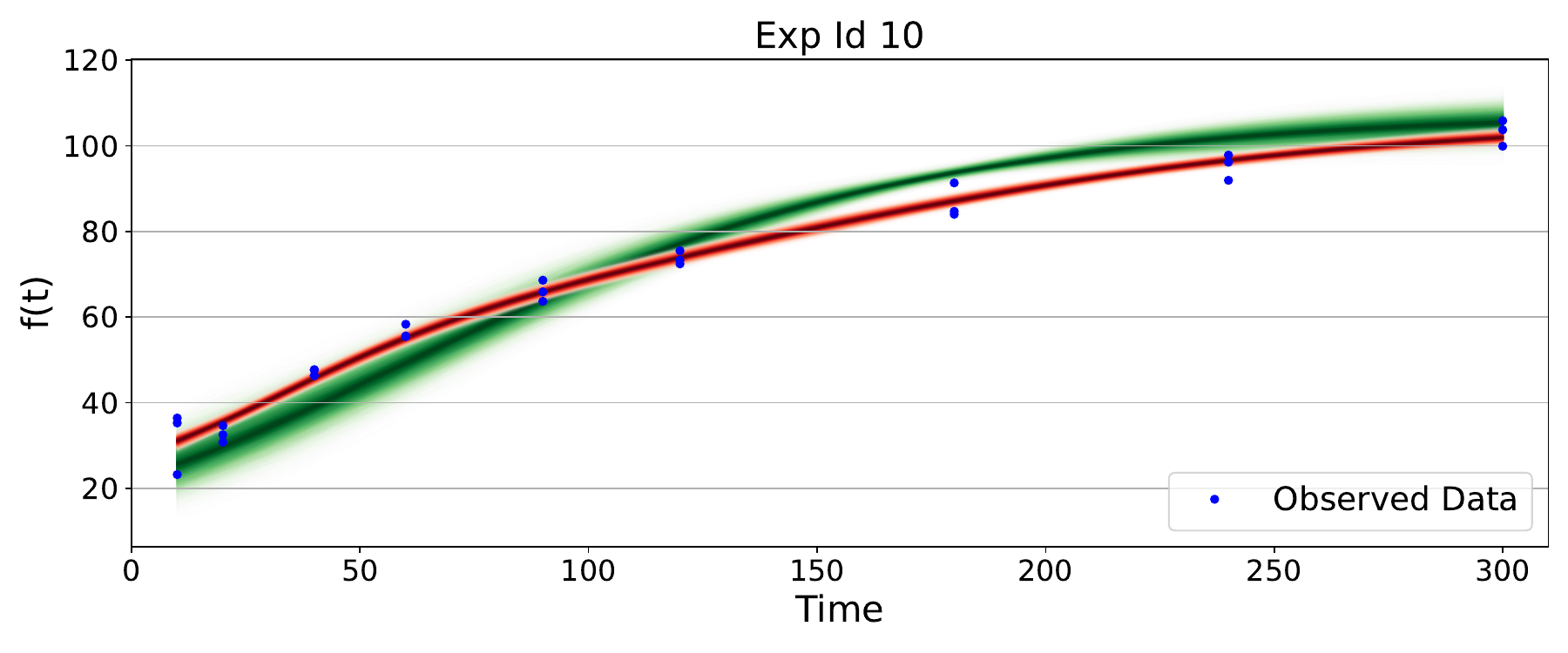}
        \includegraphics[width=0.45\textwidth]{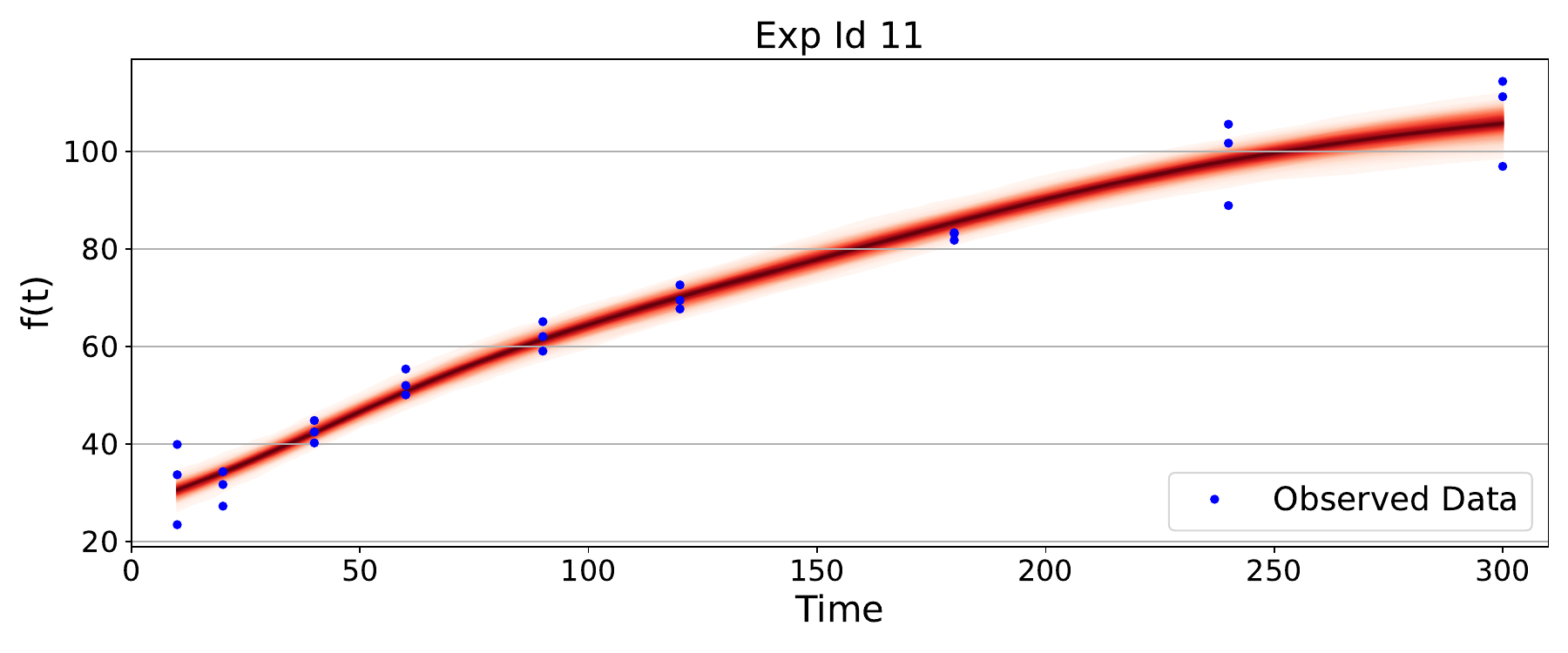}
        \includegraphics[width=0.45\textwidth]{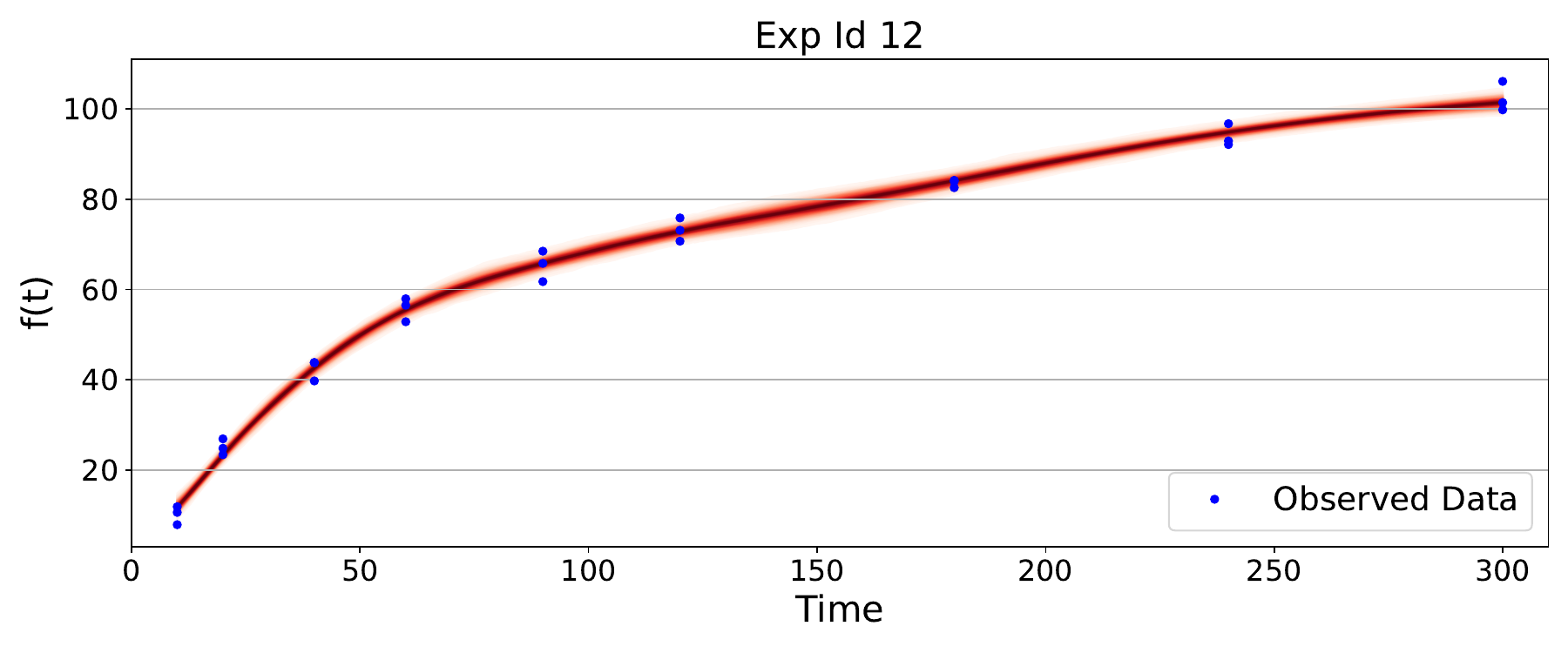}
        \caption{Predictions, $f(t)$, and $95\%$ credible regions, shown in red, generated by the LSGP model for each experiment of the real data as reported in Table \ref{tab:covariates}. For experiments $1, 5$ and $10$, the predictions and $95\%$ credible regions generated by the LSGP model, which has been fitted on all the experiments excluding the experiment being predicted (i.e. $1, 5$ or $10$), are shown in green.}
        \label{real_data_preds}
    \end{figure}

In Figure \ref{real_data_preds}, we plot the posterior mean dissolution curve and the $95\%$ credible region, shown in red, generated by the LSGP model for each experiment. For experiments $1$, $5$, and $10$, we also include the posterior mean and $95\%$ credible regions of the dissolution curve, shown in green, for the LSGP model trained on all experiments except the one for which predictions are being generated. This tests the LSGP model's ability to learn the dissolution curve of an experiment it has not been fitted on. We observe that the green and red predictions converge to a similar function, indicating that the model can effectively learn  dissolution patterns for new covariate combinations.

\section{Conclusions}

In this work, we introduced the LSGP model, which incorporates a monotonically increasing parametric function within both the mean and kernel of a Gaussian Process (GP) to estimate the dissolution process in dissolution testing experiments. Compared to fully non-parametric methods, this model demonstrates improved performance in estimating the underlying dissolution curves, by exploiting the generally monotonic nature of dissolution curves. The LSGP model enables the computation of the $\mathrm{f}_2$ similarity index, a standard measure for comparing dissolution profiles, and allows for uncertainty modeling. By drawing posterior samples from LSGP, we can obtain a distribution of $\mathrm{f}_2$ values and assess the probability of similarity between profiles. Additionally, the model supports the use of posterior samples to conduct a similarity test similar to the MSD test, which is also employed in dissolution testing, though its appropriateness is often debated. Finally, the model’s ability to incorporate covariates specific to dissolution experiments enhances both the understanding of the dissolution profile and the predictive performance of the model.      

As  future work, we plan to apply this model for smart design of experiments, using it to predict dissolution profiles across experiments conducted with different settings. Ultimately, we aim to employ this model to optimise formulations, which is one of the main reasons of performing dissolution testing.

\section*{Acknowledgments}
This work was conducted with the
financial support of the Science Foundation Ireland
Centre for Research Training in Artificial
Intelligence under Grant No. 18/CRT/6223 and by the research grant from Science Foundation Ireland (SFI), co-funded under the European Regional Development Fund (Grant number 12/RC/2275\_P2).

\section*{Appendix}

\subsubsection{Proof of Proposition \ref{prop:1}} The following are standard derivations for GP regression, which we include here for completeness. 

The likelihood can be rewritten as:
$$
\begin{aligned}
p(\{y^{(j)}(t_1),\dots,y^{(j)}(t_p)\}_{j=1}^{n}|f(\mathbf{t}),\boldsymbol{\theta})&=\prod_{j=1}^n\prod_{i=1}^pN(y^{(j)}(t_i),f(t_i),\sigma^2(t_i;\boldsymbol{\theta}))\\
&=\prod_{j=1}^nN(\mathbf{y}^{(j)},f(\mathbf{t}),D(\mathbf{t},\mathbf{t};\boldsymbol{\theta}))\\
&=\tfrac{1}{(|2\pi D(\mathbf{t},\mathbf{t};\boldsymbol{\theta})|)^{n/2}}e^{-\sum_{j=1}^n\frac{(\mathbf{y}^{(j)}-f(\mathbf{t}))^\top D^{-1}(\mathbf{t},\mathbf{t};\boldsymbol{\theta})(\mathbf{y}^{(j)}-f(\mathbf{t}))}{2}}
\end{aligned}
$$
where $\mathbf{y}^{(j)}=[y^{(j)}(t_1),\dots,y^{(j)}(t_p)]^\top$ and $D(\mathbf{t},\mathbf{t};\boldsymbol{\theta})$ is given in \eqref{eq:D}. Therefore, the likelihood equals to:
$$
\begin{aligned}
p(\{y^{(j)}(t_1),\dots,y^{(j)}(t_p)\}_{j=1}^{n}|f(\mathbf{t}),\boldsymbol{\theta})&=\tfrac{(|2\pi \frac{1}{n}D(\mathbf{t},\mathbf{t};\boldsymbol{\theta})|)^{1/2}}{(|2\pi D(\mathbf{t},\mathbf{t};\boldsymbol{\theta})|)^{n/2}}N\left({\bar{\bf y}};f(\mathbf{t}),\tfrac{1}{n}D(\mathbf{t},\mathbf{t};\boldsymbol{\theta})\right)\\
&e^{\frac{-\sum_{j=1}^n (\mathbf{ y}^{(j)})^\top(D(\mathbf{t},\mathbf{t};\boldsymbol{\theta}))^{-1}\mathbf{y}^{(j)}+n\bar{\bf y}^\top (D(\mathbf{t},\mathbf{t};\boldsymbol{\theta}))^{-1}\bar{\bf y}}{2}}, \\
&=\tfrac{1}{n^{\frac{p}{2}}(|2\pi D(\mathbf{t},\mathbf{t};\boldsymbol{\theta})|)^{\frac{n-1}{2}}}e^{\frac{-\sum_{j=1}^n (\mathbf{ y}^{(j)})^\top(D(\mathbf{t},\mathbf{t};\boldsymbol{\theta}))^{-1}\mathbf{y}^{(j)}+n\bar{\bf y}^\top (D(\mathbf{t},\mathbf{t};\boldsymbol{\theta}))^{-1}\bar{\bf y}}{2}}\\
&N\left({\bar{\bf y}};f(\mathbf{t}),\tfrac{1}{n}D(\mathbf{t},\mathbf{t};\boldsymbol{\theta})\right).
\end{aligned}
$$
Applying Bayes' theorem, we have that, 
$$
\begin{aligned}
&p(f(t)|\{y^{(j)}(t_1),\dots,y^{(j)}(t_p)\}_{j=1}^{n},\boldsymbol{\theta}) \propto p(\{y^{(j)}(t_1),\dots,y^{(j)}(t_p)\}_{j=1}^{n}|f(\mathbf{t}),\boldsymbol{\theta})p(f(\mathbf{t})|\boldsymbol{\theta}) \\
&\propto N\left({\bar{\bf y}};f(\mathbf{t}),\tfrac{1}{n}D(\mathbf{t},\mathbf{t};\boldsymbol{\theta})\right)N\left(f(\mathbf{t});\mu(\mathbf{t};\mathbf{\theta}), K(\mathbf{t}, \mathbf{t}; \mathbf{\theta})\right). \\
\end{aligned}
$$
We can then  rewrite the product of the  two PDFs in the last equation as:
$$
\begin{aligned}
&p(f(t)|\{y^{(j)}(t_1),\dots,y^{(j)}(t_p)\}_{j=1}^{n},\boldsymbol{\theta})\\
&\propto \frac{1}{(2\pi)^p|\tfrac{1}{n}D(\mathbf{t},\mathbf{t};\boldsymbol{\theta})|^{1/2}|K(\mathbf{t}, \mathbf{t}; \mathbf{\theta})|^{1/2}}e^{-\frac{1}{2}\left(({\bar{\bf y}} - f(\mathbf{t}))^\top (\tfrac{1}{n}D(\mathbf{t},\mathbf{t};\boldsymbol{\theta}))^{-1} ({\bar{\bf y}} - f(\mathbf{t})) \; + \; (f(\mathbf{t}) - \mu(\mathbf{t};\mathbf{\theta}))^\top K(\mathbf{t},\mathbf{t};\boldsymbol{\theta})^{-1}(f(\mathbf{t}) - \mu(\mathbf{t};\mathbf{\theta}))\right)}. \\
\end{aligned}
$$
By removing the constant terms and expanding the quadratic forms inside the exponential function we have that:
$$
\begin{aligned}
&p(f(t)|\{y^{(j)}(t_1),\dots,y^{(j)}(t_p)\}_{j=1}^{n},\boldsymbol{\theta})\\
&\propto e^{-\frac{1}{2}\left(f(\mathbf{t})^\top ((\tfrac{1}{n}D(\mathbf{t},\mathbf{t};\boldsymbol{\theta}))^{-1} + K(\mathbf{t},\mathbf{t};\boldsymbol{\theta})^{-1}) f(\mathbf{t}) \; - \; 2f(\mathbf{t})^\top((\tfrac{1}{n}D(\mathbf{t},\mathbf{t};\boldsymbol{\theta}))^{-1}{\bar{\bf y}} + K(\mathbf{t},\mathbf{t};\boldsymbol{\theta})^{-1}\mu(\mathbf{t};\mathbf{\theta})) \; + \; {\bar{\bf y}}^\top (\tfrac{1}{n}D(\mathbf{t},\mathbf{t};\boldsymbol{\theta}))^{-1} {\bar{\bf y}} \; + \; \mu(\mathbf{t};\mathbf{\theta})^\top K(\mathbf{t},\mathbf{t};\boldsymbol{\theta})^{-1}\mu(\mathbf{t};\mathbf{\theta})\right)}, \\
\label{eq:proofpost0}
&\propto e^{-\frac{1}{2}\left(f(\mathbf{t})^\top ((\tfrac{1}{n}D(\mathbf{t},\mathbf{t};\boldsymbol{\theta}))^{-1} + K(\mathbf{t},\mathbf{t};\boldsymbol{\theta})^{-1}) f(\mathbf{t}) \; - \; 2f(\mathbf{t})^\top((\tfrac{1}{n}D(\mathbf{t},\mathbf{t};\boldsymbol{\theta}))^{-1}{\bar{\bf y}} + K(\mathbf{t},\mathbf{t};\boldsymbol{\theta})^{-1}\mu(\mathbf{t};\mathbf{\theta}))\right)}, \\
\end{aligned}
$$
We now introduce the following terms:  the posterior covariance $K_p(\mathbf{t}, \mathbf{t};\boldsymbol{\theta})$ and $\mathbf{u}$, 
$$
\begin{aligned}
&K_p(\mathbf{t}, \mathbf{t};\boldsymbol{\theta}) = \left(\left(\tfrac{1}{n}D(\mathbf{t},\mathbf{t};\boldsymbol{\theta})\right)^{-1} + K(\mathbf{t},\mathbf{t};\boldsymbol{\theta})^{-1}\right)^{-1} = K(\mathbf{t}, \mathbf{t};\boldsymbol{\theta})- K(\mathbf{t}, {\bf t};\boldsymbol{\theta})V_{\boldsymbol{\theta}}^{-1}K({\bf t}, \mathbf{t};\boldsymbol{\theta}); \\
&\mathbf{u} = \left(\tfrac{1}{n}D(\mathbf{t},\mathbf{t};\boldsymbol{\theta})\right)^{-1}{\bar{\bf y}} + K(\mathbf{t},\mathbf{t};\boldsymbol{\theta})^{-1}\mu(\mathbf{t};\mathbf{\theta}), \\
\end{aligned}
$$
where we have used  the Woodbury matrix identity \citep{max1950inverting} to rewrite $K_p(\mathbf{t}, \mathbf{t};\boldsymbol{\theta}) $.

Finally, we can rewrite the posterior distribution as:
$$
\begin{aligned}
&p(f(t)|\{y^{(j)}(t_1),\dots,y^{(j)}(t_p)\}_{j=1}^{n},\boldsymbol{\theta})\\
&\propto e^{-\frac{1}{2}\left( (f(\mathbf{t})- K_p(\mathbf{t}, \mathbf{t};\boldsymbol{\theta})\mathbf{u})^\top K_p(\mathbf{t},\mathbf{t};\boldsymbol{\theta})^{-1} (f(\mathbf{t})- K_p(\mathbf{t}, \mathbf{t};\boldsymbol{\theta})\mathbf{u}) \; - \; \mathbf{u}^\top K_p(\mathbf{t}, \mathbf{t};\boldsymbol{\theta}) \mathbf{u}\right)} ,\\
&\propto e^{-\frac{1}{2}\left( (f(\mathbf{t})- K_p(\mathbf{t}, \mathbf{t};\boldsymbol{\theta})\mathbf{u})^\top K_p(\mathbf{t}, \mathbf{t};\boldsymbol{\theta})^{-1} (f(\mathbf{t})- K_p(\mathbf{t}, \mathbf{t};\boldsymbol{\theta})\mathbf{u})\right)} ,\\
&\propto  N(f(\mathbf{t}); K_p(\mathbf{t}, \mathbf{t};\boldsymbol{\theta})\left(\big(\tfrac{1}{n}D(\mathbf{t},\mathbf{t};\boldsymbol{\theta})\big)^{-1}{\bar{\bf y}} + K(\mathbf{t},\mathbf{t};\boldsymbol{\theta})^{-1}\mu(\mathbf{t};\mathbf{\theta})\right), K_p(\mathbf{t}, \mathbf{t};\boldsymbol{\theta})) \;.
\end{aligned}
$$
We can rearrange the posterior mean into the form defined by \eqref{eq:postmean}  by substituting the definition of the posterior covariance and simplifying

$$
\begin{aligned}
    &m_p(t;\boldsymbol{\theta}) = (K(\mathbf{t},\mathbf{t};\boldsymbol{\theta}) - K(\mathbf{t},\mathbf{t};\boldsymbol{\theta})(\tfrac{1}{n}D(\mathbf{t},\mathbf{t};\boldsymbol{\theta}) + K(\mathbf{t},\mathbf{t};\boldsymbol{\theta}))^{-1}K({\bf t}, \mathbf{t};\boldsymbol{\theta}))\left((\tfrac{1}{n}D(\mathbf{t},\mathbf{t};\boldsymbol{\theta}))^{-1}{\bar{\bf y}} + K(\mathbf{t},\mathbf{t};\boldsymbol{\theta})^{-1}\mu(\mathbf{t};\mathbf{\theta})\right) ~, \\
    &= \mu(\mathbf{t};\mathbf{\theta}) \; - \;  K(\mathbf{t},\mathbf{t};\boldsymbol{\theta})(\tfrac{1}{n}D(\mathbf{t},\mathbf{t};\boldsymbol{\theta}) + K(\mathbf{t},\mathbf{t};\boldsymbol{\theta}))^{-1}\mu(\mathbf{t};\mathbf{\theta}) \; + \;  \\
    &K(\mathbf{t},\mathbf{t};\boldsymbol{\theta})(\tfrac{1}{n}D(\mathbf{t},\mathbf{t};\boldsymbol{\theta}))^{-1}{\bar{\bf y}}(\mathbf{I} - K(\mathbf{t},\mathbf{t};\boldsymbol{\theta})(\tfrac{1}{n}D(\mathbf{t},\mathbf{t};\boldsymbol{\theta}) + K(\mathbf{t},\mathbf{t};\boldsymbol{\theta}))^{-1}) ~, \\
    &=\mu(\mathbf{t};\mathbf{\theta}) \; - \;  K(\mathbf{t},\mathbf{t};\boldsymbol{\theta})(\tfrac{1}{n}D(\mathbf{t},\mathbf{t};\boldsymbol{\theta}) + K(\mathbf{t},\mathbf{t};\boldsymbol{\theta}))^{-1}\mu(\mathbf{t};\mathbf{\theta}) \; + \;  K(\mathbf{t},\mathbf{t};\boldsymbol{\theta})(\tfrac{1}{n}D(\mathbf{t},\mathbf{t};\boldsymbol{\theta}) + K(\mathbf{t},\mathbf{t};\boldsymbol{\theta}))^{-1}{\bar{\bf y}} ~,\\
    &=\mu(\mathbf{t};\mathbf{\theta}) \; + \; K(\mathbf{t},\mathbf{t};\boldsymbol{\theta})(\tfrac{1}{n}D(\mathbf{t},\mathbf{t};\boldsymbol{\theta}) + K(\mathbf{t},\mathbf{t};\boldsymbol{\theta}))^{-1}({\bar{\bf y}} - \mu(\mathbf{t};\mathbf{\theta})) ~,\\
    &= \mu(\mathbf{t};\mathbf{\theta}) \; + \; K(\mathbf{t},\mathbf{t};\boldsymbol{\theta})V_{\boldsymbol{\theta}}^{-1}({\bar{\bf y}} - \mu(\mathbf{t};\mathbf{\theta})).
\end{aligned}
$$

which allows us to prove Equation \eqref{eq:postprop1}. Finally, the marginal likelihood of the observations given the hyperparameters can be obtained from:

$$
\begin{aligned}
    &p(\{y^{(j)}(t_1),\dots,y^{(j)}(t_p)\}_{j=1}^{n}|\boldsymbol{\theta})=\int p(\{y^{(j)}(t_1),\dots,y^{(j)}(t_p)\}_{j=1}^{n}|f(\mathbf{t}),\boldsymbol{\theta})p(f(\mathbf{t})|\boldsymbol{\theta})df(\mathbf{t}) \\
    &=\int \tfrac{e^{0.5(-\sum_{j=1}^n (\mathbf{ y}^{(j)})^\top(D(\mathbf{t},\mathbf{t};\boldsymbol{\theta}))^{-1}\mathbf{y}^{(j)}+n\bar{\bf y}^\top (D(\mathbf{t},\mathbf{t};\boldsymbol{\theta}))^{-1}\bar{\bf y) }}}{n^{\frac{p}{2}}(|2\pi D(\mathbf{t},\mathbf{t};\boldsymbol{\theta})|)^{\frac{n-1}{2}}} N\left({\bar{\bf y}};f(\mathbf{t}),\tfrac{1}{n}D(\mathbf{t},\mathbf{t};\boldsymbol{\theta})\right)N\left(f(\mathbf{t});\mu(\mathbf{t};\mathbf{\theta}), K(\mathbf{t}, \mathbf{t}; \mathbf{\theta})\right)df(\mathbf{t}),\\
    &= C\int e^{-\frac{1}{2}((\bar{\bf y} - f(\mathbf{t}))^\top (\frac{1}{n}D(\mathbf{t},\mathbf{t};\boldsymbol{\theta}))^{-1} (\bar{\bf y} - f(\mathbf{t})) + (f(\mathbf{t}) - \mu(\mathbf{t};\mathbf{\theta}))^\top K(\mathbf{t},\mathbf{t};\boldsymbol{\theta})^{-1} (f(\mathbf{t}) - \mu(\mathbf{t};\mathbf{\theta})))}df(\mathbf{t}),
\end{aligned}
$$
where 
$$
\begin{aligned}
    C = \tfrac{e^{0.5(-\sum_{j=1}^n (\mathbf{ y}^{(j)})^\top(D(\mathbf{t},\mathbf{t};\boldsymbol{\theta}))^{-1}\mathbf{y}^{(j)}+n\bar{\bf y}^\top (D(\mathbf{t},\mathbf{t};\boldsymbol{\theta}))^{-1}\bar{\bf y) }}}{(|2\pi D(\mathbf{t},\mathbf{t};\boldsymbol{\theta})|)^\frac{n}{2} (|2\pi K(\mathbf{t},\mathbf{t};\boldsymbol{\theta})|)^\frac{1}{2}}.
\end{aligned}
$$
The exponential function within the integral can be expanded to: 

$$
\begin{aligned}
    e^{-\frac{1}{2}(f(\mathbf{t})^\top((\frac{1}{n}D(\mathbf{t},\mathbf{t};\boldsymbol{\theta}))^{-1} + K(\mathbf{t},\mathbf{t};\boldsymbol{\theta})^{-1})f(\mathbf{t}) - 2f(\mathbf{t})^\top((\frac{1}{n}D(\mathbf{t},\mathbf{t};\boldsymbol{\theta}))^{-1}\bar{\bf y} + K(\mathbf{t},\mathbf{t};\boldsymbol{\theta})^{-1}\mu(\mathbf{t};\mathbf{\theta}))
    + \bar{\bf y}^\top(\frac{1}{n}D(\mathbf{t},\mathbf{t};\boldsymbol{\theta}))^{-1}\bar{\bf y} + \mu(\mathbf{t};\mathbf{\theta})^\top K(\mathbf{t},\mathbf{t};\boldsymbol{\theta})^{-1} \mu(\mathbf{t};\mathbf{\theta}))}.
\end{aligned}
$$
Now we define
$$
\begin{aligned}
    &\Sigma = \left(\left(\frac{1}{n}D(\mathbf{t},\mathbf{t};\boldsymbol{\theta})\right)^{-1} + K(\mathbf{t},\mathbf{t};\boldsymbol{\theta})^{-1}\right)^{-1}, \\
    &\mathbf{z} = \Sigma\left(\left((\frac{1}{n}D(\mathbf{t},\mathbf{t};\boldsymbol{\theta})\right)^{-1}\bar{\bf y} + K(\mathbf{t},\mathbf{t};\boldsymbol{\theta})^{-1}\mu(\mathbf{t};\mathbf{\theta})\right),
\end{aligned}
$$
and complete the square to get 
$$
\begin{aligned}
    C\int e^{-\frac{1}{2}(f(\mathbf{t}) - \mathbf{z})^\top \Sigma^{-1} (f(\mathbf{t}) - \mathbf{z})}e^{-\frac{1}{2}(\bar{\bf y}^\top(\frac{1}{n}D(\mathbf{t},\mathbf{t};\boldsymbol{\theta}))^{-1}\bar{\bf y} + \mu(\mathbf{t};\mathbf{\theta})^\top K(\mathbf{t},\mathbf{t};\boldsymbol{\theta})^{-1} \mu(\mathbf{t};\mathbf{\theta}) - \mathbf{z}^\top\Sigma^{-1}\mathbf{z}))} df(\mathbf{t}).
\end{aligned}
$$

The constant terms with respect to $f(\mathbf{t})$ can be factored out of the integral, leaving the integral equal to $\sqrt{(2\pi)^p|\Sigma|}$, which can be combined with $C$ and simplified to form,

$$
\begin{aligned}
&\tfrac{e^{-\frac{1}{2}(\bar{\bf y}^\top(\frac{1}{n}D(\mathbf{t},\mathbf{t};\boldsymbol{\theta}))^{-1}\bar{\bf y} + \mu(\mathbf{t};\mathbf{\theta})^\top K(\mathbf{t},\mathbf{t};\boldsymbol{\theta})^{-1} \mu(\mathbf{t};\mathbf{\theta}) - z^\top\Sigma^{-1}z))}e^{0.5(-\sum_{j=1}^n (\mathbf{ y}^{(j)})^\top(D(\mathbf{t},\mathbf{t};\boldsymbol{\theta}))^{-1}\mathbf{y}^{(j)}+n\bar{\bf y}^\top (D(\mathbf{t},\mathbf{t};\boldsymbol{\theta}))^{-1}\bar{\bf y) }} (|(\frac{1}{n}D(\mathbf{t},\mathbf{t};\boldsymbol{\theta}) + K(\mathbf{t},\mathbf{t};\boldsymbol{\theta}))^{-1}|)^\frac{1}{2}}{(|2\pi D(\mathbf{t},\mathbf{t};\boldsymbol{\theta})|)^\frac{n}{2} (|2\pi K(\mathbf{t},\mathbf{t};\boldsymbol{\theta})|)^\frac{1}{2}} \\
\end{aligned}
$$

By substituting in the original definitions of $\Sigma$ and $\mathbf{z}$ and exploiting the Woodbury Matrix Identity,
$$
\begin{aligned}
    &\left(\left(\frac{1}{n}D(\mathbf{t},\mathbf{t};\boldsymbol{\theta})\right)^{-1} + K(\mathbf{t},\mathbf{t};\boldsymbol{\theta})^{-1}\right)^{-1}\\
    &= \left(\frac{1}{n}D(\mathbf{t},\mathbf{t};\boldsymbol{\theta})\right) - \left(\frac{1}{n}D(\mathbf{t},\mathbf{t};\boldsymbol{\theta})\right)\left(\frac{1}{n}D(\mathbf{t},\mathbf{t};\boldsymbol{\theta}) + K(\mathbf{t},\mathbf{t};\boldsymbol{\theta})\right)^{-1}\left(\frac{1}{n}D(\mathbf{t},\mathbf{t};\boldsymbol{\theta})\right) \\
    &= K(\mathbf{t},\mathbf{t};\boldsymbol{\theta}) - K(\mathbf{t},\mathbf{t};\boldsymbol{\theta})\left(K(\mathbf{t},\mathbf{t};\boldsymbol{\theta}) + \frac{1}{n}D(\mathbf{t},\mathbf{t};\boldsymbol{\theta})\right)^{-1}K(\mathbf{t},\mathbf{t};\boldsymbol{\theta})
\end{aligned}
$$
we arrive at the final form of the marginal likelihood 
$$
\begin{aligned}
    &p(\{y^{(j)}(t_1),\dots,y^{(j)}(t_p)\}_{j=1}^{n}|\boldsymbol{\theta})\\
    &=e^{\frac{-\sum_{j=1}^n (\mathbf{ y}^{(j)})^\top(D(\mathbf{t},\mathbf{t};\boldsymbol{\theta}))^{-1}\mathbf{y}^{(j)}+n\bar{\bf y}^\top (D(\mathbf{t},\mathbf{t};\boldsymbol{\theta}))^{-1}\bar{\bf y}}{2}} \frac{1}{(2\pi)^\frac{p(n-1)}{2} n^\frac{p}{2}|D(\mathbf{t},\mathbf{t};\boldsymbol{\theta})|^\frac{n-1}{2} (2\pi)^\frac{p}{2} (|\frac{1}{n}D(\mathbf{t},\mathbf{t};\boldsymbol{\theta}) + K(\mathbf{t},\mathbf{t};\boldsymbol{\theta})|)^\frac{1}{2}}\\
    &e^{-\frac{1}{2}(\bar{\bf y} - \mu(\mathbf{t};\mathbf{\theta}))^\top(\frac{1}{n}D(\mathbf{t},\mathbf{t};\boldsymbol{\theta})+K(\mathbf{t},\mathbf{t};\boldsymbol{\theta}))^{-1}(\bar{\bf y} - \mu(\mathbf{t};\mathbf{\theta}))}\\
    &=\frac{1}{(2\pi)^\frac{p(n-1)}{2} n^\frac{p}{2}|D(\mathbf{t},\mathbf{t};\boldsymbol{\theta})|^\frac{n-1}{2}} N({\bar{\bf y}};\mu(\mathbf{t};\boldsymbol{\theta}),V_{\boldsymbol{\theta}})e^{\frac{-\sum_{j=1}^n (\mathbf{ y}^{(j)})^\top(D(\mathbf{t},\mathbf{t};\boldsymbol{\theta}))^{-1}\mathbf{y}^{(j)}+n\bar{\bf y}^\top (D(\mathbf{t},\mathbf{t};\boldsymbol{\theta}))^{-1}\bar{\bf y}}{2}}.
\end{aligned}
$$

\setcounter{table}{0}
\begin{table}[!h]
\caption{Real dissolution dataset 1, including the reference and test groups, as reported in \citep{ocana_using_2009}.}
\centering
  \begin{tabular}{|c|cccccccc|c|}
    \hline
    \multicolumn{9}{|c|}{\bf{Real Dissolution Dataset 1}} \\[5pt]
    \hline
    Group & Time = 1 & 2 & 3 & 4 & 5 & 6 & 7 & 8 \\[5pt]
    \hline
    Ref. &    19.78&	37.61&	48.53&	60.62&	63.34&	68.72&	75.76&	83.42\\
    Ref. & 22.05&	36.74&	46.12&	53.9&	61.35&	67.35&	72.85&	77.88\\
    Ref. & 21.5&	34.58&	44.88&	53.46&	61.09&	66.83&	73.11&	77.28\\
    Ref. & 20.24&	35.23&	50.27&	66.49&	75.87&	79.84&	83.33&	87.57\\
    Ref. & 17.66&	33.51&	47.97&	59.29&	67.52&	70.05&	77.04&	80.88\\
    Ref. & 18.16&	33.75&	49.02&	58.26&	71.63&	74.98&	80.38&	84.33\\
    Ref. & 20.94&	33.31&	47.40&	59.26&	68.65&	75.82&	79.98&	82.96\\
    Ref. & 19.43&	31.4&	44.18&	56.2&	67.36&	75.03&	79.7&	83.98\\
    Ref. & 19.12&	32.93&	47.79&	60.19&	69.71&	76.26&	80.84&	83.95\\
    Ref. & 19.76&	31.67&	44.92&	56.79&	66.83&	74.41&	78.77&	82.30\\
    Ref. & 20.64&	32.42&	45.80&	58.11&	69.46&	76.81&	81.57&	85.83\\
    Ref. & 19.22&	32.14&	46.60&	58.16&	64.18&	75.79&	81.49&	83.94\\
    \hline
    Test & 35.70&	48.81&	57.05&	65.01&	71.92&	75.32&	80.64&	83.52\\
    Test & 36.10&	52.88&	61.79&	67.54&	74.47&	78.71&	80.71&	83.23\\
    Test & 41.66&	52.52&	63.22&	70.21&	76.71&	80.00&	83.25&	85.61\\
    Test & 35.49&	50.39&	58.99&	65.68&	72.59&	76.58&	80.62&	83.00 \\
    Test & 36.06&	52.09&	63.46&	69.84&	75.23&	78.57&	81.53&	84.37\\
    Test & 35.53&	51.63&	62.59&	68.68&	75.64&	79.43&	82.99&	85.81\\
    Test & 32.70&	46.85&	57.51&	65.00&	71.26&	75.24&	79.18&	81.21\\
    Test & 33.85&	44.89&	53.64&	60.87&	67.34&	71.18&	73.51&	76.42\\
    Test & 32.33&	46.58&	56.46&	64.38&	70.3&	75.17&	77.83&	79.86\\
    Test & 32.80&	46.37&	56.66&	65.37&	70.04&	75.33&	77.11&	80.45\\
    Test & 33.86&	46.72&	57.45&	65.4&	71.05&	75.78&	76.22&	79.38\\
    Test & 32.46&	45.60&	55.04&	62.78&	68.54&	73.06&	75.73&	77.68\\
    \hline
    \end{tabular}
    \label{real_diss_1}
\end{table}

\begin{table}[!h]
\caption{Real dissolution dataset 2, including the reference and test groups, as reported in \citep{ocana_using_2009}.}
\centering
  \begin{tabular}{|c|cccccccc|c|}
    \hline
    \multicolumn{9}{|c|}{\bf{Real Dissolution Dataset 2}} \\[5pt]
    \hline
    Group & Time = 1 & 2 & 3 & 4 & 5 & 6 & 7 & 8 \\[5pt]
    \hline
    Ref. &    31.35&	43.79&	57.27&	64.37&	68.13&	72.11&	75.26&	77.25\\
    Ref. & 32.20&	41.67&	49.95&	58.45&	65.80&	71.06&	75.20&	76.24\\
    Ref. &32.87&	44.35&	55.16&	61.33&	66.11&	71.99&	74.17&	76.83\\
    Ref. &31.32&	44.59&	56.16&	62.77&	66.11&	71.48&	76.00&	78.52\\
    Ref. &32.00&	48.04&	57.03&	62.72&	67.27&	70.66&	75.01&	76.35\\
    Ref. &33.55&	47.61&	57.51&	64.62&	70.11&	74.16&	76.54&	78.52\\
    Ref. &33.68&	49.83&	57.98&	63.54&	67.95&	71.67&	74.83&	75.66\\
    Ref. &32.04&	41.80&	49.58&	57.14&	63.32&	67.54&	72.76&	74.65\\
    Ref. &34.08&	45.89&	55.13&	63.11&	68.88&	73.44&	77.77&	79.95\\
    Ref. &33.79&	42.13&	52.82&	58.95&	66.03&	70.00&	74.75&	76.95\\
    Ref. &36.46&	47.08&	56.19&	64.46&	70.95&	75.52&	79.79&	80.11\\
    Ref. &33.92&	52.18&	60.85&	66.76&	70.41&	74.78&	79.10&	79.68\\
    \hline
    Test & 33.55&	47.61&	57.51&	64.62&	70.11&	74.16&	76.54&	78.52\\
    Test &32.37&	46.18&	55.08&	61.03&	66.07&	69.77&	71.96&	73.69\\
    Test &34.92&	47.14&	55.53&	63.19&	69.27&	73.64&	75.14&	77.85\\
    Test &33.87&	47.10&	59.04&	67.46&	72.05&	75.87&	79.33&	81.69\\
    Test &33.15&	47.34&	58.49&	65.03&	71.18&	75.92&	77.34&	80.47\\
    Test &33.76&	48.01&	57.88&	65.00&	70.49&	74.49&	76.99&	78.76\\
    Test &32.24&	49.57&	62.25&	70.56&	76.29&	79.21&	81.20&	82.74\\
    Test &36.43&	48.82&	57.34&	64.42&	70.81&	75.03&	76.54&	78.73\\
    Test &32.30&	46.89&	55.87&	60.89&	66.78&	71.05&	72.55&	75.00\\
    Test &30.65&	47.56&	60.08&	68.45&	73.67&	76.20&	78.00&	79.44\\
    Test &36.74&	54.20&	62.46&	69.38&	73.20&	77.87&	80.62&	81.70\\
    Test &33.01&	45.16&	57.69&	64.05&	66.68&	71.38&	76.14&	78.48\\
    \hline
    \end{tabular}
    \label{real_diss_2}
\end{table}

\bibliographystyle{apalike} 
\bibliography{refs}

\end{document}